\documentclass{article}[12pt]
\usepackage{jheppub}
\usepackage[utf8]{inputenc}
\usepackage{amsmath}
\usepackage{amssymb}
\usepackage{amsfonts}
\usepackage{amsthm}
\usepackage{dutchcal}
\usepackage{xcolor}
\usepackage{rotating}
\usepackage{array}
\usepackage{appendix}
\usepackage{bbold}
\usepackage{textcomp}
\def\sc{\textit{sc}}

\def\SO{\textit{SO}}

\def\msj{\iota}

\makeatletter
\def\moverlay{\mathpalette\mov@rlay}
\def\mov@rlay#1#2{\leavevmode\vtop{%
   \baselineskip\z@skip \lineskiplimit-\maxdimen
   \ialign{\hfil$\m@th#1##$\hfil\cr#2\crcr}}}
\newcommand{\charfusion}[3][\mathord]{
    #1{\ifx#1\mathop\vphantom{#2}\fi
        \mathpalette\mov@rlay{#2\cr#3}
      }
    \ifx#1\mathop\expandafter\displaylimits\fi}
\makeatother

\title{Universal Spinning Casimir Equations and Their Solutions}

\author[a]{Ilija Buri\'c}
\author[b]{and Volker Schomerus}
\affiliation[a]{Department of Physics, University of Pisa, Largo Bruno Pontecorvo 3, I-56127 Pisa, Italy}
\affiliation[b]{DESY, Notkestra\ss e 85, D-22607 Hamburg, Germany}

\emailAdd{ilija.buric@df.unipi.it}
\emailAdd{volker.schomerus@desy.de}
\abstract{Conformal blocks are a central analytic tool for higher dimensional conformal field theory. 
We employ Harish-Chandra's radial component map to construct universal Casimir differential equations 
for spinning conformal blocks in any dimension $d$ of Euclidean space. Furthermore, we 
also build a set of differential ``shifting'' operators that allow to construct solutions of the Casimir 
equations from certain seeds. In the context of spinning four-point blocks of bulk conformal field theory, 
our formulas provide an elegant and far reaching generalisation of existing expressions to arbitrary tensor 
fields and arbitrary dimension $d$. The power of our new universal approach to spinning blocks is further
illustrated through applications to defect conformal field theory. In the case of defects of co-dimension 
$q=2$ we are able to construct conformal blocks for two-point functions of symmetric traceless bulk tensor 
fields in both the defect and the bulk channel. This opens an interesting avenue for applications to the 
defect bootstrap. Finally, we also derive the Casimir equations for bulk-bulk-defect three-point functions 
in the bulk channel.}

\begin{document}
\maketitle

\section{Introduction}
\addtolength{\baselineskip}{2pt}
Conformal partial wave (block) expansions of correlation functions are a standard analytical tool in conformal field theory (CFT) that is fundamental for the conformal bootstrap program. While the most basic field in a CFT is the stress tensor, which is a field of spin $l=2$, much of the initial theory was developed for conformal blocks of four scalar fields \cite{Dolan:2003hv,Dolan:2011dv}. Extensions to spinning four-point functions \cite{Costa:2011dw,Costa:2011mg} were driven by the revival of the conformal bootstrap program, see \cite{Poland:2018epd} and references therein. Today, spinning four-point blocks can be evaluated quite efficiently through the use of weight shifting technology \cite{Karateev:2017jgd}, by reducing them recursively to scalar blocks. After some early contributions in \cite{McAvity:1995zd,Liendo:2012hy}, correlation functions of non-local operators, such as defects, interfaces 
and boundaries, have also received increasing attention as interesting probes of non-perturbative dynamics in higher dimensional CFT. Surprisingly little is known about the blocks for correlation functions involving spinning bulk-local fields in the presence of bulk-local and non-local operators, see however \cite{Lauria:2018klo}. Our goal in this work is to advance and simplify the theory of spinning bulk and defect blocks, with a particular focus on defects of co-dimension $q=2$.

Since Dolan and Osborn's influential work on scalar four-point blocks, it is common to characterise and investigate blocks through the differential equations they satisfy. Examples of Casimir differential equations have been worked out for a large number of setups, including spinning four-point functions in $d=3$ and $d=4$ dimensions, see e.g. \cite{Iliesiu:2015akf,Echeverri:2016dun}, as well as defect two-point functions, see \cite{Billo:2016cpy,Gimenez-Grau:2019hez,Gimenez-Grau:2020jvf,Gimenez-Grau:2021wiv}. 

Later, it has been pointed out that the Dolan-Osborn equations are equivalent to an integrable Schr\"odinger problem, namely the Calogero-Sutherland model associated to the root system $BC_2$, \cite{Isachenkov:2016gim}. The link between the two systems goes in two steps, where both partial waves and Calogero-Sutherland wavefunctions are related to the same class of harmonic functions on the conformal group. The appearance of Calogero-Sutherland models in harmonic analysis was the subject of the classic work \cite{Olshanetsky:1983wh}, while the connection to conformal partial waves was understood in \cite{Schomerus:2016epl,Schomerus:2017eny,Isachenkov:2018pef}. Wavefunctions of scalar Calogero-Sutherland Hamiltonians associated with root systems were constructed by Heckman and Opdam starting with \cite{Heckman-Opdam} (see \cite{Isachenkov:2017qgn} for many more details).
\smallskip

The present work starts with the observation that all the systems mentioned above admit a \textit{universal} extension in spin. This vastly generalises and simplifies constructions of \cite{Schomerus:2016epl,Schomerus:2017eny}, that also derived Calogero-Sutherland Hamiltonians for spinning fields, though for selected spin assignments only. The associated 
eigenvalue equations could be mapped to the Casimir equations of \cite{Iliesiu:2015akf,Echeverri:2016dun}. 
Our generalisation is rooted in the harmonic analysis interpretation of partial waves, in terms of so-called spherical functions. The latter have been recognised to be of central importance in harmonic analysis on Lie groups and symmetric spaces since the early works of Gelfand, Harish-Chandra, Godement and others, \cite{Gelfand-spherical,
Godement1952TheoryOS,HarishChandra,Berezin-Karpelevic}. The key property of spherical functions for our purposes is that they are naturally vector-valued (i.e. spinning) and satisfy explicit differential equations. These equations have a simple universal dependence on the spin, thanks to Harish-Chandra's radial component map, \cite{HarishChandra} (resulting generalisations of the Calogero-Sutherland Hamiltonian have been considered in \cite{Stokman:2020bjj,Reshetikhin:2020wep} and termed 'spinning Calogero-Moser' models). With this preparation, we can state the main results of the present work: After deriving universal Casimir equations for any spin assignment and any dimension $d$ from Harish-Chandra's map, we will develop a general and explicit solution theory for these models through an algebra of weight-shifting operators. Namely, eigenfunctions of spinning models will be constructed by applying weight-shifting operators to well-understood scalar eigenfunctions. 
\smallskip 

In the context of CFTs, we may view this result as the completion of the program initiated in \cite{Isachenkov:2016gim} to generate arbitrary four-point conformal blocks by exploiting the underlying integrable structure. The main new CFT setup to which we want to apply our advances concerns spinning bulk two-point functions of symmetric traceless tensors in the presence of a $p$-dimensional defect. Focusing on defects of co-dimension, $q=d-p=2$, we will derive explicit expressions for conformal blocks both in the \textit{bulk channel} and the \textit{defect channel}. Finally, the first part of our analysis, namely the compact expressions for Casimir equations, applies to a much wider class of higher-point correlation functions. In this work, we will derive Casimir equations for what is probably the simplest system of this kind, the correlation function of two bulk fields together with a defect of co-dimension two and a further local field on it. Solution theory for this and other multipoint systems, which are formally very similar to spinning models solved in this work, is left for future research.
\medskip 

\subsection{Results on spinning Casimir and shifting operators}  

Let us now describe the main new results of this work in some more detail. We shall begin with the general and more formal results, leaving a description of the main new applications in defect 
CFT to the next subsection. As we have stated before, the universal formula for the spinning Casimir operators is key to our advances. This formula expresses the action of the Laplace-Beltrami operator on a Lie group $G$ on spherical functions. The latter are defined as vector-valued functions on $G$ which have definite covariance properties under the left and right action of a subgroup $K \subset G$. The pair of groups $(G,K)$ is not arbitrary, but should be a \textit{Gelfand pair}, i.e. the Lie algebra $\mathfrak{k}$ has to be the fixed point set of an involutive automorphism of the Lie algebra $\mathfrak{g}$. Given a Gelfand pair, $G$ admits a Cartan decomposition and spherical functions are fully determined by their dependence on $\text{rank}(G,K)$ variables, rather than $\text{dim}(G)$ of them. The corresponding reduced Laplacian is written in an explicit way in terms of the root system of $(\mathfrak{g},\mathfrak{k})$, see eq. \eqref{Casimir-radial-part}.

In the example relevant for CFTs, the expression we shall derive is surprisingly simple. After an appropriate factor has been split off, eigenfunctions of the spinning Casimir operator take the 
form of wavefunctions for some matrix valued two-particle Hamiltonian $H$. The two coordinates of the particles are denoted by $t_1$ and $t_2$. These two variables $t_i$ are related to conformally invariant cross ratios. In order to construct the operator-valued potential, we shall employ some representation matrices of the conformal Lie 
algebra $\mathfrak{so}(d+1,1)$. Let us denote the generators of this algebra by $L_{\alpha\beta}= -L_{\beta\alpha}$ with $\alpha,\beta = 0,1,\dots,d+1$ with $\alpha = 0$ corresponding to the 
timelike direction, as usual. The generators of spatial rotation are given by $M_{\mu\nu} = L_{\mu\nu} \in \mathfrak{so}(d)$ for $\mu,\nu = 2, \dots, d+1$. In addition, we shall need the 
generator $D = L_{01} \in \mathfrak{so}(1,1)$. Spherical functions are associated with the choice of two finite-dimensional representations $\rho_l,\rho_r$ of the Lie algebra $\mathfrak{k} 
= \mathfrak{so}(d) \oplus \mathfrak{so}(1,1)$. In an abuse of notations, we shall denote the generators of $\mathfrak{k}$ in the representation $\rho_r^\ast$ by $M_{\mu\nu}$, $D$. When we 
evaluate the same generators in $\rho_l$, on the other hand, the representations operators are $-M'_{\mu\nu}$ and $-D'$. With these notations we are now ready to spell out the matrix
valued Schr\"odinger operator $H$. It takes the form 
\begin{align}
     H^{\rho_l,\rho_r} &= \partial_{t_1}^2 + \partial_{t_2}^2 +\frac{1 - D'^2_+ + 2\cosh(t_1+t_2)
     D'_+D_+ -  D^2_+}{2\sinh^2(t_1+t_2)}
     + \frac{1 - D'^2_- + 2\cosh(t_1-t_2)D'_-D_- - D^2_-}{2\sinh^2(t_1-t_2)}\nonumber\\[2mm]
       & + \frac{M'_{2a}M'_{2a}-2\cosh t_1 M'_{2a}M_{2a}+M_{2a}M_{2a}-\frac14(d-2)(d-4)}{\sinh^2 t_1}
       \label{universal-Hamiltonian}\\[2mm]
       & + \frac{M'_{3a}M'_{3a}-2\cosh t_2 M'_{3a}M_{3a}+M_{3a}M_{3a} - \frac14 (d-2)(d-4)}{\sinh^2 t_2}-
       \frac12 L^{ab}L_{ab}- \frac{d^2-2d+2}{2}\ .\nonumber
\end{align}
Here $D_\pm = D \pm i M_{23}$ and the indices $a,b=4, \dots, d+1$ are summed over. The Hamiltonian $H$ acts on the tensor product $W_l \otimes W_r$ of the carrier spaces $W_l$ and $W_r$ of the representations $\rho_l$ and $\rho_r^\ast$. Since the spaces $W_l$ and $W_r$ carry representations of the rotation generators $M'$ and $M$, we may think of them as describing the spin degrees of freedom of our two particles. The potential energy of each particle depends on the spin. In addition, there are also interaction terms that couple the spin matrices $M'$ of the first particle to those of the second. In case the spin representations are both trivial, we can set all the matrices $M= 0 = M'$ to zero. The resulting Hamiltonian is the usual Schr\"odinger operator of the hyperbolic Calogero-Sutherland model for the root system $\textit{BC}_2$. For this scalar case, the complete solution theory is known, see e.g. \cite{Heckman-Opdam,Isachenkov:2017qgn}. Solutions 
of the spinning problem have not been constructed in general. But as we will show below, large classes of wavefunctions can be constructed from those of the scalar model by the application of certain weight-shifting  operators. 

There are two constructions to build solutions of spinning Calogero-Sutherland models that we shall explore. The first one exploits left and right invariant vector fields. Since such vector fields do not commute with left and right action of the spherical subgroup $K$, respectively, their application does modify the covariance properties of the function on the conformal group. On the other hand, left and right invariant vector fields commute with the Laplacian. Hence, the action of these first order differential operators does respect the decomposition into eigenfunctions of the Laplacian. After reduction, the vector fields must therefore turn into matrix valued first order differential operators in $t_1$, $t_2$ that map eigenfunctions of the 
Hamiltonian \eqref{universal-Hamiltonian} to eigenfunctions of the same universal operator with same eigenvalue, but with $\rho_{l/r}$ replaced by $\rho_{l/r} \otimes \pi$. Here $\pi$ denotes the restriction of the adjoint representation of $\mathfrak{g}$ to the subalgebra $\mathfrak{k}\subset\mathfrak{g}$. In particular, the decomposition of $\pi$ into irreducible components determines the number of shifting operators that we shall construct. Our method realises the suggestion made in \cite{Schomerus:2017eny} to obtain shifting operators from vector fields. We will turn these into concrete first order matrix differential operators in the variables $t_1$, $t_2$ using the Harish-Chandra radial component map.

The differential shift operators discussed in the previous paragraph shift the external parameters, or more concretely the Calogero-Sutherland potential, while leaving 
the eigenvalues unaltered. There exists a second type of operators that allow to shift the eigenvalues while keeping the potential invariant. These are obtained with the 
help of so-called zonal spherical functions. By definition, zonal-spherical functions are spherical functions with trivial left and right representations $\rho_l$ and $\rho_r$, i.e. eigenfunctions of the scalar Calogero-Sutherland Hamiltonian. For special discrete choices of the eigenvalues which are associated with (non-unitary) finite dimensional representations of $\mathfrak{g}$, these functions are polynomial. Now, given any eigenfunction of the universal spinning Casimir operator \eqref{universal-Hamiltonian} with eigenvalue/energy $\varepsilon$, its product with a polynomial zonal spherical function turns out to decompose into a finite sum of eigenfunctions of the spinning Hamiltonian with the same representations $\rho_{l,r}$ but different eigenvalues. The number $\varpi$ of different energies that appear and their precise values $\varepsilon_k$, $k=1, \dots, \varpi$, is determined by elementary group theory. Note that these data depends on $\rho_{l,r}$, the eigenvalue $\varepsilon$ and the choice of zonal spherical function. Once $\varepsilon_k$ are known, one can form the following operators of order $2K-2$, 
\begin{equation}\label{eq:weightshifting}
 S^{\rho_l,\rho_r}_m := \prod_{k\neq m} \frac{H^{\rho_l,\rho_r} - \varepsilon_m}
 {\varepsilon_k - \varepsilon_m} \ .      
\end{equation}
When these operators are applied to the product of an eigenfunction of the universal $H$ with the zonal spherical functions, it returns an eigenfunction of $H^{\rho_l,\rho_r}$
with eigenvalue $\varepsilon_m$. Whenever $\varepsilon_m \neq \varepsilon$ the operator $S_m$ shifts the eigenvalue without altering the external parameters/potential. Note 
that all it takes to construct these internal or weight shifting operators is our expression for the universal spinning Casimir operators \eqref{universal-Hamiltonian}, 
along with some simple group theory. 

So far we thought of the universal spinning Casimir operators as well as the shifting operators as taking values in various spaces of matrices. In practise, however, we shall often 
realise the representation matrices $M_{\mu\nu}$ of the Lie algebra $\mathfrak{so}(d)$ as differential operators that act on some space of polynomials. To construct irreducible 
symmetric traceless tensor representations of $\mathfrak{so}(d)$, for example, one can start with a set of $d$ complex coordinates $\zeta_\mu$, $\mu = 1, \dots, d$, and impose 
the constraint $\zeta^2 = \sum_{\mu=1}^d \zeta_\mu^2 = 0$. It is well known that the carrier space $W_J$ for the symmetric traceless tensors of rank $J$ can be realised on 
the space $W_J$ of homogeneous polynomials of order $J$, restricted to the submanifold $\zeta^2=0$. The constraint can be used, for example, to ensure the polynomials are of 
the form  $p(\zeta) = p_0(\zeta') + \zeta_1 p_1(\zeta')$ where $p_0$ and $p_1$ are homogeneous polynomials in the variables $\zeta_2, \dots \zeta_d$ of order $J$ and $J-1$, 
respectively. With such a realisation of the generators $M_{\mu\nu}$ as differential operators in $\zeta_\mu$ in mind, the eigenfunctions of the Hamiltonian should also 
be thought of as polynomials in the variables $\zeta_\mu$ rather then vector valued objects. It is these realisations that will allow us to write reductions of Casimir elements and invariant vector fields as differential operators in a small number of 'invariant spin variables', see e.g. \eqref{eq:HCSMST3scalars}, \eqref{shift-q-bulk-channel}, \eqref{shift-p-bulk-channel}. As a result, reduced Casimir and shifting operators all act on the same space and form an algebraic structure given by exchange relations such as \eqref{bulk-channel-shifting}, \eqref{relation-shifting}. 

\subsection{A guided tour to applications in defect CFT}  

In applications to CFT, the eigenfunctions $\Psi$ of the Hamiltonian $H$ are related to the building blocks of a correlation function $G(x_i)$ as 
\begin{equation} \label{eq:GXiPsi}
G(x_i) = \Xi(x_i) \Psi(t_1(x_i),t_2(x_i))\ . 
\end{equation} 
Here $x_i$ denote insertion points of some local fields. These are acted upon by the conformal Lie algebra $\mathfrak{so}(d+1,1)$ and $t_1(x_i)$, $t_2(x_i)$ are two conformally invariant cross 
ratios one can build from $x_i$. The precise functional dependence of $t_1$, $t_2$ on $x_i$ depends on the particular correlation function we consider, as does the precise form of the matrix valued prefactor $\Xi(x_i)$. In early work on the harmonic analysis approach to conformal blocks, see \cite{Schomerus:2016epl,Schomerus:2017eny}, the cross ratios and the factor $\Xi(x_i)$ could only be fixed in cases in which the Casimir equations for conformal blocks had been worked out already. This changed with \cite{Buric:2019dfk,Buric:2020buk}, where a systematic group theoretic construction of the cross ratios $t_i$ and the matrix factor $\Xi(x_i)$ was developed. Here we shall adapt this approach to all the cases under consideration. In the case of spinning bulk four-point functions, the variables $t_i$ are constructed in eq.\ \eqref{eq:ztcoordinates4pt} and the prefactor $\Xi$ can be read off from eq.\ \eqref{eq:4ptGFnew}. The derivation we review below is essentially taken from our previous work \cite{Buric:2020buk}.
\medskip 

Most of our new results on the explicit construction of spinning conformal blocks concern 
correlation functions of two spinning bulk fields in the presence of a defects of co-dimension 
$q=2$. In this case, the Harish-Chandra radial component map will allow us to construct all 
blocks for two-point functions of spinning bulk fields in symmetric traceless tensor 
representations. To set the stage, we consider a conformal defect of dimensions $p=d-2$ in 
a $d$-dimensional Euclidean space along with two spinning bulk fields $\varphi_1$ and 
$\varphi_2$ with conformal weights $\Delta_1$ and $\Delta_2$ that are inserted at points 
$x_1, x_2 \in \mathbb{R}^d$, respectively. We shall assume the bulk fields to transform 
as a symmetric traceless tensor of spin $J_i$. The associated correlation function reads 
\begin{equation} \label{eq:2pointdefect} 
G_{2,0}(x_i) = \langle \varphi_1(x_1) \varphi_2(x_2) \mathcal{D}^{(q=2)}(\mathcal{X})\rangle\ .
\end{equation} 
The fields $\varphi_i$ take values in the finite dimensional carrier space $W_{J_i}$ of the 
irreducible symmetric traceless tensor representation for spin $J$. As we recalled 
above, one way to realise this vector space $W_J$ is through the space of homogeneous polynomials of order $J$ in $d$ variables $\zeta_\mu$ subject to the constraint 
$\zeta^2=0$.  
 
The correlation function \eqref{eq:2pointdefect} can be evaluated in two different ways. In the so-called defect channel, one first performs the bulk-to-defect operator product 
expansion of the two bulk fields. The resulting two-point function of defect fields is determined by conformal symmetry. For the bulk channel, on the other hand, one 
expands the product of the two spinning bulk fields using the bulk operator product expansion. Once again, each term in this expansion is then fixed 
by conformal symmetry, up to one constant prefactor. These two channels give rise to two different expansions in terms of defect and bulk blocks. Our goal is to find 
explicit formulas for these blocks. We shall describe the two channels separately now. 
\medskip 

\noindent 
\paragraph{Guide to defect channel blocks.} Let us address the defect channel blocks first. If the bulk fields are symmetric traceless tensor fields, the defect fields that 
can appear in the bulk-to-defect operator product are symmetric traceless tensor fields of the defect rotation group $\SO(d-2)$. In addition, they can carry 
arbitrary transverse spin $s$, i.e. they can transform in any of the irreducible representations of group $\SO(2)$ of transverse rotations. Hence the defect fields 
$\hat\varphi$ carry three quantum numbers $(\hat \Delta, \hat \ell, s)$. The bulk to defect operator product for the setup we consider involves the choice of a tensor 
structure which one can label by an integer $l = \ell, \dots, J$. So, in total we expect the relevant blocks to depend on five quantum numbers. As we shall show below, these blocks can be factorised as 
\begin{equation} 
G^{l,l'}_{\hat \Delta, \hat \ell, s} (x_i) = 
\Xi^{l,l'}_{\hat \Delta, \hat \ell, s}(x_i) 
\psi^{l,l'}_{\hat \Delta,\hat \ell} (\lambda,y) 
e^{i \kappa s}\,,
\end{equation} 
into an $x_i$-dependent prefactor $\Xi$ and a function of three cross ratios.  
The two cross ratios $\lambda$ and $\kappa$ are obtained from the insertion points 
$x_i$ through eq.\ \eqref{eq:lambdakappa}. To discuss the remaining invariant
we note that the relevant spherical functions $\psi$ take values in the irreducible 
representations space $W_{l}$ and $W_{l'}$ of $\SO(d-1)$. These representations spaces
can be realised through polynomials in $d-3$ variables $\xi_A$ and $\xi'_A$ with $A= 
3,\dots,d-1$. Explicit formulas for the action of the Lie algebra $\mathfrak{so}(d-1)$ 
on the coordinates $\xi_A$ are given in eqs. \eqref{eq:rhol1}, \eqref{eq:rhol2}. Consistency 
actually requires that $\psi$ takes values in the subspace of $\SO(d-2)$-invariants 
within $W_{l} \otimes W_{l'}$. When translated into the dependence on the variables
$\xi_A$ and $\xi'_A$, the requirements we just described imply that $\psi$ 
can only depend on the combination 
\begin{equation} \label{eq:defy}
 y = \frac{(\xi^2+1)(\xi^{'2}+1)-4\xi\cdot \xi'}
 {(\xi^{2}-1)(\xi^{'2}-1)} \ .  
\end{equation} 
The prefactor $\Xi$ is constructed in section 4.1.1 below, using ideas and 
constructions from \cite{Isachenkov:2018pef} and \cite{Buric:2020zea}, properly
extended to spinning bulk fields. The final formula is stated in eq.\ 
\eqref{eq:G20spherical}. 
\smallskip 

Here we mostly want to focus on the construction of the special functions $\psi^{l,l'}_{\hat\Delta,\hat \ell}(\lambda,y)$. These are eigenfunctions 
of the second order (Hamiltonian) differential operators spelled out in eqs.\ \eqref{defect-channel-Laplacian} and \eqref{Gegenbauer-diff-op}
for eigenvalue \eqref{eigenvalue-defect-Casimir}. The Hamiltonian is the image of the second order Casimir element of the defect conformal group 
$\SO(d-1,1)$ under the Harish-Chandra radial component map. For $l=l'=\hat\ell=0$ the relevant solution can be constructed easily in terms of 
Gauss' hypergeometric functions, see eq.\ \eqref{zonal-spherical-defects}. 

The solutions for non-vanishing $l,l'$ and $\hat\ell$ require a bit more work. 
To obtain these we introduce two differential shift operators in eq.\ 
\eqref{defect-channel-weight-shifting}. These operators may be considered 
as images of the left and right derivatives on the defect conformal group 
$\SO(d-1,1)$ under the radial component map. As we shall show, these two 
operators allow to raise the labels $l,l'$ by one unit, i.e. 
\begin{equation} 
q^{l,l'} \psi^{l,l'}_{\hat \Delta,\hat \ell}(\lambda,y) = \psi^{l+1,l'}_{\hat\Delta,\hat\ell}(\lambda,y) \ , \quad 
\bar q^{l,l'} \psi^{l,l'}_{\hat \Delta,\hat \ell}(\lambda,y) = \psi^{l,l'+1}_{\hat\Delta,\hat \ell}(\lambda,y)\ .  
\end{equation} 
These two raising operators suffice to construct all the special functions $\psi$ from the `ground states' $\psi^{\hat \ell,\hat \ell}_{\hat \Delta,\hat\ell}$ where the indices $l,l'$ assume the minimal allowed value $\hat\ell$.

To obtain these ground states we employ the idea that was sketched before eq.\ \eqref{eq:weightshifting}. In the present setup, we start from products of the form 
\begin{equation} \label{eq:product} 
\psi^{\hat \ell,\hat \ell}_{\hat \Delta,0}(\lambda,y) \psi^{0,0}_{-\hat \ell,0}(\lambda)\ .
\end{equation} 
Note that the first factor can be obtained from the known ground states $\psi^{0,0}_{\hat \Delta,0}$ by applications of $\hat \ell$ raising operators $q,\bar q$. The second factor, on the other hand, is a zonal spherical function $\psi^{0,0}_{\hat \Delta, 0}(\lambda)$ that is continued to $\hat \Delta = - \hat \ell$. The product turns out to be finite a linear combination of eigenfunctions of the Hamiltonian and one of the summands is the desired ground state $\psi^{\hat \ell,\hat \ell}_{\hat \Delta,\hat \ell}$. We can project to the latter using an operator of the form \eqref{eq:weightshifting}, see our discussion around eq.\ \eqref{internal-projection} for more detail. This concludes the construction of the special functions $\psi$. As far as we know, the construction of the spherical functions $\psi$ we carry out here was not described in the mathematical literature before. The techniques we employ are closely related to the differential 
operators and weight shifting techniques in CFT, though everything is carried out for functions of the cross ratios rather than functions of the insertion points $x_i$ and requires no Clebsch-Gordan coefficients and the like. 
\smallskip 

\noindent 
\paragraph{Guide to bulk channel blocks.} Let us now turn attention to the bulk channel. As we mentioned above, the second way to evaluate the correlator \eqref{eq:2pointdefect} is to perform the operator product of the two bulk fields first. A priory, the resulting fields sit in mixed symmetry tensors of the rotation group $\SO(d)$. But only fields in a symmetric traceless tensor representation of $\SO(d)$ can couple to a defect of co-dimension $q=2$. Hence the exchanged bulk fields carry two quantum numbers $(\Delta,J)$ only. These are complemented by three integers that characterise the choice of a tensor structure for three-point functions of STT fields. We denote these by three integers $(j,\msj)$ and $m$ where $(j,\msj)$, $j \geq \msj,$ labels an 
irreducible representation of $\SO(d)$ that can appear in the tensor product $(J_1)\otimes(J_2)$ and $m=0, \dots, j-\msj$. As before, we split the associated blocks as 
\begin{equation} 
G^{j,\msj}_{\Delta,J, m} (x_i,\zeta_i) = \Xi^{j,\msj}_{\Delta, J, m}(x_i,\zeta_i) \psi^{j,\msj}_{\Delta,J,m} (t_1,t_2,X)\ . 
\end{equation} 
The two cross ratios $t_1$ and $t_2$ are obtained from the insertion points through eq.\ \eqref{eq:ztcoordinates}. Their relation to the cross ratios we $\lambda$ and $\kappa$ used for the defect channel can be found in eq. \eqref{coordinates-ti}. To construct the remaining invariant $X$ we realise the representation space $W_{j,\msj}$ of mixed symmetry tensors in the space of polynomials in $z_A$ and $w_A$ with $A=4, \dots, d+1$. A complete description 
along with formulas for the action of $\mathfrak{so}(d)$ on such polynomials can be found in section 4.2.2, see in particular eqs.\  \eqref{eq:rhobulk1} 
to \eqref{eq:rhobulk3}. The invariant $X$ that appears in the argument of the special function $\psi$ is simply $X = \sum_A z_A^2$. Once again the prefactor 
$\Xi$ is constructed explicitly, see section 4.2.1, using ideas and constructions from \cite{Isachenkov:2018pef} and \cite{Buric:2020zea}. The final formula is 
stated in eq.\ \eqref{eq:G20sphericalbulk}. 

\noindent
The most novel part of our construction concerns again the spherical functions $\psi^{j,\msj}_{\Delta,J,m}(t_1,t_2,X)$. These are eigenfunctions of 
the second order (Hamiltonian) differential operators spelled out in eqs.\ \eqref{eq:HCSMST3scalars} and \eqref{eq:Lll}. This operator represents the 
image of the quadratic Casimir element of the $d$-dimensional conformal group $\SO(d+1,1)$ under the Harish-Chandra radial component map. The 
associated eigenvalue is determined by the quantum numbers $\Delta$ and $J$ through $C_2(\Delta,J) = \Delta(\Delta-d) + J(J+d-2)$. For $j=\msj$ the 
eigenfunctions of the associated Hamiltonian are well known. In fact, they are given by certain scalar conformal blocks in dimension $d+2j$ and hence 
they are close relatives of Heckman-Opdam hypergeometric functions for the root system $BC_2$, see discussion around eq.\ \eqref{eq:Hll} and 
\cite{Heckman-Opdam,Isachenkov:2016gim,Isachenkov:2017qgn,Isachenkov:2018pef} 
for details. 

The solutions for non-vanishing $j > \msj$ can then be constructed through application of 'commuting' differential shifting operators $q$ and $p$ that are given in eqs.\ \eqref{bulk-channel-shifting}. As we shall show, the operators $p$ and $q$ allow to raise the labels $j$ by one unit, 
\begin{equation} 
q_{j,\msj} \psi^{j,\msj}_{\Delta,J,m}(t_1,t_2,X) = \psi^{j+1,\msj}_{\Delta,J,m}(t_1,t_2,X) \ , \quad 
p_{j,\msj} \psi^{j,\msj}_{\Delta,J,m}(t_1,t_2,X) = \psi^{j+1,\msj}_{\Delta,J,m+1}(t_1,t_2,X)\ .  
\end{equation} 
Together, $p$ and $q$ suffice to construct all the special functions $\psi$ from the `ground states' $\psi^{j,j}_{\Delta,J,0}(t_1,t_2)$. This concludes our construction of the special functions $\psi$.

\subsection{Plan of the paper} 
The plan of this paper is as follows. In the next section we will introduce the main mathematical background, and in particular the Harish-Chandra radial component map. This will allow us to compute the spinning Casimir operators for the $d$-dimensional conformal group with any spin assignment. In addition, we also construct the two types of differential shifting operators that we described above. These can be used to build solutions of the Casimir differential equations from simpler seeds, very much in the same spirit as constructions in the CFT literature by Costa et al. \cite{Costa:2011dw} and in \cite{Karateev:2017jgd}. Applications of the general theory to CFT correlation functions are discussed in sections 3 and 4. These require to uncover the precise relation \eqref{eq:GXiPsi} between correlation and spherical functions. We shall illustrate this in section 3, where we review this relation, and in particular the construction of $\Xi$ from group theory, for spinning four-point functions in any dimension $d$ from \cite{Buric:2019dfk,Buric:2020buk}. With the appropriate factor $\Xi$, the wave functions $\Psi$ eigenfunctions of the universal Casimir operators that were derived in Section 2. We shall also review briefly how these universal Casimir operators for spinning four-point functions were used recently in \cite{Buric:2021kgy} to study the OPE limit of multipoint functions with more that $N=4$ scalar field insertions. This application illustrates nicely the power of universality. In section 4, we discuss new applications to defect CFTs for a defect of co-dimension $q = 2$. Starting with the defect channel, we construct the factor $\Xi$ that uplifts a spinning bulk-bulk two-point function in the presence of a defects to spherical function in the first subsection. Then we construct the associated spherical functions explicitly, as outlined in the previous subsection, using many of the general constructions and results of section 2. The second subsection addresses a similar problem for the bulk channel. Once again we uplift spinning bulk-bulk two-point correlations in the presence of the defect to spherical functions in the harmonic analysis of the bulk conformal group. The Casimir equations that characterise these spherical functions are very closely related to those for spinning four-point functions with two scalar and two spinning fields. This extends similar relations for scalar correlators in the presence of defects with $q=2$, see e.g.\ \cite{Isachenkov:2018pef} and references therein. Once again, solutions for spinning bulk channel blocks are constructed explicitly through a set of differential shifting operators by acting on the seeds that were built in \cite{Isachenkov:2017qgn}. In comparison to usual CFT treatments, see e.g. \cite{Costa:2011dw,Costa:2011mg} and \cite{Karateev:2017jgd}, our differential shifting operators act in the cross ratios, which makes them rather compact and easy to use. Finally, the third part of section 4 develops a intriguing application to three-point functions of two bulk and one defect local field that is inserted along a $q=2$ defect. We shall show that the Casimir equation for such a system in the bulk channel is once again controlled by the universal spinning Casimir operators introduced in section 2. The associated defect channel blocks were constructed recently in \cite{Buric:2020zea}. The paper concludes with a brief summary and a list of interesting future directions. 

\section{Universal Spinning Casimir and Shifting Operators}

This section is devoted to the main mathematical background that allows us to write universal Casimir and shifting operators for spinning four-point and other types of conformal blocks: 
the Harish-Chandra radial component map. In the first subsection we shall illustrate the main ingredients of constructions to follow at the example of the 1-dimensional conformal group 
$\SO(1,2)$. In particular, we shall explain the notions of spherical functions and Cartan decomposition in this case, compute the Laplacian on spherical functions and describe its 
relation to Calogero-Sutherland Hamiltonians. In this context we shall also meet the first simple instance of the radial component map, as well as shift operators. Then we shall dive into the general theory, with the group $\SO(d,2)$ providing a recurring key example. The second subsection contains all the relevant background concerning Cartan decompositions and spherical functions. The Harish-Chandra radial component map is then introduced at the beginning of the third subsection before it is used in the forth subsection to calculate the universal Laplacian and two types ('external' and 'internal') of weight-shifting operators. Some background on Lie algebras and our notation is collected in appendix A. Our conventions mostly follow \cite{Warner2} (see also \cite{10.1215/S0012-7094-82-04943-2} 
for a related discussion).

\subsection{Illustration: Casimir operators for \SO(1,2)}

Before entering the somewhat technical discussion of the Harish-Chandra radial component map and its 
applications below, we want illustrate the main concepts and constructions at the example of the group 
$G = \SO(1,2)$ of rank one. We work with the usual basis $\{H,E_+,E_-\}$ for its Lie algebra 
$\mathfrak{g} = \mathfrak{so}(1,2)$ whose Lie brackets take the form 
\begin{equation} \label{eq:so12commutation} 
    [H,E_+] = E_+, \quad [H,E_-]=-E_-, \quad [E_+,E_-] = 2H\ .
\end{equation}
The group $\SO(1,2)$ has a number of interesting subgroups. Here we shall focus on the maximal compact subgroup $K 
\cong U(1) \subset G$. The latter is generated by the element $Y =\frac12(E_+ - E_-)$. Elements of the 
subgroup $K$ take the form $k = \exp (\varphi Y)$, i.e. they are parametrised by an angle $\varphi$. 
Once we have fixed our subgroup $K$ we can naturally introduce the following spaces of $K$-$K$ covariant 
functions
\begin{equation*}
    \Gamma_{m,n} = \{ f : G\xrightarrow{}\mathbb{C}\ |\ 
    f(e^{\varphi Y} g e^{\psi Y}) = e^{i(m\varphi - n\psi)}f(g)\}\,,
\end{equation*}
where $m,n \in \mathbb{Z}$ are two integers. Since the group $G$ is 3-dimensional and we have imposed 
covariance conditions under left and right translations with elements of the 1-dimensional subgroup 
$K$, these $K$-$K$ covariant functions depend effectively on a single coordinate.  More precisely, if 
we parametrise elements $g \in  G$ as  
\begin{equation} \label{eq:SO12coordinates}
g(\varphi,t,\psi) = e^{\varphi Y}e^{t H}e^{\psi Y}\,,
\end{equation} 
then the values $f(\exp(t H))$ clearly determine $f$ uniquely, due to the covariance properties that 
define the subspace $\Gamma_{m,n}$. Elements $f \in \Gamma_{m,n}$ are known as $K$-spherical 
functions.

Our goal is to compute the restriction of the Laplacian on the group $G$ to the subspace of $K$-spherical 
functions. Note that the full Laplacian $\Delta$ on $G$ commutes with left and right regular actions and 
therefore acts within the space $\Gamma_{m,n}$. To find the action of $\Delta$ on functions $F(t) = 
f(\exp(t H))$ is straightforward. First we can write the Laplacian in the coordinates $\varphi,t,\psi$ 
which we have introduced in eq.\ \eqref{eq:SO12coordinates}, 
\begin{equation}\label{Laplacian-SL(2,R)}
    \Delta = \partial_t^2 + \coth t\ \partial_t + \frac{1}{\sinh^2 t} \left(\partial_\varphi^2 -
    2\cosh t\partial_\varphi \partial_\psi  + \partial_\psi^2\right)\ .
\end{equation}
In these coordinates, the restriction to $K$-$K$ covariant functions can now be implemented by the simple 
substitutions $\partial_\varphi\xrightarrow{} im$ and $\partial_\psi\xrightarrow{}-in$ due to covariance 
properties of our functions $f$. The resulting differential operator on $K$ spherical functions reads
\begin{equation} \label{eq:radialLaplacian}
\Delta_{m,n} = \partial_t^2 + \coth t\ \partial_t - \frac{1}{\sinh^2 t} \left(m^2 + 2mn\cosh t + n^2\right)\ .
\end{equation}
To derive eq.\ \eqref{Laplacian-SL(2,R)}, one observes that the Laplace-Beltrami operator on $G$ coincides 
with the quadratic Casimir built out of invariant vector fields. One may use either left- or right-invariant 
fields - both prescriptions lead to the same operator. Invariant vector fields are in turn encoded in the 
Maurer-Cartan form.
\medskip

While the calculation of the Laplacian on $K$-spherical functions we have just performed was rather 
simple, it may at least seem a bit unnatural that we had to choose some specific coordinates on the subgroup 
$K$ and write the Laplace operator for $G$ before we descended to the $K$-$K$ covariant functions. Note that 
the final formula for the Laplacian on spherical functions does not remember the choice of coordinates on 
$K$. The only information that matters is our choice of the two representations of $K$ which we parametrised 
by the integers $m,n$. Here lies the key to the understanding of the Harish-Chandra radial component map. In 
fact, we can observe that we could have obtained a very close cousin of our formula \eqref{Laplacian-SL(2,R)} 
directly in the universal enveloping algebra $U(\mathfrak{so}(2,1))$ of $\mathfrak{g}$. Given some fixed 
element $h = \exp(tH)$ of $G$ we can pass to a basis $(H,Y',Y)$ of the Lie algebra $\mathfrak{g}$ that 
consists of the Cartan generator $H$ along with the two elements
\begin{equation}
Y'  =  h^{-1} Y h = \frac12(e^{-t}E_+ - e^t E_-) \quad \textit{and} \quad Y = \frac12(E_+-E_-)\ .
\end{equation}
Note that the choice of the basis depends on the parameter $t$ which needs to be sufficiently generic in 
order for $H,Y'$ and $Y$ to be linearly independent. We can now rewrite the quadratic Casimir element $C_2$ 
of the conformal Lie algebra in terms of the new generators. Some simple manipulations give
\begin{equation}\label{Casimir-second-expression}
    C_2 = H^2 + \frac12 \{E_+,E_-\} = H^2 + \coth t\ H + \frac{1}{\sinh^2 t}\left(Y'^2-2\cosh t\ Y' Y + 
    Y^2\right)\ .
\end{equation}
This is called the \textit{radial decomposition} of the quadratic Casimir element $C_2$. In deriving the 
expression we imposed an ordering prescription that instructs us to move all the generators $Y$ to 
the right of the generators $Y'$. The theorem of Harish-Chandra allows to directly write down the 
restrictions $\Delta_{m,n}$ once the radial decomposition is known. Namely, it asserts that the Casimir 
operator \eqref{Casimir-second-expression} is turned into $\Delta_{m,n}$ through the substitutions
\begin{equation}\label{HC-substitutions}
    H\to \partial_t, \quad Y' \to im, \quad Y \to -in \ .
\end{equation}
Comparison with our previous formula \eqref{Laplacian-SL(2,R)} shows that this claim holds true, at least 
in this example. Conceptually, the substitution rules replace $Y'$ and $Y$ by the characters that govern 
the left and right covariance laws of functions in $\Gamma_{m,n}$. 

This seems like a good place to briefly illustrate the relation of the radial Laplacian spelled out in eq.\ 
\eqref{eq:radialLaplacian} with the associated hyperbolic Calogero-Sutherland Hamiltonians we have mentioned
in the introduction. For the special case at hand, the Hamiltonian acts on functions in a single variable 
$t$ only and it can be identified with the hyperbolic P\"oschl-Teller Hamiltonian. The latter is given by 
the following one-dimensional Schr\"odinger operator
\begin{equation}\label{Poschl-Teller-Hamiltonain-hyperbolic}
    H^{(a,b)}_{PT} = -\partial_t^2 + V^\textit{PT}_{(a,b)}(t) = 
    - \partial_t^2 - \frac{ab}{\sinh^2\frac{t}{2}} + \frac{(a+b)^2-\frac14}{\sinh^2 t}\ .
\end{equation}
It is easy to verify that the two operators $\Delta_{m,n}$ and $H^{(a,b)}_{PT}$ can be mapped to one another 
through conjugation with the function $\delta(t) \equiv \sqrt{\sinh t}$,
\begin{equation*}
    \delta \Delta_{m,n} \delta^{-1} = -H^{(im,in)}_{PT} - \frac14\,,
\end{equation*}
where the coupling constants $a=im$, $b=in$ in the P\"oschl-Teller potential on the right hand side are 
determined by the parameters $m,n$ that enter through the covariance law of our spherical functions.
\smallskip 
We have actually not yet defined the radial component map for our example, though it was lurching in 
the back when we evaluated the quadratic Casimir. In fact, The radial component map $\Pi$ is defined 
on the entire universal enveloping algebra $U(\mathfrak{g})$ and it sends elements in $X \in 
U(\mathfrak{g})$ to differential operators in a single variable $t$ with coefficients that are built 
out of $Y',Y$ and $t$, i.e. the coefficients can be regarded as $U(\mathfrak{k}) \otimes U(\mathfrak{k})$
valued functions in the variable $t$. We will give a formal construction of the map in the general 
case below. Here it suffices to have some operative understanding of how the assignment works. Given 
any element $X$ of $U(\mathfrak{g})$ we first express it in terms of the basis $H$, $Y'$, $Y$, treating 
$t$ as a formal variable rather than just a number. Once this is done, we order the basis generators my moving all 
factors $Y$' to the far left and all factors $Y$ to the far right. Finally, we replace $H$ by the differentiation 
with respect to $t$. In our derivation of the radial Laplacian, we have applied this map to the 
quadratic Casimir element $C_2 \in U(\mathfrak{g})$. In the construction of the differential 
shifting operators, the Harish-Chandra radial component map is applied to the generators $X$ of 
the Lie algebra $\mathfrak{g}$ so that we obtain some first order operators. 
\medskip

Even though it is certainly easy to construct eigenfunctions of the P\"oschl-Teller Hamiltonian in 
terms of the hypergeometric function $\ _2F_1$, we do want to briefly discuss the construction of 
shifting operators in this simple example. As was described in \cite{Schomerus:2017eny} already, 
left/right invariant vector fields on the conformal group provide us with a set of differential 
operators which move between spaces $\Gamma_{m,n}$ with different values of $m$ and $n$. Indeed, 
acting with invariant fields typically changes covariance laws. The radial component map allows 
to "project" these vector fields to operators in the single variable $t$ much in the same way as 
in the case of the Casimir element. To discuss the details, let us introduce the following 
elements in the complexification $\mathfrak{g}_c$ of $\mathfrak{g}$, 
\begin{equation*}
   F_+ =\frac12(E_+ + E_- + 2i H)\ , \quad F_- = \frac12(E_+ + E_- - 2iH)\ .
\end{equation*}
It is easy to verify that the generators $(i Y,F_+,F_-)$ possess the same Lie brackets as our 
original generators $(H,E_+,E_-)$, see eq. \eqref{eq:so12commutation}. Note that the element $Y$ 
plays a distinguished role in our discussion since it appears in the covariance law that 
characterises our spherical functions $f \in \Gamma_{m,n}$. In terms of the infinitesimal 
action of $Y$ on functions, the covariance law implies the following first order differential 
equations for spherical functions $f\in\Gamma_{m,n}$,
\begin{equation*}
    \mathcal{L}_Y f = - i n f, \quad \mathcal{R}_Y f = i m f\ .
\end{equation*}
Here $\mathcal{L}_X$ and $\mathcal{R}_X$ denote the left and right invariant vector fields 
associated with elements $X \in \mathfrak{g}$, respectively, i.e. they describe infinitesimal 
actions obtained from the right and left multiplication in the group $G$. By acting with vector 
fields $\mathcal{L}_{F_\pm}$ and $\mathcal{R}_{F_\pm}$ the covariance properties of $f$ are 
altered. For example
\begin{equation*}
    \mathcal{L}_Y \mathcal{L}_{F_+} f = 
    ([\mathcal{L}_Y,\mathcal{L}_{F_+}] + \mathcal{L}_{F_+} \mathcal{L}_Y) f = 
    (\mathcal{L}_{[Y,F_+]}- \mathcal{L}_{F_+} in) f = -i(n+1)\mathcal{L}_{F_+} f\ .
\end{equation*}
Since $\mathcal{R}_Y$ and $ \mathcal{L}_{F_+}$ commute, the right covariance of 
$\mathcal{L}_{F_+}f$ is the same as that of $f$. As a consequence, we conclude that 
$\mathcal{L}_{F_+}f\in\Gamma_{m,n-1}$. Continuing along these lines one finds
\begin{equation} 
 \mathcal{L}_{F_\pm} f \in \Gamma_{m,n\pm1} \quad , 
 \quad \mathcal{R}_{F_\pm} f \in \Gamma_{m\pm 1,n}\ . 
 \end{equation} 
So far, we shown that the operators $\mathcal{L}_{F_\pm}$ and $\mathcal{R}_{F_\pm}$ map spaces 
of spherical functions to each other, but what about the eigenfunctions of the differential 
operators $\Delta_{m,n}$ that act on spherical functions? Recall that the Laplacian $\Delta_{m,n}$ 
has been obtained by restricting the Laplacian on the conformal group. Since the latter commutes 
with all left and right invariant vector fields, we conclude that eigenfunctions of $\Delta_{m,n}$ 
are actually mapped onto each other, i.e. the action of $\mathcal{L}_{F_\pm}$ and 
$\mathcal{R}_{F_\pm}$ on spherical functions respects the decomposition with respect to 
eigenfunctions of the Laplacian. In order to obtain explicit expressions for the restriction of 
these first order differential operators on spherical functions we simply apply the Harish-Chandra
map $\Pi$ that we described above. In the first step, the map instructs us to express $F_\pm$ in 
terms of the basis $(H,Y',Y)$. This gives 
\begin{equation}\label{weight-shifting-1d}
   F_+ =\coth t\ Y -\frac{1}{\sinh t} Y' + i H, \quad F_- = \coth t\ Y -\frac{1}{\sinh t} Y' - iH \ .
\end{equation}
Since $F_\pm$ is linear in the Lie algebra generators, we do not need to reorder anything and can 
simply apply the substitutions \eqref{HC-substitutions}, This results in a set of first order 
operators that act on functions in a single variable $t$ as
\begin{equation*}
    p^+_{m,n} := i \left(\partial_t - n\coth t - \frac{m}{\sinh t} \right), \quad
    p^-_{m,n} := i \left(-\partial_t - n\coth t -\frac{m}{\sinh t} \right) \ .
\end{equation*}
It is instructive to verify explicitly that these operators satisfy the following exchange relations
with the Laplacians $\Delta_{m,n}$,
\begin{equation}
    \Delta_{m,n+1} p_{m,n}^+ = p_{m,n}^+ \Delta_{m,n}\ , \quad
    \Delta_{m,n-1} p_{m,n}^- = p_{m,n}^- \Delta_{m,n}\ .
\end{equation}
Operators $q^\pm$ that shift the left representation $m$ may be constructed similarly.
We shall refer to $p_{m,n}^\pm$ and $q_{m,n}^\pm$ as \textit{differential shifting operators}. As we 
have shown in detail, they indeed shift the weights of the representations of $K$ that characterise 
the covariance law of spherical functions.
\smallskip

This concludes our discussion of Casimir operators and the radial component map in the case of the
1-dimensional conformal group $\SO(1,2)$. In the remainder of this section, we will describe how the above
discussion generalises to higher-dimensional non-compact Lie groups. It will turn out that we can find the
action of the Laplacian on $K$-spherical functions once again by writing the radial decomposition of the
quadratic Casimir element. In order to do so, we need to introduce suitable generators $(h,y,y')$ which
are defined similarly to the generators $(H,Y,Y')$ we introduced for $\mathfrak{so}(1,2)$ above. To
obtain the restriction of $\Delta$ to a space of spherical functions, one replaces generators $y'$ and
$y$ by representation operators from the left and right covariance laws. Weight-shifting operators,
which intertwine between radial parts of the Casimir on different spaces of spherical functions, are
similarly obtained from radial decompositions of sets of elements of $\mathfrak{g}$ which form a
representation of $\mathfrak{k}$ under the adjoint action. It is always possible to eliminate first
order derivatives from radial parts of the Laplacian and thus bring it to the form of a
Schr\"odinger operator. The P\"oschl-Teller problem from above generalises to spinning
Calogero-Sutherland models.

\subsection{Relevant group theoretical background}

The main goal of this subsection is to collect all the group theoretical background, both in terms
of concepts and notations, that is needed to discuss the radial component map. As we explained
in the introduction, spherical functions are in the very centre of our considerations. So, we shall
introduce and discuss these first before we gradually zoom into the cases that appear in the
context of CFT.

\subsubsection{Spherical functions}

To begin with, let $G$ be any group and $K$ some subgroup. In the most standard setup, $G$ is 
assumed to be a real Lie group and $K$ its maximal compact subgroup, but we do not have to make 
these assumptions throughout most of our discussion and prefer to keep the discussion a bit more
general. Let us stress that in most applications to CFT, neither the group $G$ 
nor the subgroup $K$ is compact. As part of the general setup we pick two irreducible representations 
$\rho_l$ and $\rho_r$ of the subgroup $K$. We shall denote their carrier spaces by $W_{l}$ and $W_r$, 
respectively. The space of $K$-spherical functions is defined as
\begin{equation}\label{K-spherical-functions}
    \Gamma_{\rho_l,\rho_r} = \{ f : G \xrightarrow{} \text{Hom}(W_r,W_l) \ |\ f(k_l g k_r) =
    \rho_l(k_l) f(g) \rho_r(k_r); \  g\in G\, , \ k_{l,r}\in K \ \} \ .
\end{equation}
In order to make the covariance law that defines spherical functions even more explicit, we choose
a basis $\{e_a\}$ of $W_l$ and similarly a basis $\{e_\alpha\}$ of $W_r$. We shall use the Dirac
notation and write basis elements of $\text{Hom}(W_r,W_l)\cong W_l\otimes W_r^\ast$ as
$|e_a\rangle\langle e^\alpha|$. Given such a choice of basis, the covariance properties of
spherical functions $f$ may be written as
\begin{equation*}
    f^a_{\ \alpha} (k_l g k_r) = \rho^a_{l;b}(k_l)  f^b_{\ \beta}(g) \rho^{\beta}_{r;\alpha}(k_r)\ .
\end{equation*}
Examples of spherical functions can be easily found among various matrix elements of irreducibles
of the group $G$. Indeed, let $\pi$ be a unitary irreducible representation of $G$ on the carrier
space $V$ with a basis $\{e_i\}$ and assume that the $K$-modules $W_l$ and $W_r$ both appear in
the restriction $\pi$ to the subgroup $K$. Then we have
\begin{equation*}
    \pi^a_{\ \alpha}(k_l g k_r) = \langle e^a|\pi(k_l)|e_i\rangle
    \langle e^i|\pi(g)|e_j\rangle\langle e^j|\pi(k_r)|e_\alpha\rangle\ .
\end{equation*}
Note that even though we started with a very small subset of the matrix elements $\pi^i_{\ j}$ such
that $e_i = e_a \in W_l$ and $e^j = e^\alpha \in W_r$, the sums on the right hand side can involve
many more matrix elements in general. In order for the right hand side to involve sums over basis
elements of $W_{l,r}$ only, we shall assume that the restriction of $\pi$ to $K$ has simple
spectrum, i.e. that every irreducible of $K$ appears at most once in the restriction of $\pi$.
If the restriction to $K$ of any unitary irreducible $\pi\in\hat G$ has simple spectrum, $K$ is 
said to be {\it big} in $G$.\footnote{A subgroup $K\leq G$ is big iff and only
if the convolution algebra of functions $f:G\xrightarrow{}\mathbb{C}$ which satisfy $f(k g k^{-1})
= f(g)$ is commutative, \cite{Kirillov}.} By orthogonality of matrix elements, we obtain non-zero
contributions only from $i=b$ and $j=\beta$,
\begin{equation}
    \pi^a_{\ \alpha} (k_l g k_r) = \rho^a_{l;b}(k_l)  \pi^b_{\ \beta}(g) \rho^{\beta}_{r;\alpha}(k_r)\ .
\end{equation}
We conclude that, in case $K$ is big in $G$, the collection of matrix elements $\{\pi^a_{\ \alpha}\}$
provides us with a $K$-spherical function.
\smallskip

Among various pairs of a group $G$ and a subgroup $K$, especially interesting are the so-called
\textit{Gelfand pairs}. To introduce this notion, we consider the convolution product
\begin{equation}\label{convolution}
    (f_1\ast f_2)(g) = \int_G dg_1 \ f_1(g_1) f_2(g_1^{-1}g)\,,
\end{equation}
of two $K$-spherical functions $f_1\in\Gamma_{\rho_l,1}$, $f_2\in\Gamma_{1,\rho_r}$. Here, $dg$
denotes the Haar measure on $G$ and $1$ the trivial representation of $K$. The right-covariance 
of $f_1\ast f_2$ is immediate
\begin{equation*}
    (f_1^a\ast f_{2\alpha})(gk) = \int_G dg_1\ f_1^a(g_1) f_{2\alpha}(g_1^{-1} g k)=
    \int_G dg_1\ f_1^a(g_1) f_{2\beta}(g_1^{-1} g) \rho^\beta_{r; \alpha}(k) =
    (f_1^a\ast f_{2\beta})(g)\rho^\beta_{r;\alpha}(k)\ .
\end{equation*}
For the left-covariance, we use properties of the Haar measure to deduce
\begin{align*}
    (f_1^a\ast f_{2\alpha}) (kg) &= \int_G dg_1\ f_1^a(g_1) f_{2\alpha}(g_1^{-1}kg)
    = \int_G dg_2\ f_1^a(k g_2) f_{2\alpha}(g_2^{-1}g)\\
    & = \int_G dg_2\ \rho_l(k)^a_{\ b} f_1^b(g_2) f_{2\alpha}(g_2^{-1}g)
    = \rho_l(k)^a_{\ b} (f_1^b\ast f_{2\alpha})(g)\ .
\end{align*}
Therefore, the convolution product $f_1\ast f_2$ is a $(\rho_l,\rho_r)$-spherical function,
i.e. $f_1\ast f_2\in \Gamma_{\rho_l,\rho_r}$. In particular, the space $\Gamma_{1,1}$ of
functions $f:G\xrightarrow{}\mathbb{C}$ that are bi-invariant under $K$ is closed under
convolutions. If this convolution algebra is commutative, $K$ is said to be a spherical
subgroup of $G$ and $(G,K)$ is called a Gelfand pair. There are several other equivalent
ways to define Gelfand pairs. One of them, under suitable assumptions on $G$ and $K$, is
to require that the geometric representation of $G$ on the space of functions on $G/K$
has simple spectrum. There is also a simple useful criterion for identifying Gelfand
pairs: if there exists an anti-automorphism $s$ of the group $G$ such that for all its
elements $g\in G$ one can factorise $s(g) = k_l g k_r$ with $k_{l,r} \in K$, then $(G,K)$
is a Gelfand pair. Finally, if $G$ is a simple real Lie group and $K$ its maximal compact
subgroup, it can be shown that $(G,K)$ is a Gelfand pair. This is the context in which
Gelfand pairs appear here. For this particular case we now want to derive a new 
characterisation of the space $\Gamma$ of spherical functions. This requires a bit 
of preparation.
\medskip

When $G$ is a simple real Lie group and $K$ its maximal compact subgroup, every element
of $G$ may be factorised as $g = k_l h k_r$ where $k_l,k_r\in K$. Here $h\in A_p$ is an
element of an abelian subgroup $A_p$. In the case $G=\SO(n,m)$, this abelian subgroup $A_p$
has dimension $\textit{min}(m,n)$. We shall discuss its infinitesimal generators in the
next subsection. Such a factorisation of $g$ into a product involving $k_{l,r}$ and $h$ is
called the Cartan decomposition and we write $G=K A_p K$. Looking at our definition of 
$K$-spherical functions, we see that elements $f \in \Gamma_{\rho_l,\rho_r}$ are uniquely 
fixed by their restriction $F = f|_{A_p}$ to the abelian subgroup $A_p$. On the other hand, 
not every $V=\textrm{Hom}(W_l,W_r)$-valued function $F$ on $A_p$ can arise through the 
restriction of a $K$-spherical function. Actually there are two issues to consider. On the 
one hand, an orbit $KhK$ with $h \in A_p$ can intersect $A_p$ multiple times. If that 
happens, the values the restriction $F$ takes on the various intersection points are not 
independent. On the other hand, group elements $g$ may possess different Cartan 
decompositions. Whenever this happens, we have several ways to extend a K-spherical 
function from $A_p$ to $g$ which of course have to coincide in order to obtain a function 
on $G$. This imposes constraints on the values $F$ takes. In order to make a rigorous 
statement about the relation between $\Gamma_{\rho_l,\rho_r}$ and $\textrm{Fun}(A_p,V)$ 
we need a bit of preparation. 

Given our Gelfand pair $(G,K)$, we can introduce two important subgroups of $K$. The 
first one is the the normaliser of $A_p$ in $K$, i.e.  
\begin{equation} \label{eq:N_KMdef}
    N_K(A_p) \equiv \{ k\in K\ |\ k h k^{-1} \in A_p \ \textit{ for all } \ h \in A_p\}\ .
\end{equation}
Note that the adjoint action of $N_K(A_p)\subset K$ on $K$ leaves $A_p$ invariant 
as a subgroup, but it acts non-trivially in elements $h \in A_p$. The centraliser $M$ 
of $A_p$ in $K$ is defined by the stricter condition 
\begin{equation} 
M = \{\,  k\in K\ |\ k h k^{-1} = h\ \textit{for all} \ h \in A_p\, \}  
\ \subset \ N_K(A_p)\ .
\end{equation} 
It is easy to see that $M$ is normal in $N_K(A_p)$, i.e.\ that the adjoint action of 
$N_K(A_p)$ leaves the subgroup $M$ invariant. Hence their quotient  
\begin{equation} \label{eq:restrictedWeyl}
W = N_K(A_p)/M\,,
\end{equation}
is a group as well. This group $W$ is referred to as the restricted Weyl group of the
of the pair $(G,K)$. By construction, the restricted Weyl group $W$ acts on the abelian 
subgroup $A_p$. This is because the action of the normaliser $N_K(A_p)$ on $A_p$ is 
stabilised by the centraliser $M$. 

Now we can give a more precise statement concerning the two issues we mentioned above. On the one hand, it is easy to see that two elements $h,h' \in A_p$ are in the same 
orbit, i.e. $h' \in K h K$ if and only if $h'$ can be obtained from the $h$ by acting with some element of the restricted Weyl group $W$. The second issue that is caused by 
the non-uniqueness of the Cartan decomposition can also be understood easily. Note that the freedom in the Cartan decomposition is described by the centraliser $M$. In 
fact given one decomposition $g = k_l h k_r$ we can pass to $g = k_l m^{-1} h m k_r$ for any $m \in M$. Hence, if we want a $V$-valued function $F$ on $A_p$ to possess a 
single valued $K \times K$ covariant extension to $G$, we need to ensure that $F$ takes values in the space $V^M$ of $M$-invariant elements in $V$. 

In order to summarise our findings we note that the space of $V^M$-valued functions 
$\textrm{Fun}(A_p,V^M)$ on $A_p$ admits an action of the normaliser subgroup $N_K(A_p)$
which is defined by 
\begin{equation}\label{twisted-action-NKA}
    (n\cdot F)(h) = \rho(n) F(n^{-1} h n)\ .
\end{equation}
Here $\rho$ denotes the action of $N_K(A_p)$ on $V^M$. Since elements $m \in M 
\subset N_K(A_p)$ act trivially on the base $A_p$ and the fibre $V^M$, the action 
\eqref{twisted-action-NKA} descends to an action of the restricted Weyl group $W$ 
on $\textrm{Fun}(A_p,V^M)$. We thereby conclude that restriction map $f\mapsto F 
= f|_{A_p}$ gives an isomorphism $\pi$ of vector spaces
\begin{equation} \label{eq:GammaL1}
\pi:\, \Gamma_{\rho_l,\rho_r}\xrightarrow{}\textrm{Fun}(A_p,V^M)^W \ .
\end{equation}
In this way we have obtained a new characterisation of the space of spherical functions
for a pair $(G,K)$ of a real simple Lie group $G$ and its maximal compact subgroup $K$.
According to the statement \eqref{eq:GammaL1}, in this case spherical functions can be
regarded as $W$-invariant functions on the abelian subgroup $A_p \subset G$ that take
values in the the space $V^M$ of $M$-invariants in $V = \textrm{Hom}(W_l,W_r)$. Note 
that elements of the restricted Weyl group $W$ act on both $A_p$ and $V^M$.

Given the isomorphism \eqref{eq:GammaL1}, we can now formulate the problem that is 
solved by the Harish-Chandra radial component map. Note that the space $\Gamma$ on the 
left hand side consists of functions on the group $G$. As such, elements of $\Gamma$ 
can be acted upon with left- and right-invariant differential operators. For those 
operators that preserve $\Gamma$, one may now wonder how the action passes through the 
isomorphism $\pi$, i.e. whether one can compute their action directly in terms of 
differential operators on $\textrm{Fun}(A_p,V^M)^W$, without passing to the group $G$. 
The answer is assertive and the Harish-Chandra radial component map provides a constructive 
solution to the challenge. Before we get there, however, we need to introduce some more 
concepts and notations related to Lie algebras and Cartan decompositions.

\subsubsection{Lie algebra and Cartan decomposition}

In this subsection, we shall continue to assume that $G$ is a real non-compact simple
Lie group and $K$ its maximal compact subgroup as we turn attention from spherical
functions to vector fields, i.e. discuss some central notions and notations concerning
the Lie algebra $\mathfrak{g}$ of $G$ and various Lie subalgebras thereof. Our notations
regarding the root system of $\mathfrak{g}$ are collected in appendix A. We consider the
\textit{Cartan decomposition} of $\mathfrak{g}$, i.e. the decomposition into the Lie
subalgebra $\mathfrak{k}$ and its orthogonal complement
\begin{equation}\label{Cartan-decomposition-Lie-algebra}
    \mathfrak{g} = \mathfrak{k} \oplus \mathfrak{p}\ .
\end{equation}
The subalgebra $\mathfrak{k}$ and the subspace $\mathfrak{p}$ of $\mathfrak{g}$ can be
characterised as eigenspaces of an involutive isomorphism $\theta$ acting on $\mathfrak{g}$
which is known as the \textit{Cartan involution}. By definition, elements of $\mathfrak{k}$
are left invariant by the action of $\theta$, i.e.\ $\theta(X) = X$ for $X \in \mathfrak{k}$,
while elements $X \in \mathfrak{p}$ are sent to $-X = \theta(X)$.

Let $\mathfrak{a}_p$ be a maximal (ad-diagonalisable) abelian subspace of $\mathfrak{p}$.
The dimension of $\mathfrak{a}_p$ is called the \textit{real rank} of $G$. We will denote
the real rank by $l$ and pick some basis $\{H_1,...,H_l\}$ for $\mathfrak{a}_p$. The space
$\mathfrak{g}$ carries the adjoint action of $\mathfrak{a}_p$ and can be decomposed into
joint eigenspaces of $\text{ad}_{H_i}$,
\begin{equation}\label{restricted-root-decomposition}
    \mathfrak{g} = \mathfrak{a}_p \oplus \mathfrak{m} \oplus \sum_{\lambda\in\Sigma} \mathfrak{g}^\lambda\ .
\end{equation}
Here, $\mathfrak{m}$ is the centraliser of $\mathfrak{a}_p$ in $\mathfrak{k}$. The first two
summands make up the subspace of $\mathfrak{g}$ on which all $\text{ad}_{H_i}$ vanish. We regard
the objects $\lambda$ that we sum over in the final term as linear maps on $\mathfrak{a}_p$ and
define $\mathfrak{g}^\lambda \subset \mathfrak{g}$ as the subspace on which $[H,X]= \lambda(H)X$
for all $X\in \mathfrak{g}^\lambda$ and $H \in \mathfrak{a}_p$. Linear functionals $\lambda$ are
referred to as \textit{restricted roots} and $\mathfrak{g}^\lambda$ are the associated restricted
root spaces. The set of restricted roots is denoted by $\Sigma$. When the meaning is clear from
the context, we will drop the adjective "restricted" and refer to the objects $\lambda$ simply as
roots etc. We observe that $\theta(\mathfrak{g}^\lambda) = \mathfrak{g}^{-\lambda}$ since all
elements in $\mathfrak{p}$ have eigenvalue $-1$ under the Cartan involution $\theta$ and
$\mathfrak{a}_p$ is a subspace of $\mathfrak{p}$.

The validity of the decomposition \eqref{restricted-root-decomposition} rests on the following
observation. By the properties of the Cartan decomposition \eqref{Cartan-decomposition-Lie-algebra},
the following bilinear form on $\mathfrak{g}$ (where $\kappa$ is the Killing form)
\begin{equation}\label{positive-inner-product}
    (x,y)_\theta \equiv -\kappa(x,\theta(y))\,,
\end{equation}
is positive-definite and turns $\mathfrak{g}$ into a Hilbert space. It is not difficult to verify
that, with respect to this inner product, one has $\text{ad}_X^\ast = -\text{ad}_{\theta(X)}$ for
all $X\in\mathfrak{g}$. Given that $\theta(H_i) = - H_i$, this implies that the operators
$\text{ad}_{H_i}$ are hermitian and hence they possess real eigenvalues. We conclude that the
decomposition \eqref{restricted-root-decomposition} holds and is orthogonal with respect to
the scalar product $(,)_\theta$.  Furthermore, these remarks allow to define positive restricted
roots with respect to the basis $\{H_i\}$ (and the ordering of basis vectors) in the usual way:
we say that $\lambda$ is positive if the first non-zero entry of the sequence $(\lambda(H_i))$
is positive. The set of positive restricted roots is denoted by $\Sigma_+$.

In contrast with the ordinary root spaces of complex simple Lie algebras, restricted root spaces
$\mathfrak{g}^\lambda$ may not be one-dimensional. The dimension of the space $\mathfrak{g}^\lambda$
is denoted by $m(\lambda)$ and called the multiplicity of the root $\lambda$. The half-sum of positive
roots, counted with multiplicities, is denoted by $\rho$,
\begin{equation}\label{restricted-Weyl-vector}
    \rho \equiv \frac12 \sum_{\lambda\in\Sigma_+} m(\lambda) \lambda\ .
\end{equation}
It seems appropriate to refer to $\rho$ as the restricted Weyl vector. The Weyl group of the root system $\Sigma$ is the restricted Weyl group \eqref{eq:restrictedWeyl}.

The decomposition \eqref{restricted-root-decomposition} along with the notion of positive restricted
roots also provides us with a \textit{Gauss decomposition}
\begin{equation}\label{Gauss-decomposition-Lie-algebra}
    \mathfrak{g} = \mathfrak{a}_p \oplus \mathfrak{m} \oplus \mathfrak{n} \oplus \bar{\mathfrak{n}}\
    \quad \textit{where} \quad \mathfrak{n} \equiv \sum_{\lambda \in \Sigma_+} \mathfrak{g}^\lambda\,,
\end{equation}
and $\bar{\mathfrak{n}} = \theta(\mathfrak{n})$ are obtained from the partial sums over restricted
positive and negative roots $\lambda$, respectively. Another closely related decomposition is the
Iwasawa decomposition which is discussed in appendix $A$.

It is possible to identify elements in the set $\Sigma$ of restricted roots with those (ordinary)
roots of $\mathfrak{g}$ that do not vanish on the subspace $\mathfrak{a}_p$. Furthermore, we can even
split such roots $\alpha$ of $\mathfrak{g}$ into positive and negative roots. We write $\alpha \in
P_+$ if the restriction of $\alpha$ to $\mathfrak{a}_p$ belong to the set $\Sigma_+$ of restricted
positive roots. Now let $\alpha\in P_+ \cup (-P_+)$ be any root of $\mathfrak{g}$ with a
non-vanishing restriction to $\mathfrak{a}_p$. Then we can decompose the associated element
$e_{\alpha}$ as\footnote{We write $\mathfrak{g}_c = \mathfrak{g}\otimes\mathbb{C}$, and similarly for all
other Lie groups and Lie algebras under consideration.}
\begin{equation}\label{y-and-z-definition}
    e_\alpha = y_\alpha + z_\alpha, \quad \textit{where} \quad
    y_\alpha = \frac{e_\alpha + \theta(e_\alpha)}{2}\in\mathfrak{k}_c,
    \quad z_\alpha = \frac{e_\alpha - \theta(e_\alpha)}{2}\in\mathfrak{p}_c\ .
\end{equation}
In order to write down an explicit expression of the quadratic Casimir element, we introduce a 
basis $\{X_1,...,X_m\}$ of $\mathfrak{m}_c$ and set $g_{pq} = \kappa(X_p,X_q)$. Similarly,
$h_{ij}=\kappa(H_i,H_j)$ is the restriction of the Killing form to $\mathfrak{a}_p$. Inverses
of $g_{pq}$ and $h_{ij}$ are denoted by $g^{pq}$ and $h^{ij}$, as usual. It is a standard result
that the quadratic Casimir element of $\mathfrak{g}$ takes the form
\begin{equation}\label{quadratic-Casimir-first-expression}
    C_2 = g^{pq}X_p X_q + h^{ij}H_i H_j + \sum_{\alpha\in P_+} \{e_\alpha, e_{-\alpha}\}\ .
\end{equation}
This form of the quadratic Casimir is adapted to the Gauss decomposition
\eqref{Gauss-decomposition-Lie-algebra} and will be our starting point for the evaluation of the
Laplacian on $K$-spherical functions in the next subsection.
\smallskip

Before we close this subsection, let us briefly illustrate the relevant Lie theoretic constructions 
on the example of the Lie algebra $\mathfrak{g} = \mathfrak{so}(d,2)$. The Lie bracket for the 
generators $\{L_{\alpha\beta}\}$, $\alpha,\beta = 0,1,...,d+1$ is formally similar to that of 
the usual Lorentz algebra,
\begin{equation}\label{brackets-so(d,2)}
    [L_{\alpha\beta},L_{\gamma\delta}] = \eta_{\beta\gamma} L_{\alpha\delta}
    - \eta_{\alpha\gamma} L_{\beta\delta} + \eta_{\beta\delta} L_{\gamma\alpha}
    - \eta_{\alpha\delta} L_{\gamma\beta}\,,
\end{equation}
except that the metric is now given by $\eta_{\alpha\beta}=\text{diag}(-1,1,...,1,-1)$, i.e.\ it has
two timelike directions. In the $(d+2)$-dimensional vector representation, the Lorentz generators are
given by
\begin{equation}\label{generators-so(d,2)}
    L_{\alpha\beta} = \eta_{\alpha\gamma} E_{\gamma\beta} - \eta_{\beta\gamma} E_{\gamma\alpha}\,,
\end{equation}
where $(E_{\alpha\beta})_{ij} = \delta_{\alpha i} \delta_{\beta j}$ are the usual elementary matrices.
The Cartan decomposition of $\mathfrak{g}$ involves the following two summands $\mathfrak{k}$ and
$\mathfrak{p}$ which are given by
\begin{equation}\label{k-and-p-so(d,2)}
    \mathfrak{k} = \text{span}\{L_{0,d+1}, L_{ij}\} \cong \mathfrak{so}(2) \oplus \mathfrak{so}(d),
    \quad \mathfrak{p} = \text{span}\{L_{0i},L_{i,d+1}\}, \quad i,j = 1,...,d\ .
\end{equation}
Indeed, we see that $\mathfrak{k}$ is the Lie algebra of the maximal compact subgroup $K\sim \SO(2)
\times \SO(d)$ of the group $G\sim \SO(d,2)$. The corresponding Cartan involution $\theta$ may be 
realised as conjugation by the Weyl inversion
\begin{equation}\label{Weyl-so(d,2)}
    w = \text{diag}(-1,1,...,1,-1)\ .
\end{equation}
Clearly, the real rank of $G$ is equal to two. We will chose a basis for $\mathfrak{a}_p$ that
consists of the two elements 
\begin{equation} \label{eq:H1H2} 
H_1 = L_{01}\quad , \quad H_2 = L_{d,d+1}\ . 
\end{equation} 
For the centraliser $\mathfrak{m}$ of the abelian subalgebra $\mathfrak{a}_p$ in $\mathfrak{k}$ 
one finds
\begin{equation} \label{eq:m} 
    \mathfrak{m} = \text{span}\{L_{ab}\}, \quad a,b = 2,...,d-1\ .
\end{equation}
Obviously, the centraliser $\mathfrak{m} \cong \mathfrak{so}(d-2)$ is isomorphic to the $(d-2)$-dimensional
rotation algebra. Unless stated otherwise, indices $\alpha,\beta...$, $i,j...$ and $a,b...$ will always
assume the range as in the previous equations. The pair $(\mathfrak{g},\mathfrak{k})$ has four restricted
roots that possess a $(d-2)$-dimensional root space.\footnote{We wish to bring to reader's attention that,
although the pair $(\lambda,a)$ behaves like an $\alpha$, vectors $e^a_\lambda$ are not root vectors of
$\mathfrak{g} = \mathfrak{so}(d,2)$. Indeed we have $[e^a_\lambda,\tau(e^a_{\lambda})] = [e^b_\lambda,
\tau(e^b_{\lambda})]$ for $a\neq b$. But, in case of roots, $h_\alpha = h_\beta$ implies that
$\alpha = \beta$.} These are given by
\begin{align} \label{eq:restrictedroot1}
    & \mathfrak{g}^{(1,0)} = \text{span}\{e^a_{(1,0)} \equiv L_{0a} + L_{1a}\} , \quad
    & \mathfrak{g}^{(-1,0)} = \text{span}\{e^a_{(-1,0)} \equiv L_{0a} - L_{1a}\}, \\[2mm]
    & \mathfrak{g}^{(0,1)} = \text{span}\{e^a_{(0,1)} \equiv L_{ad} - L_{a,d+1}\} , \quad
    & \mathfrak{g}^{(0,-1)} = \text{span}\{e^a_{(0,-1)} \equiv -L_{ad} - L_{a,d+1}\}\ .
\end{align}
In addition, there also exist four restricted roots for which the associated root space is 1-dimensional.
The associated root vectors are given by
\begin{align}
    & e_{(1,1)} = \frac{1}{\sqrt{2}}(L_{0d} + L_{1d} - L_{0,d+1} - L_{1,d+1}), \quad
    & e_{(1,-1)} =\frac{1}{\sqrt{2}}(L_{0d} + L_{1d} + L_{0,d+1} + L_{1,d+1}),\\[2mm]
    & e_{(-1,1)} =\frac{1}{\sqrt{2}}(L_{0d} - L_{1d} - L_{0,d+1} + L_{1,d+1}), \quad
    & e_{(-1,-1)} =\frac{1}{\sqrt{2}}(L_{0d} - L_{1d} + L_{0,d+1} - L_{1,d+1})\ .
\label{eq:restrictedroot4}
\end{align}
We have chosen the signs such that the above vectors behave as $\theta(e^{(a)}_\lambda) =
-e^{(a)}_{-\lambda}$ under the action of the Cartan involution $\theta$. Here and in the following, the
upper index $(a)$ is used when we want to deal simultaneously with the vectors listed in eqs. 
\eqref{eq:restrictedroot1}-\eqref{eq:restrictedroot4}. In case the restricted root $\lambda$ 
corresponds to one of the 1-dimensional root spaces, the upper index $(a)$ should be omitted. 
Furthermore, we have normalised our root vectors such that
\begin{equation}
    [e^{(a)}_\lambda,e^{(a)}_{-\lambda}] = 2 \lambda\cdot H, \quad\quad H = (H_1,H_2)\ .
\end{equation}
We proclaim the positive restricted roots to be
\begin{equation}
    \Sigma_+ = \{(1,0),(0,1),(1,1),(1,-1)\}\ .
\end{equation}
Finally, given that the Cartan involution is obtained by conjugation with the Weyl inversion
\eqref{Weyl-so(d,2)}, it is easy to read off the elements defined in eqs.\ \eqref{y-and-z-definition}. 
For the positive restricted roots, these are given by
\begin{align} \label{eq:yalambda1}
    & y^a_{(1,0)} = L_{1a}, \quad y^a_{(0,1)} = L_{ad},\quad  z^a_{(1,0)} = L_{0a}, 
    \quad z^a_{(0,1)} = -L_{a,d+1},\\[2mm] 
    & y_{(1,\pm1)} = \frac{1}{\sqrt{2}}(L_{1d} \mp L_{0,d+1}), \quad z_{(1,\pm1)} = 
    \frac{1}{\sqrt{2}}(L_{0d}\mp L_{1,d+1})\ .
      \label{eq:yalambda2}
\end{align}
The corresponding expressions for negative restricted roots are obtained using the symmetry properties
$y_{-\lambda}^{(a)} = -y_\lambda^{(a)}$ and $z_{-\lambda}^{(a)} = z_\lambda^{(a)}$. This concludes our
review of the group theoretical background that is needed for the discussion of the Harish-Chandra
radial component map.

\subsection{Harish-Chandra's radial component map}
\def\Gmap{\Lambda}

After the extensive preparation in the previous subsection we are now in position to 
introduce the main actor of this work: Harish-Chandra's radial component map. All required 
notations have also been introduced above. In particular, $G$ continues to denote a real 
simple non-compact Lie group and $K$ its maximal compact subgroup. The construction of the 
radial component maps is based on the Cartan decomposition $G = K A_p K$ of $G$. Recall 
that $A_p$ is an abelian Lie group whose dimension is called the real rank of $G$. As we 
stated before, Harish-Chandra's radial component map $\Pi$ sends elements of the 
complexified universal enveloping algebra $U(\mathfrak{g}_c)$ to differential operators 
on $A_p$ whose coefficients take values in $ U(\mathfrak{k}_c) \otimes U(\mathfrak{k}_c)$, 
\begin{equation}\label{radial-component-map}
    \Pi : U(\mathfrak{g}_c) \xrightarrow{} \textrm{Fun}(A_p)\otimes U(\mathfrak{a}_{p_c}) \otimes 
    U(\mathfrak{k}_c) \otimes U(\mathfrak{k}_c)\ .
\end{equation}
Let us point out that the first two tensor factors on the right hand side form to the 
algebra of complex valued differential operators on $A_p$.

The essential step in constructing the map $\Pi$ is an infinitesimal version of the $K A_p 
K$ decomposition that we will now describe. As is clear from our discussion in the previous
subsection, the Lie algebra $\mathfrak{k}$ of $K$ admits the following decomposition
\begin{equation}
    \mathfrak{k} = \mathfrak{m} \oplus \mathfrak{q}\ .
\end{equation}
Here, $\mathfrak{m}$ is spanned by all elements of $\mathfrak{k}$ that commute with 
$\mathfrak{a}_p$ and the orthogonal decomposition is performed with respect to the 
quadratic form \eqref{positive-inner-product}. The complexification $\mathfrak{q}_c$ 
is spanned by elements $y_\alpha$ with $\alpha\in P_+$ which we have introduced in 
eq.\ \eqref{y-and-z-definition}. For almost any element $h\in A_p$ we can now  
obtain a decomposition of the complexification $\mathfrak{g}_c$ as 
\begin{equation}\label{KAK-Lie-algebra}
    \mathfrak{g}_c = \mathfrak{a}_{p,c} \oplus h^{-1} \mathfrak{q}_c h \oplus 
    \mathfrak{k}_c\ .
\end{equation}
It is important to restrict the second summand to the space $h^{-1} \mathfrak{q}_c h$, 
i.e. to remove the centraliser $\mathfrak{m}_c$ of $\mathfrak{a}_{p,c}$, in order to 
fix the inherent gauge freedom of the Cartan decomposition, see our discussion around eq.\ 
\eqref{eq:N_KMdef}. For a simple proof of the decomposition \eqref{KAK-Lie-algebra} 
see \cite{Warner2}. One can think of eq.\ \eqref{KAK-Lie-algebra} as an infinitesimal 
version of the $K A_p K$ decomposition of $G$. 

A direct sum decomposition of $\mathfrak{g}_c$ corresponds to a factorisation of the 
universal enveloping algebra $U(\mathfrak{g}_c)$, 
\begin{equation} \label{eq:isom}
U(\mathfrak{g}_c) \cong U(\mathfrak{a}_{p_c}) \otimes \mathcal{K}_2 \ , 
\end{equation} 
where we have introduced
\begin{equation} 
\quad \mathcal{K}_2 \equiv U(\mathfrak{k}_c) \otimes_{U(\mathfrak{m}_c)} U(\mathfrak{k}_c) \cong S(\mathfrak{q}_c) \otimes U(\mathfrak{k}_c)\ .
\end{equation}  
The isomorphism of vector spaces on the two sides of eq. \eqref{eq:isom} is not canonical. In fact, we can easily write down an entire family of isomorphisms $\Gmap^l_h$ that is 
parametrised by elements $h \in A_p$ 
\begin{equation}\label{Gamma-map-left}
    \Gmap^l_h : U(\mathfrak{a}_{p_c}) \otimes \mathcal{K}_2 \xrightarrow{} U(\mathfrak{g}_c),\quad \Gmap^l_h(H\otimes x \otimes y) = h^{-1} x h\ H\ y\ .
\end{equation}
Here, elements of $\mathcal{K}_2$ are parametrised by pairs of elements $x,y \in 
U(\mathfrak{k}_c)$ subject the equivalence relation that comes from the freedom to 
move a factor $m \in U(\mathfrak{m}_c)$ from the first to the second component. Let us 
briefly check that $\Gmap^l_h$ is indeed well defined, i.e. that it is independent of 
the choice of representative $x \otimes y$. So, we assume that $x = x'm$ with $m \in 
U(\mathfrak{m}_c)$ to show 
\begin{equation*}
    \Gmap^l_h(H\otimes x' m\otimes y) = h^{-1} x' m h\ H\ y =
    h^{-1} x' h\ H\ m y = \Gmap^l_h(H \otimes x' \otimes my)\ .
\end{equation*}
In the process we have used that, by definition, elements $m \in U(\mathfrak{m}_c)$ 
commute with $H$ and with $h$. Hence the image $\Gmap^l_h$ assigns to its argument is 
indeed independent of the choice of representative. Clearly, we could define a similar 
map $\Gmap^r$ that conjugates $y$ by $h$ rather than $x$. We shall mostly focus on 
$\Gmap^l$ and shall therefore also drop the superscript $l$, i.e. we simply write 
$\Gmap = \Gmap^l$.
\smallskip 

Let us stress that $\Gmap_h$ provides an isomorphism of linear spaces for generic values 
of $h$. We can think of the collection $\Gmap = (\Gmap_h)$ as a map that takes a pair 
$(h,Z)$ with $h \in A_p$ and $Z \in U(\mathfrak{a}_{p_c}) \otimes \mathcal{K}_2$ to the 
element 
$$ \Gmap( h,Z) = \Gmap_h(Z) \in U(\mathfrak{g}_c)\ . $$  
Actually, we prefer to work with maps between linear spaces and therefore want to extend 
$\Gmap$ linearly from the group $A_p$ to its group algebra \textrm{Fun}$(A_p)$. We shall 
denote the associated map by $\Gmap$ as well, i.e.  
\begin{equation}\label{tilde-Gamma-map}
    \Gmap : \textrm{Fun}(A_p)\otimes U(\mathfrak{a}_{p_c}) \otimes \mathcal{K}_2  \to
    U(\mathfrak{g}_c)\ .
\end{equation}
We are finally in a position to fully define Harish-Chandra's (left) radial component 
map $\Pi$. In fact, 
\begin{equation} 
\Pi: U(\mathfrak{g}_c)\xrightarrow{}\textrm{Fun}(A_p)\otimes U(\mathfrak{a}_{p_c})\otimes \mathcal{K}_2\,,
\end{equation} 
is defined such that for almost all $h\in A_p$ and all $u\in U(\mathfrak{g}_c)$ we have 
$\Gmap(\Pi(u)) = u$. In other words, the radial component map sends elements in the 
universal enveloping algebra $U(\mathfrak{g}_c)$ to differential operators on the 
abelian group $A_p$ whose coefficients are valued in $\mathcal{K}_2$. The most important 
property of the radial component map $\Pi$ is that, for $f\in\Gamma_{\rho_l,\rho_r}$, 
\begin{equation}\label{Harish-Chandra-theorem}
    (u f)|_{A_p} = \Pi(u)(f|_{A_p}) , \quad \forall u\in U(\mathfrak{g}_c)\ .
\end{equation}
Here, the action of $u$ on $f$ is defined through the the left regular action. A similar formula exists for the right regular action but with $\Pi^r$ instead of $\Pi = \Pi^l$. 
In dealing with $K$-spherical functions, this is a very useful result. By definition, a $K$-spherical function on the group $G$ is entirely determined by the values it takes 
on $A_p$. If we want to act with left or right invariant vector fields on a spherical function, we first need to extend these from $A_p$ to $G$ and apply the vector fields. 
The radial component map allows us to shortcut this procedure and act directly on the restriction to $A_p$, without the need to extend to whole of $G$. Let us note that 
the action of vector fields on spherical functions does not preserve the covariance properties of a spherical function, in general. In other words, given $f \in 
\Gamma_{\rho_l,\rho_r}$, the function $uf$ is not in $\Gamma_{\rho_l,\rho_r}$ unless e.g.\ $u$ is a Casimir element so that it commutes with all $k \in K$. In case $u$ 
does have this property, the associated differential operators $\Pi(u)$ does act with the space on the right hand side of eq.\ \eqref{eq:GammaL1} and we have 
$\pi(u f) = \Pi(u) \pi(f)$.

\subsection{Applications of the radial component map}

As an application of the radial component map, we now wish to compute the radial part of the quadratic Casimir, $\Pi(C_2)$, in terms of simple Lie-algebraic data. As we 
have pointed out, the left-hand side of eq.\ \eqref{radial-component-map} may be identified with the space of left-invariant differential operators on $G$ and the 
right hand side with differential operators on $A_p$ with coefficients in $U(\mathfrak{k}_c)\otimes U(\mathfrak{k}_c)$. By replacing the two copies of 
$U(\mathfrak{k}_c)$ in eq.\ \eqref{radial-component-map} by representations of $K$, 
the radial component map provides us with the restriction of the Laplacian to the appropriate space of $K$-spherical functions on $G$. In this sense, $\Pi$ from above 
is universal and captures simultaneously all spaces of $K$-spherical functions. Having obtained the universal spinning Casimir, we will explain the construction of differential shifting operators. The latter are obtained in the second subsection by applying the radial component map to certain vector fields. Differential shifting operators are an important tool for the construction of eigenfunctions of spinning Casimir operators. In combination with internal weight-shifting operators, described in the last subsection, they often suffice to obtain all eigenfunctions of the spinning problem from those of the scalar case in which both $\rho_l$ and $\rho_r$ are characters. 

\subsubsection{The universal spinning Casimir}

In order to compute the radial part of the quadratic Casimir, i.e. the quantity 
$\Pi(C_2)$, let us begin by fixing a generic point $h = e^{t_i H_i}$ of $A_p$. 
Our first goal is to rewrite the quadratic Casimir operator in terms of the 
generators of $U(\mathfrak{a}_{p_c}) \otimes U(\mathfrak{k}_c) \otimes 
U(\mathfrak{k}_c)$. Given our choice of $h$ we introduce the shorthand $y'_\alpha 
= h^{-1} y_\alpha h$ for the restricted root vectors that are obtained from 
$y_\alpha \in \mathfrak{k}_c$, $\alpha\in P_+ \cup (-P_+)$ by conjugation with 
$h$. More explicitly these take the form
\begin{equation}\label{yp-equation}
     y'_\alpha = \frac12 h^{-1} (e_\alpha + \theta(e_\alpha)) h =
     \frac12 \left(e^{-\alpha\cdot t}e_\alpha + e^{\alpha\cdot t}\theta(e_\alpha)\right) =
     \cosh(\alpha\cdot t) y_\alpha - \sinh(\alpha\cdot t) z_\alpha\ .
\end{equation}
In the first and the third equality we used the definition of $y_\alpha$ and 
$z_\alpha$, see eq. \eqref{y-and-z-definition}. The second equality follows from 
$\theta(\mathfrak{g}^\lambda) = \mathfrak{g}^{-\lambda}$. If one uses the last 
line to express $z_\alpha$ in terms of $y_\alpha, y'_\alpha$ and then substitutes 
the result in $e_\alpha = y_\alpha + z_\alpha$, one arrives at
\begin{equation}\label{ealpha-y-yp}
    e_\alpha = \frac{1}{\sinh(\alpha\cdot t)}\left(e^{\alpha\cdot t}y_\alpha - y'_\alpha\right)\ .
\end{equation}
Having found an expression for the restricted root vectors $e_\alpha$ in terms of $y_\alpha$ 
and $y'_\alpha$ we can now turn to the Casimir operator \eqref{quadratic-Casimir-first-expression} 
and evaluate the products
\begin{align*}
    & e_\alpha e_{-\alpha} = \frac{-1}{\sinh^2(\alpha\cdot t)}(y_\alpha y_{-\alpha} + y'_\alpha y'_{-\alpha}
    - e^{\alpha\cdot t} y_\alpha y'_{-\alpha} - e^{-\alpha\cdot t}y'_\alpha y_{-\alpha})\\[2mm]
    & = \frac{-1}{\sinh^2(\alpha\cdot t)}(y_\alpha y_{-\alpha} + y'_\alpha y'_{-\alpha}
    - e^{\alpha\cdot t} y'_{-\alpha}y_\alpha - e^{-\alpha\cdot t}y'_\alpha y_{-\alpha}
    - e^{\alpha\cdot t}[y_\alpha,\cosh(\alpha\cdot t)y_{-\alpha} + \sinh(\alpha\cdot t) z_{-\alpha}])\ .
\end{align*}
To get to the second line, we commuted $y'_{-\alpha}$ past $y_\alpha$ to the left and used 
the equation \eqref{yp-equation}. Applying $\theta$ to the last expression transforms $y'_\alpha$ 
to $h y_\alpha h^{-1}$ and changes the sign of the term that involves $z$. Replacing $\alpha$ by 
$-\alpha$ and simultaneously $h$ by $h^{-1}$ we find
\begin{align*}
    &\theta(e_{-\alpha} e_{\alpha}) =\\
    &=\frac{-1}{\sinh^2(\alpha\cdot t)}\left(y_{-\alpha} y_\alpha + y'_{-\alpha} y'_\alpha- e^{\alpha\cdot t}
    y'_\alpha y_{-\alpha} - e^{-\alpha\cdot t}y'_{-\alpha} y_\alpha - e^{\alpha\cdot t} [y_{-\alpha},
    \cosh(\alpha\cdot t) y_{\alpha} - \sinh(\alpha\cdot t) z_{\alpha}] \right)\ .
\end{align*}
Summing the last two equations leads to
\begin{align*}
    e_\alpha e_{-\alpha} + \theta(e_{-\alpha}e_\alpha) = -\frac{\{y_\alpha,y_{-\alpha}\} + \{y'_\alpha,y'_{-\alpha}\}
    - 2\cosh(\alpha\!\cdot\!t) (y'_\alpha y_{-\alpha}\!+\!y'_{-\alpha} y_\alpha)}{\sinh^2(\alpha\cdot t)}
    - \frac{e^{\alpha\cdot t}([y_{-\alpha},z_\alpha]-[y_\alpha,z_{-\alpha}])}{\sinh(\alpha\cdot t)}.
\end{align*}
Notice that the first term on the right-hand side is invariant under $\alpha\mapsto-\alpha$. Consequently, 
one may deduce
\begin{align*}
    & \sum_{\alpha\in P_+}\{e_\alpha, e_{-\alpha}\} = \frac12 \sum_{\alpha\in P_+} \Big( \{e_\alpha, e_{-\alpha}\}
    + \theta\left(\{e_\alpha, e_{-\alpha}\}\right) \Big)\\[2mm]
    & = - \sum_{\alpha\in P_+} \left( \frac{\{y_\alpha,y_{-\alpha}\} + \{y'_\alpha,y'_{-\alpha}\}
    - 2\cosh(\alpha\cdot t) (y'_\alpha y_{-\alpha} + y'_{-\alpha} y_\alpha)}{\sinh^2(\alpha\cdot t)}
    + \coth(\alpha\cdot t) ([y_{-\alpha},z_\alpha]-[y_\alpha,z_{-\alpha}])\right)\\[2mm]
    & = - \sum_{\alpha\in P_+} \left(\frac{\{y_\alpha,y_{-\alpha}\} + \{y'_\alpha,y'_{-\alpha}\}
    - 2\cosh(\alpha\cdot t) (y'_\alpha y_{-\alpha} + y'_{-\alpha} y_\alpha)}{\sinh^2(\alpha\cdot t)}
    - \coth(\alpha\cdot t) h_{\tilde\alpha}\right),
\end{align*}
where $h_{\tilde\alpha}$ is defined by $\tilde\alpha(H)=\kappa(h_{\tilde\alpha},H)$ for all 
$H\in\mathfrak{a}_p$. In arriving at this formula we have also used that the sum on the left 
side is actually invariant under the action of the Cartan involution $\theta$. To get to the 
last line we used that $h_{\tilde\alpha}$ is also  the $\mathfrak{p}_c$-part of $h_\alpha$. 
Therefore, by taking the $\mathfrak{p}_c$-part of the relation $[e_\alpha,e_{-\alpha}] = 
h_\alpha$ we obtain 
\begin{equation}
    [y_\alpha,z_{-\alpha}] + [z_\alpha,y_{-\alpha}] = h_{\tilde\alpha}\,,
\end{equation}
which was substituted in the last step of the above manipulation. From these auxiliary results it is now
simple to read off the radial part of the quadratic Casimir operator \eqref{quadratic-Casimir-first-expression}. 
Writing $y^{(i)}_\alpha$ for the element that equals $y_\alpha$ on the $i$-th tensor factor of $U(\mathfrak{k}_c)$, 
we conclude
\begin{equation}\label{Casimir-radial-part}
    \Pi(C_2) = \Pi(C_2^{(\mathfrak{m})}) + h^{ij}H_i H_j + \sum_{\alpha\in P_+} \coth(\alpha\cdot t) 
    h_{\tilde\alpha}\ - 2\sum_{\alpha\in P_+}\frac{y^{(1)}_\alpha y^{(1)}_{-\alpha} - 2\cosh(\alpha\cdot t) 
    y^{(1)}_\alpha y^{(2)}_{-\alpha} + y^{(2)}_\alpha y^{(2)}_{-\alpha}}{\sinh^2(\alpha\cdot t)}\ .
\end{equation}
This is the main result of the present section. To turn the universal radial part into a differential operator acting on $K$-spherical functions, we make the replacements $H_i\xrightarrow{}\partial_{t_i}$. The resulting differential operator with coefficients in $U(\mathfrak{k}_c)\otimes U(\mathfrak{k}_c)$ will be denoted by  the same letter as 
the element \eqref{Casimir-radial-part}. If one furthermore replaces the abstract generators $y_\alpha^{(i)}$ by matrices in representations $\rho_l$ and $\rho_r$, one gets the the reduction of the Laplacian $\Pi(C_2)_{\rho_l,\rho_r}$ to the space of $K$-spherical functions $\Gamma_{\rho_l,\rho_r}$. Notice that $\Pi(C_2^{\mathfrak{m}})_{\rho_l,\rho_r}$ is simply a matrix of constants. 

As claimed before, we can bring the radial part of the quadratic Casimir $C_2$ into the form of a Schr\"odinger 
operator. Indeed, this is achieved by conjugating the differential operator $\Pi(C_2)$ with the function
\begin{equation}\label{Haar-measure}
    \delta : A_p \xrightarrow{}\mathbb{C}, \quad \delta(a) = \prod_{\alpha\in P_+} \sqrt{\sinh(\alpha\cdot t)} \ .
\end{equation}
It is straightforward to verify that after conjugation with $\delta$  the first order terms are removed from 
the differential operator $\Pi(C_2)$. 
\medskip 

Up to now, our calculation was carried our for arbitrary Gelfand pairs. We will now specialise it to the pair $G= \SO(d,2)$
and $K$ its maximal compact subgroup. As we explained in the previous subsection, the real rank of this group 
is two and we have already selected the two generators $H_1$ and $H_2$ of the 2-dimensional abelian group 
$A_p$, see eq. \eqref{eq:H1H2}. Given some choice of $h = h(t_1,t_2)$ we can decompose the vectors 
$e^{(a)}_\lambda$ as
\begin{equation}
    e^{(a)}_\lambda = \frac{1}{\sinh(\lambda\cdot t)}
    \left(e^{\lambda\cdot t}y^{(a)}_\lambda - y'^{(a)}_\lambda\right), \quad \lambda\in\Sigma\ .
\end{equation}
Following the general discussion this allows us to write the quadratic Casimir of $\SO(d,2)$ in the
following form
\begin{equation}\label{Casimir-so(d,2)}
    C_2 = H_1^2 + H_2^2 + \sum_{\lambda\in\Sigma_+} m(\lambda)
    \coth(\lambda\cdot t)\lambda\cdot H + \frac{y^{(a)}_\lambda y^{(a)}_\lambda -
    2\cosh(\lambda\cdot t) y'^{(a)}_\lambda y^{(a)}_\lambda + y'^{(a)}_\lambda y'^{(a)}_\lambda}
    {\sinh^2(\lambda\cdot t)}-\frac12 L^{ab}L_{ab}\ .
\end{equation}
Here, we have written $t=(t_1,t_2)$ and $H=(H_1,H_2)$ and the summation over $a$ is understood. 
Now we can apply the Harish-Chandra radial component map. After conjugation with $\delta$, we 
obtain the following Schr\"odinger operator
\begin{align}\label{main-formula}
    H & = \partial_{t_1}^2 + \partial_{t_2}^2 + \frac12 \left(\frac{1}{\sinh^2(t_1+t_2)} +
    \frac{1}{\sinh^2(t_1-t_2)}\right) - \frac{(d-2)(d-4)}{4}\left(\frac{1}{\sinh^2 t_1} +
    \frac{1}{\sinh^2 t_2}\right)\\
    & - \frac{d^2-2d+2}{2} + \sum_{\lambda\in\Sigma_+} \frac{y^{(a)}_\lambda y^{(a)}_\lambda
    - 2\cosh(\lambda\cdot t) y'^{(a)}_\lambda y^{(a)}_\lambda + y'^{(a)}_\lambda y'^{(a)}_\lambda}
    {\sinh^2(\lambda\cdot t)} -\frac12 L^{ab}L_{ab}\nonumber\ .
\end{align}
By construction, this operator describes the action of the quadratic Casimir operator on $K$-spherical
functions where the latter are regarded as Weyl invariant vector valued functions on the abelian group
$A_p$ with the help of the isomorphism \eqref{eq:GammaL1}. The formula \eqref{universal-Hamiltonian} we
spelled out in the introduction is a close cousin of eq.\ \eqref{main-formula}. Namely, one starts from the pair 
$(G,K) = (\SO(d+1,1),\SO(1,1)\times \SO(d))$ and performs the radial decomposition of the Casimir. From the 
discussion above, it is clear that the statement of Harish-Chandra's theorem applies to such a case as well.

\subsubsection{The universal differential shifting operators} 

To construct spherical functions with high dimensional representations $\rho_l$ and $\rho_r$, it is 
often useful to apply differential shifting operators to spherical functions in a space $\Gamma
_{\rho^0_l,\sigma^0_r}$ with simpler (lower dimensional) representations $\rho^0_l$ and $\rho^0_r$. 
By definition, a differential shifting operator is a covariant differential operator which maps spherical 
functions in one space $\Gamma_{\rho_l^0,\rho_r^0}$ to spherical functions in another $\Gamma_{\rho_l,
\rho_r}$ while commuting with all Casimir elements. In this section, we will explain how to construct 
a set of differential shifting operators by looking at radial parts of invariant vector fields on $G$.

The space $\mathfrak{g}$ carries a representation of $\mathfrak{k}$ under the adjoint action. Let us
pick any one irreducible component $\pi$ of this representation and write its basis as $\{X^\mu\}$
\begin{equation}
    [k,X^\mu] = \pi(k)^\mu_{\ \nu} X^\nu, \quad k\in\mathfrak{k}\ .
\end{equation}
Any element $X\in\mathfrak{g}$ gives rise to left- and right-invariant vector fields on $G$, denoted $\mathcal{L}_X$ and $\mathcal{R}_X$. Before going on, let us make a few remarks about left
and right regular actions. By left and right Maurer-Cartan forms, we will understand the Lie
algebra valued forms
\begin{equation}
    g^{-1} dg = dx^i C^L_{ij} X^j,\quad dg g^{-1} = dx^i C^R_{ij} X^j\,,
\end{equation}
respectively. Here we introduced the coefficients $C^L_{ij}$ and $C^R_{ij}$ of these forms
with respect to the coordinate system $(x^i)$ on the group and a choice of basis elements
$X^i$ of the Lie algebra. Left- and right-invariant vector fields in coordinates $x^i$
are computed as
\begin{equation}
    \mathcal{L}_{X^i} = C^{ik}_L \partial_{x^k},\quad \mathcal{R}_{X^i} = C^{ik}_R \partial_{x^k}\,,
\end{equation}
where $C^{ij}_{L,R}$ are inverses of $C^{L,R}_{ij}$. Note that left-invariant vector fields
generate the right regular action
\begin{equation}
    f(e^{\varepsilon X}g) = f(g) + \varepsilon \left(\mathcal{R}_X f\right)(g)
    + o(\varepsilon),
    \quad f(g e^{\varepsilon X}) =
    f(g) + \varepsilon \left(\mathcal{L}_X f\right)(g) + o(\varepsilon), \quad \varepsilon\to0\ .
\end{equation}
With the above conventions, left-invariant fields satisfy the same commutation relations
as Lie algebra generators, while right-invariant vector fields satisfy the opposite brackets.

Having fixed the conventions, we return to $K$-spherical functions. Vector fields act on these
functions component-wise. Infinitesimally, the covariance conditions \eqref{K-spherical-functions}
read
\begin{equation}
    \mathcal{R}_k f^a_{\ \alpha} = \rho^a_{l; b}(k) f^b_{\ \alpha},
    \quad \mathcal{L}_k f^a_{\ \alpha} = f^a_{\ \beta}
    \rho^\beta_{r; \alpha}(k), \quad k\in \mathfrak{k}\ .
\end{equation}
Let us consider the the set of functions $\mathcal{L}_{X^\mu}f$ that are obtained from some
$K$-spherical function $f$ by acting with a Lie derivative. Upon action with a vector field
$\mathcal{L}_k$, $k \in \mathfrak{k}$, these functions behave as
\begin{eqnarray}
    \mathcal{L}_k \mathcal{L}_{X^\mu} f^a_{\ \alpha}
    & = &
    ([\mathcal{L}_k,\mathcal{L}_{X^\mu}] + \mathcal{L}_{X^\mu}\mathcal{L}_k ) f^a_{\ \alpha}
    = \pi(k)^\mu_{\ \nu}\mathcal{L}_{X^\nu} f^a_{\ \alpha} +
    \mathcal{L}_{X^\mu} f^a_{\ \beta}\rho^\beta_{r,\alpha}(k)\nonumber \\[2mm]
    & = &
    (\pi(k)^\mu_{\ \nu}\delta^\beta_\alpha + \delta^\mu_\nu\rho^\beta_{r,\alpha}(k))
    \mathcal{L}_{X^\nu} f^a_{\ \beta}\ .
\end{eqnarray}
Similarly, we can also evaluate the action of right invariant vector fields $\mathcal{R}_k$
with $k\in \mathfrak{k}$,
\begin{equation*}
    \mathcal{R}_k \mathcal{L}_{X^\mu} f^a_{\ \alpha} = \mathcal{L}_{X^\mu} \mathcal{R}_k
    f^a_{\ \alpha} = \rho^a_{l,b}(k) \mathcal{L}_{X^\mu} f^b_{\ \alpha}\ .
\end{equation*}
What we conclude from these two short calculations is that the set of functions $\mathcal{L}_{X^\mu}f$ belongs to the space $\Gamma_{\rho_l\otimes 1,\rho_r\otimes\pi}$. Recall that in this subsection $\pi$ denotes any irreducible component of the representation of the maximal compact subgroup $K$ obtained from the adjoint representation of $G$ by restriction. Going through the same type of analysis for the set of functions $\mathcal{R}_{X^\mu}f$ obtained through the action of right-invariant vector fields, one may show that these are elements the space $\Gamma_{\rho_l\otimes\pi,\rho_r\otimes1}$. Hence, the Lie derivatives $\mathcal{L}_{X^\mu}$ and $\mathcal{R}_{X^\mu}$ act as {\it shifting operators} in the sense that they change covariance properties of $f$. Since invariant vector fields commute with the Laplacian, they map eigenfunctions of the Laplacian to eigenfunctions of the Laplacian.

By applying the radial component map $\Pi = \Pi^l$ to the Lie algebra generators $X^\mu \in\mathfrak{g}$, we obtain differential operators $\Pi^{l}(X^\mu)$ on $A_p$ with coefficients in
$\mathcal{K}_2$. Harish-Chandra's theorem \eqref{Harish-Chandra-theorem} tells us the these operators intertwine between radial parts of Laplacians reduced to $\Gamma_{\rho_l,\rho_r}$ 
and  $\Gamma_{\rho_l\otimes\pi,\rho_r\otimes1}$. In the example of $\SO(d,2)$, there exist two sets of such generators, namely, 
\begin{equation}\label{vector-shift-generators}
     A_i = L_{0i} + L_{i,d+1}\ ,\quad B_i = L_{0i} - L_{i,d+1}\ .
\end{equation}
Both sets transform in the vector representation of $\mathfrak{so}(d)\subset\mathfrak{k}$. The decomposition of $\mathfrak{g}$ over $\mathfrak{k}$ contains also the third irreducible component, namely $\mathfrak{k}$ itself. However, by definition, generators of $\mathfrak{k}$ act on spherical functions simply as matrices rather than differential operators. In terms of elements $y^{(a)}_\lambda$ and $y'^{(a)}_\lambda$, generators \eqref{vector-shift-generators} read
\begin{align}
    & A_1 = H_1 + \frac{1}{\sqrt{2}} \left(\coth(t_1-t_2) y_{(1,-1)}
    - \frac{1}{\sinh(t_1-t_2)}y'_{(1,-1)} - \coth(t_1+t_2) y_{(1,1)}
    + \frac{1}{\sinh(t_1+t_2)}y'_{(1,1)}\right)\nonumber \\[2mm]
    & B_1 = H_1 - \frac{1}{\sqrt{2}} \left(\coth(t_1-t_2) y_{(1,-1)}
    - \frac{1}{\sinh(t_1-t_2)}y'_{(1,-1)} - \coth(t_1+t_2) y_{(1,1)}
    + \frac{1}{\sinh(t_1+t_2)}y'_{(1,1)}\right)\nonumber \\[2mm]
    & A_a  = \coth t_1 y^a_{(1,0)} - \frac{1}{\sinh t_1}y'^a_{(1,0)}
    - \coth t_2 y^a_{(0,1)} + \frac{1}{\sinh t_2}y'^a_{(0,1)}\nonumber 
     \\[2mm]
    & B_a  = \coth t_1 y^a_{(1,0)} - \frac{1}{\sinh t_1}y'^a_{(1,0)}
    + \coth t_2 y^a_{(0,1)} - \frac{1}{\sinh t_2}y'^a_{(0,1)}
    \label{eq:shift} \\[2mm]
    & A_d = H_2 + \frac{1}{\sqrt{2}} \left((\coth(t_1-t_2) y_{(1,-1)}
    - \frac{1}{\sinh(t_1-t_2)}y'_{(1,-1)} + \coth(t_1+t_2) y_{(1,1)}
    - \frac{1}{\sinh(t_1+t_2)}y'_{(1,1)}\right)\nonumber \\[2mm]
    & B_d = -H_2+\frac{1}{\sqrt{2}} \left(\coth(t_1-t_2) y_{(1,-1)}
    - \frac{1}{\sinh(t_1-t_2)}y'_{(1,-1)} + \coth(t_1+t_2) y_{(1,1)}
    - \frac{1}{\sinh(t_1+t_2)}y'_{(1,1)}\right)\nonumber 
\end{align}
As in our discussion of the quadratic Casimir, it is now straightforward to apply the radial component map and to obtain first order differential operators in the variables
$t_1,t_2$. After conjugation with $\delta$ these provide shifting operators for the associated spinning Calogero-Sutherland Hamiltonians. The operators we have
discussed here shift the spin on the left and leave the right representation $\rho_r$ unchanged. The construction of operators that shift the right spin by using the appropriate variant of the radial component map $\Pi^r$ is entirely analogous.

\subsubsection{Weight-shifting operators}

In this final subsection, we will construct another set of operators which, used together with differential shifting operators from above, generate a very large class of spherical functions.

To appreciate the problem at hand, let us go back to the interpretation of spherical functions as matrix elements, see section 2.2.1. It is then observed that differential shifting operators change {\it external} representations $\rho_{l,r}$, but keep the {\it internal representation} $\pi$ fixed. For this reason, we shall also refer to these operators as external weight-shifting operators. Since the complexity of matrix elements depends both on $\rho_{l,r}$ and $\pi$, one would wish to obtain analogous internal weight-shifting operators which change $\pi$ and keep $\rho_{l,r}$ fixed.

Our construction makes use of the fact that any space $\Gamma_{\rho_l,\rho_r}$ is a module over $\Gamma_{1,1}$. Elements of the latter are known as zonal spherical functions. Indeed, it is directly observed that if $f_0$ is a zonal spherical function and $f^a{}_\alpha\in\Gamma_{\rho_l,\rho_r}$, the product $f_0 f^a{}_\alpha$ again belongs to $\Gamma_{\rho_l,\rho_r}$. Furthermore, matrix elements satisfy
\begin{equation}
    \pi'^0{}_0 \pi^a{}_\alpha = \sum_{\pi_i\subset\pi'\otimes\pi} c_i\ \pi_i^a{}_\alpha\ .
\end{equation}
The $c_i$-s are Clebsch-Gordan coefficients of $G$, but this will not be important for the argument below. Assume that the tensor product $\pi\otimes\pi'$ decomposes over a finite number of irreducibles $\pi_1,\dots,\pi_n$, each with multiplicity one. If all these representations have different values of the quadratic Casimir, $C_2(\pi_i)$, then we have
\begin{equation}\label{internal-weight-shifing}
    (\Delta-C_2(\pi_1))\dots(\Delta - C_2(\pi_{n-1})) \left(\pi'^0{}_0 \pi^a{}_\alpha\right) = c\ \pi_n^a{}_\alpha\,,
\end{equation}
for some constant $c$. Should $\pi_n$ have the same value of the quadratic Casimir as some other representations $\pi_j$ appearing in the decomposition, the same argument is repeated with higher higher Casimirs until all representations except for $\pi_n$ are projected out. One can apply the radial component map to both sides of eq. \ \eqref{internal-weight-shifing} and possible additional projectors. Due to properties of this map, it is applied to each term in the composition on the left hand side separately, thus giving a comparatively simple equation.

If all the above assumptions are satisfied, we have succeeded in constructing $\pi_n^a{}_\alpha$ from $\pi'^0{}_0$ and $\pi^a{}_\alpha$. In practice, representation $\pi$ is taken to be simple so that $\pi^a{}_\alpha$ are known by external weight-shifting. The representation $\pi'$ can be complicated, with $\pi'^0{}_0$ still known, because this is an eigenfunction of the much-studied scalar Calogero-Sutherland model. It is the complexity of $\pi'$ that allows to have complicated representation $\pi_n$ appearing in the tensor product.

Finally, let us comment on the assumption about finiteness of the decomposition of $\pi\otimes\pi'$ into irreducibles. If $G$ is a non-compact group, typically this assumption does not hold. However, it is satisfied for compact groups $G$. In some applications therefore, our strategy will be to compute spherical functions for a Gelfand pair with both $G$ and $K$ compact and then analytically continue results to cases when either $G$ or both $G$ and $K$ become non-compact. Details of this process will be given in examples below.

\section{Applications to Bulk Conformal Field Theory}

The first application of the universal spinning Casimir operators we constructed in the previous 
section concerns the study of spinning four-point correlators in a $d$-dimensional bulk CFT. 
Spinning four-point functions and the associated blocks have been the subject of intense studies 
since the influential work \cite{Costa:2011dw,Costa:2011mg}. A very successful theory of spinning 
conformal blocks has been developed in the meantime that constructs spinning blocks from scalar 
ones with the help of weight-shifting operators \cite{Karateev:2017jgd}. Below, we will explain 
how the Casimir equations for spinning four-point blocks are related to the universal spinning 
Calogero-Sutherland models constructed in the previous section. In fact, in order to establish 
the relation all that remains to be discussed is the map between bulk four-point and 
$K$-spherical functions. This map was derived in \cite{Buric:2019dfk,Buric:2020buk} - we shall 
state a precise formula in eq.\ \eqref{eq:4ptGFnew}. As we have explained, along with the 
Calogero-Sutherland formulation of Casimir equations comes a system of internal and external 
weight-shifting operators. This differ from their relatives in the CFT literature in that they 
act directly on the cross ratios. But just as their CFT cousin, our shift operators can be used 
to construct eigenfunctions of the Calogero-Sutherland Hamiltonians. The shifting operators 
will only be spelled out in the context of defect correlators, see section 4. Note however, 
that the same formulas can be used for bulk four-point functions since both Casimir equations
are described by the same Calogero-Sutherland model.

\subsection{Review: From four-point functions to spherical functions}

Given four spinning fields that transform in representations $\sigma_i$ of the group $K = 
\SO(1,1)\times\SO(d)$ and that are inserted at four points $x_i\in\mathbb{R}^d$, the desired 
relation between four-poinr correlators and spherical fuunctions takes the form
\begin{equation}\label{eq:4ptGF}
G_4(x_i;\sigma_i) = \Xi(x_i;\sigma_i) \, f(t_1,t_2;\rho_l,\rho_r)\ .
\end{equation}
On the left hand side, $f$ is a restriction to $A_p$ of a $K$-spherical function with 
representations $\rho_l$ and $\rho_r$ of $K = \SO(1,1) \times \SO(d)$ which are determined 
by the weights and spins of the four fields. In a channel in which we pair up the fields 
$i=1,2$ and those with index $i=3,4$, these representations are determined as
\begin{eqnarray} \label{eq:rhol}
\rho_l(D) = \Delta_1 - \Delta_2  \quad  & , \quad &
\rho_l(r) = \sigma_1(r) \otimes \tilde \sigma_2(r)\,,  \\[2mm]
\rho_r(D) = \Delta_4 - \Delta_3 \quad & , \quad &  \rho_r(r) = \sigma_3(r) \otimes \tilde \sigma_4(r)\,,
\label{eq:rhor}
\end{eqnarray}
where $D$ denotes the generator of dilations, $r \in \SO(d)$ is some element in the rotation 
group and $\tilde\sigma_i$ denotes some appropriate conjugate of $\sigma_i$, see below. The 
abelian group $A_p$ is generated by $H_1 = \frac12(P_2+K_2)$ and $H_2 = \frac{-i}{2}(P_3-K_3)$ 
and parametrised as usual, $h = e^{t_i H_i}$. In relation \eqref{eq:4ptGF} these coordinates 
should be regarded as functions of the insertion points $x_i$. The precise relation is actually 
not that difficult to spell out
\begin{equation} \label{eq:ztcoordinates4pt}
\left(\cosh t_1 - \cosh t_2\right)^2 = \frac{4}{z_1 z_2}\,, \quad 
\left(\cosh t_1 + \cosh t_2\right)^2 = \frac{4(1-z_1)(1-z_2)}{z_1 z_2}\,,
\end{equation}
where $z_1$ and $z_2$ parametrise the standard cross ratios $u = z_1 z_2$ and $v = (1-z_1)(1-z_2)$ 
one can form from the four insertion points $x_i$.

It remains to discuss the prefactor $\Xi$ that relates the four-point correlator $G_4(x_i)$ with 
the $K$-spherical function in eq. \eqref{eq:4ptGF}. Initially, when the relation between the two
types of objects was first explored, such prefactors were determined indirectly through a 
comparison of spinning Casimir differential equations. In fact, there are a number of cases
in which these differential equations for ordinary conformal blocks had been worked out in 
the CFT literature, see \cite{Iliesiu:2015akf} and \cite{Echeverri:2016dun} for the main 
examples in $d=3,4$, respectively. For exactly those cases, the associated $K$-spherical 
functions and the Casimir differential equations they obey were constructed in 
\cite{Schomerus:2016epl} and \cite{Schomerus:2017eny}. By comparing the two sets of differential 
equations it was then possible to infer the prefactor $\Xi$ for this limited set of cases in 
which blocks and $K$-spherical functions had both been studied. An independent group theoretic
construction of $\Xi$ was developed later in \cite{Buric:2019dfk}. The resulting expression for 
$\Xi$ were evaluated in case of the 3- and 4-dimensional conformal group. Remarkably, the 
formulas for $\Xi$ are also universal in spin, in the same sense in which the spinning Casimir 
equations for $K$-spherical functions are universal. The group theoretic analysis of $\Xi$ was 
later streamlined a bit further in the context of superconformal algebras \cite{Buric:2020buk}. 
Since this construction of $\Xi$ is crucial to understand the precise relation between conformal 
blocks and $K$-spherical eigenfunctions of the spinning Casimir equations, we will briefly review 
the main ideas here.

In order to do so, let us introduce a bit of notation. Group theoretic decompositions, such as 
the Cartan and Gauss decomposition, have been discussed extensively already in section 2.
Now we shall focus on the factorisation $g=mnk$ of an element $g$ in the conformal group as a 
product of a translation $m$, a special conformal transformation $n$ and an element $k$ from 
the subgroup $K$ that is generated by dilations and rotations. Let us parametrise translations 
$m$ and special conformal transformations $n$ by elements $x \in \mathbb{R}^d$ such that
\begin{equation}
    m(x) = e^{x \cdot P} \quad , \quad n(x) = w e^{x\cdot P} w^{-1}\ .
\end{equation}
Here, $P$-s denote the infinitesimal generators of translations as usual, and $w = 
e^{\pi\frac{K_{d+1} - P_{d+1}}{2}}$ is the Weyl inversion. Given an insertion point $x$ we can 
then look at the following factorisation formula for the products $w m(x) \in G$ with the Weyl 
inversion,
\begin{equation} \label{eq:factorization}
    w m(x) = m(y(x))\,  n(z(x)) \, k(t(x)) \ .
\end{equation}
Through the factorisation of $w m(x) \in \SO(d+1,1)$ we have introduced the functions $y(x)$, 
$z(x)$ and $t(x)$ that determine how the parameters of the factors $m$, $n$ and
$k$ depend on the choice of $x \in \mathbb{R}^d$. Simple expressions for these functions 
may be found in \cite{Dobrev:1977qv,Buric:2021yak}. Once they are known we introduce
\begin{equation}
y_{ij} = y(x_{ij})   \ , \quad  z_{ij}= z(x_{ij}) \,, \quad t_{ij} = t(x_{ij})\,,
\end{equation}
for $x_{ij} = x_i-x_j \in \mathbb{R}^d$ and $i,j = 1, \dots, 4$. By definition we therefore have
\begin{equation} \label{eq:factorizationij}
w m(x_{ij}) = m(y_{ij}) \, n(z_{ij}) \, k(t_{ij}) \ .
\end{equation}
With this notation introduced, we are now in a position to state the main result of \cite{Buric:2019dfk}. It 
states that, given a correlation function $G_4(x_i)$ of four fields that transform in representations 
$\sigma_i$ of the subgroup $K \subset G$ as described before, there exists a unique $K$-spherical 
function $F$ such that
\begin{eqnarray}
G_4(x_i) & = & \Big(1\otimes\sigma_2(k(t_{21}))^{-1}\otimes1\otimes
\sigma_4(k(t_{43}))^{-1}\Big) F(g(x_i))\, ,  \label{magic-formula}\\[2mm]
& & \textit{where}\ g(x_i) =  n(y_{21})^{-1} m(x_{31}) n(y_{43})\ .  \label{eq:gxi}
\end{eqnarray}
The covariance laws of $F$ are governed by the two representations $\rho_l$ and $\rho_r$ that we 
introduced in eqs.\ \eqref{eq:rhol} and \eqref{eq:rhor}, i.e. $F\in\Gamma_{\rho_l,\rho_r}$.
A proof of this remarkable formula was originally given in \cite{Buric:2019dfk}. For the 
reader's convenience we have included a more elegant derivation in Appendix D. Once we 
accept the formula \eqref{magic-formula}, it is not difficult to obtain eq.\ \eqref{eq:4ptGF}. 
All this requires is to apply the Cartan factorisation to the argument $g(x_i)$ of $F$,
\begin{equation}\label{Cartan-factors}
   g(x_i) = k_l(x_i) a (x_i) k_r(x_i) \ .
\end{equation}
The formula \eqref{magic-formula} and the covariance properties of $F$ give
\begin{equation}\label{eq:4ptGFnew}
G_4(x_i)  = \sigma_1(k_l)\sigma_2\left(k(t_{21})^{-1} 
k_l^w\right)\sigma_3(k_r^{-1})\sigma_4\left(k(t_{43})^{-1} (k^{-1}_r)^w\right) F(a)\ . 
\end{equation}
For simplicity, we dropped the dependence of Cartan factors on the insertion points, 
i.e.\ for example $k_l = k_l(x_i)$. The concrete functional dependence has been worked 
out in \cite{Buric:2019dfk}. We now see that $G_4(x_i)$ indeed has the form we claimed 
it to have in eq.\ \eqref{eq:4ptGF} with a function $f(t) = F(a(x_i))$ that is determined 
by the values the $K$-spherical function $F$ takes on the 2-dimensional abelian subgroup
$A_p$, i.e. a function of cross ratios only. Note that the calculation of the matrix 
prefactor $\Xi$ is reduced to a group theoretic decomposition that determines the factors 
$k(t_{ij})$, $k_l$ and $k_r$ in terms of the insertion points $x_i$. For the conformal
group in $d=3$ and $d=4$ dimensions, this computation has been carried out in
\cite{Buric:2019dfk}. Once the relevant group theoretic factors are known,
they must be evaluated in representations $\sigma_i$ that depend on the
weights and spins of the involved fields. 

\subsection{Application: OPE limits of six-point blocks}

Now that we have reviewed how four-point functions $G_4(x_i)$ of local operators 
may be written in terms if $K$-spherical functions $F$ with $K = \SO(d) \times 
\SO(1,1)$, we can fully appreciate the importance of the radial component map in 
the context of CFT. In fact, we can now conclude that the differential equations 
satisfied by the blocks are derived from the radial decomposition of the quadratic 
Casimir element. After conjugation with the factor $\delta$ introduced in eq.\ 
\eqref{Haar-measure} we obtain the spinning Calogero-Sutherland Hamiltonian 
\eqref{universal-Hamiltonian}. It seems quite remarkable that, with our choice of 
coordinates and `gauge factor' $\Omega = \Xi \delta$, the spinning Casimir equations 
take such a compact form.
\medskip 

Nevertheless, one may wonder to which extend our new insights go beyond what was known 
about spinning four-point blocks in the CFT literature. Part of the answer lies in the 
difference between universal and recursive. The most powerful existing weight-shifting 
techniques allow to construct conformal blocks recursively, e.g. increasing the rank 
of the involved tensors step by step. This is useful for explicit evaluations of the 
blocks but rather cumbersome when it comes to e.g. establish their more general features 
such as behaviour in certain limits etc. In the next section, we will construct 
weight-shifting operators for spinning Calogero-Sutherland Hamiltonians, which provide 
a recursive construction of the blocks, very much in the spirit of 
\cite{Costa:2011dw,Karateev:2017jgd}. For the remainder of this section, we will focus 
on an application in the context of multipoint blocks, \cite{Buric:2021kgy}, which 
illustrates the usefulness of universality.

Conformal blocks for correlation functions of $N > 4$ scalar fields have received quite 
a bit of attention recently, see \cite{Rosenhaus:2018zqn,Pal:2020dqf,Goncalves:2019znr,Anous:2020vtw,Poland:2021xjs,Fortin:2022grf} and references therein, though a complete theory similar to the one for $N=4$ is not yet within reach. In \cite{Buric:2020dyz} it was proposed to 
characterise scalar multipoint blocks through a set of differential equations. The 
number of equations that are needed to fully determine the blocks (along with appropriate 
boundary conditions) must agree with the number of conformal cross ratios. It was shown 
that a complete set of such differential operators could be obtained from the commuting 
Hamiltonians of a Gaudin integrable system for the conformal group. Setting up the latter 
requires to fix $N$ points on the a 2-dimensional sphere. For each configuration of these 
points one obtains a sufficiently large number of mutually commuting Hamiltonians. In 
certain special limits of the Gaudin model, these Hamiltonians include the Casimir 
operators of an $N$-point function. The precise set of Casimir operators depends on the 
choice of an operator product channel. Any such channel can be reached by colliding 
punctures on the 2-sphere on which the Gaudin model is defined, see \cite{Buric:2021ywo} 
for details. 

\begin{figure}[thb]
    \centering
    \includegraphics{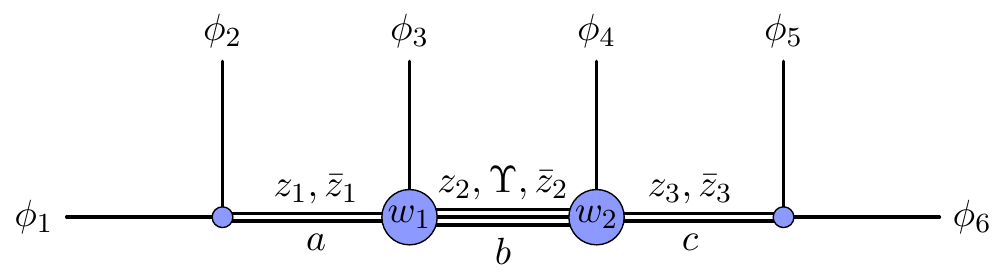}
    \caption{Six point function with external scalars in the comb channel. 
    The $z_i$, $\bar{z}_i$, $w_i$ and $\Upsilon$ type of cross ratios are 
    naturally associated with one particular internal leg or vertex of the 
    OPE diagram, see \cite{Buric:2021kgy} for details.} 
    \label{fig:six_point_comb}
\end{figure}

\begin{figure}[ht]
    \centering
    \includegraphics{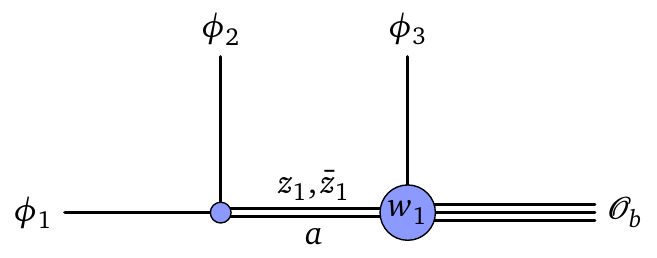}
    \caption{One of the four-point functions obtained in the OPE limit for the middle leg in a six-point 
    function in comb channel. The rightmost field is a mixed symmetry tensor with two spin indices and the 
    exchanged field is a symmetric traceless tensor.}
    \label{fig:one_MST_four_point}
\end{figure}

Conformal block expansions of multipoint correlators are only useful if one has sufficient 
control over the OPE limit of the individual blocks so that one can distinguish the dynamical 
OPE coefficients from the values the blocks take in these limits. Within the integrability 
based approach to multipoint blocks we have just sketched, the issue of OPE limits was 
addressed in \cite{Buric:2021kgy}. In this work, a new set of multipoint cross ratios was 
introduced that extends the familiar variables $z,\bar z$ Dolan and Osborn introduced for 
$N=4$ point correlators. The new set is adapted to taking OPE limits in the sense that, 
when written in these new cross ratios, the power series expansion of 
the blocks reproduces the sum over descendants in each intermediate exchange. What one had 
to show, however, was that the leading term in the OPE limits factorises into the appropriate 
product of lower-point spinning blocks. An example of a scalar block with $N=6$ insertions 
is depicted in Figure \ref{fig:six_point_comb}. Upon taking the OPE limit 
in the three variables that are attached to the intermediate channel in the middle, one 
expects the six-point block to split into a product of two four-point blocks with three 
scalar and one spinning field of arbitrarily high spin, see Figure \ref{fig:one_MST_four_point}.  
This is the context in which the differential operators became an ideal and powerful 
tool. Before taking the OPE limit, the multipoint blocks are characterised by the set 
of Gaudin Hamiltonians we described above. Once these were expressed in terms of the 
new variables, it was possible to take the OPE limit directly on the level of the 
differential operators and thereby to derive a set of differential equations that 
are satisfied by the OPE limit of the $N=6$-point block. If the latter was to split into 
a product of spinning four-point blocks, the limiting differential operators of the 
six-point blocks should coincide with the universal spinning Casimir operators of the 
left and the right four-point function, after an appropriate change of variables. This 
is indeed what was established in \cite{Buric:2021kgy}. Note that a six-point function 
in $d\geq 4$ depends on nine cross ratios. The OPE limit in the middle internal leg, in 
which a mixed symmetry tensor is exchanged, corresponds to a limiting point for three of 
these nine cross ratios. Hence, the leading term only depends on six cross ratios. Three 
of these are associated with the four-point block on the right, the other three with the 
left factor. These three cross ratios can indeed be mapped to the variables of a universal 
spinning Casimir operator, where two variables are $t_1$ and $t_2$ while the third one 
enters through the construction of the mixed symmetry tensor, see \cite{Buric:2021kgy} 
for an extensive discussion. The Hamiltonian will also be written in eq.\ 
\eqref{eq:HCSMST3scalars} below, as the exact same operator appears in the context 
of defect CFTs.

This discussion illustrates nicely how useful a universal characterisation of conformal 
blocks can be. We will also discuss other applications in the concluding section, including 
some comments on a universal solution theory that could eventually complement the recursive 
constructions through weight-shifting operators. But for the time being we want to move on 
from he discussion of spinning four-point blocks, which are quite well understood already, 
and address another set of blocks in defect CFT about which less is known and which 
turns out to be characterised by the same universal spinning Casimir operators. 

\section{Applications to Defects of Co-dimension \textit{q} = 2}

We now turn our attention to applications of the universal spinning Casimir operators to the study of 
CFTs in the presence of a defect of co-dimension $q=2$. Our task consists of two parts - firstly to 
establish the precise relation between CFT correlation functions and appropriate spherical functions, 
similar to eq.\ \eqref{eq:4ptGFnew}, and secondly to construct wavefunctions of the resulting 
Calogero-Sutherland Hamiltonian, i.e. conformal blocks. We shall address these challenges in two 
related cases. In the first two subsections we study correlation functions involving insertions of 
two spinning bulk fields in the presence of a co-dimension $q=2$ defect. Conformal partial waves for 
both the defect and the bulk channel expansion of such a correlator can be mapped to appropriate 
spherical functions of rank one and two, respectively. We will construct general wavefunctions 
through application of weight-shifting operators, for traceless symmetric tensor fields in the 
bulk of arbitrary dimension $d$. Universal spinning Casimir operators in the bulk channel coincide 
with those relevant for spinning four-point functions. Discussions of blocks in the two cases only 
differ in analysis of boundary conditions and the prefactor that relates the CFT correlator and the 
spherical function. These facts will be manifest in our analysis and extend observations of 
\cite{Billo:2016cpy,Isachenkov:2017qgn} to include spin. Furthermore, in our approach to solution 
theory it is sufficient to determine the boundary conditions in the scalar case as they are 
respected by weight-shifting operators. Our results in spinning two-point functions provide
the complete input for the analysis of the associated crossing symmetry constraints. 

In the third subsection we briefly describe one additional step and consider three-point functions 
involving an additional field $\hat\varphi$ that is localised on the defect. Once again we will be 
able to establish a precise relation between such three-point functions and spherical functions. 
This leads to very compact Casimir equations for the bulk-channel conformal blocks with the help 
of the universal spinning Casimir operators we described in section 2. For this setup involving 
two  bulk and one defect field, the corresponding defect blocks were constructed recently in 
\cite{Buric:2020buk} at least for scalar bulk fields are scalar. The work here brings us a step 
closer to constructing the associated bulk channel blocks and thereby to studying crossing 
equations for the three-point function. 
\smallskip

\noindent
{\bf Some Comments on Notation.} Throughout this section, we will be using what has become the 
standard notation in the defect CFT literature. A point in the bulk spacetime $M = \mathbb{R}^d
\cup\{\infty\}$ is denoted by $x=(x^\mu)$. Projections of $x$ to the defect subspace and its 
orthogonal complement are denoted by $\hat x = x_\parallel = (x^a)$ and $x_\perp = (x^i)$, 
respectively. The dimension of the defect is denoted by $p$, and its co-dimension by $q=d-p$. 
As said above, we fix $q=2$. In particular, with our conventions for bulk indices from the 
previous sections, we agree that 
\begin{equation}
    \mu,\nu = 2,\dots,d+1, \quad a,b = 2,\dots,d-1, \quad i,j = d,d+1\ .
\end{equation}
Fields in the bulk are written as $\varphi$, while fields localised on the defect carry a hat, 
$\hat\varphi$. The transformation properties of a bulk field are encoded in a representation 
$\sigma$ of $\SO(1,1)\times\SO(d)\equiv K_d$, with the carrier space $W$. A defect field is 
characterised by a representation $\hat\sigma$ of $\SO(1,1)\times\SO(p)\times \SO(q)$ with 
the carrier space $\hat W$. Finally, $G_d\sim SO(d+1,1)$ will stand for the conformal group 
and $G_{d,p}\sim \SO(p+1,1)\times \SO(q)$ for the defect conformal group, i.e. the group of 
those conformal transformations that preserve the defect subspace.

\subsection{Bulk two-point functions - the defect channel}

In this and the next subsection, we will be looking at two bulk primary fields $\varphi_i$, 
$i=1,2$, with conformal weights $\Delta_i$. We will allow these two fields to carry arbitrary 
spin and study correlation functions of the form \eqref{eq:2pointdefect}. The defect is 
preserved by the subgroup $G_{d,p=d-2} = \SO(d-1,1) \times \SO(2)$ of the conformal group. 
Here, $\SO(d-1,1)$ is the conformal group of the $(d-2)$-dimensional defect and elements 
in the second factor $\SO(2)$ are rotations around the defect. The insertion of such a 
defect is characterised by $2d$ parameters which one can think of as parametrising points 
$\mathcal{X}$ on the coset manifold $\mathcal{X} \in G_d/G_{d,d-2}$.

In the defect channel, each of the fields $\varphi_i$ is expanded in defect primaries 
$\hat\varphi_i$. In order to have a non-vanishing two-point function on the defect, 
the two fields $\hat\varphi_i$ need to have equal quantum numbers. Associated conformal 
blocks are therefore labelled by irreducibles of $\SO(1,1)\times\SO(p)\times\SO(q)$ 
(together with additional quantum numbers detailed below). They are eigenfunctions 
of all Casimir operators of the defect conformal group acting at the point $x_1$.

As in the case of four-point blocks discussed in the previous section the first important step is to establish the precise relation between correlators and spherical functions. For the defect channel of bulk two-point functions with scalar bulk insertions, this relation was worked out in \cite{Buric:2020zea}. In the case of spinning fields, the appropriate generalisation will be derived below. After establishing it, we compute wavefunctions of the resulting spinning Calogero-Sutherland model, and thereby the blocks.

\subsubsection{Construction of the lift}

In this subsection we want to map the defect conformal blocks to appropriate spherical harmonics on the defect conformal group $G_{d,p}$. As we shall show, this can be done with the rank one Gelfand pair $(\SO(d-1,1)\times\SO(2),\SO(d-1)\times\SO(2))$. The basic strategy is adopted from our previous work \cite{Buric:2020zea} on scalar bulk insertions. The main step is to lift an individual (spinning) bulk primary field $\varphi:\mathbb{R}^d\to W$ to a function $f_\varphi : G_{d,p}\to W$ on the defect conformal group. Such an uplift then also allows to map the two-point function \eqref{eq:2pointdefect} to a function on two copies of the defect conformal group, i.e. to an object $F_{2,0}:G_{d,p}^2\to W_1\otimes W_2$. The conditions \eqref{eqn-iota-Phi} we impose on the lift of the individual primary fields ensure that (1) the correlator can be recovered from the function $F(g) = F_{2,0}(e,g)$ on the defect conformal group, (2) $F(g)$ is a spherical function with respect to the subgroup $K=\SO(d-1)\times\SO(2)$ and (3) the action of defect Casimir operators at the point $x_1$ on the correlation function $G_{2,0}(x_i)$ is carried to the action of Laplace operators on $F$.

In order to lift an individual bulk primary to the defect conformal group, the first step is to embed the space $\mathbb{R}^d$ of bulk insertion points into $G_{d,p}$. Following \cite{Buric:2020zea}, this is done according to
\begin{equation} \label{eq:defect_embedding}
    g_d : M \to G_{d,p}, \quad g_d(x) = e^{x^a P_a} |x_\perp|^D e^{\phi^i M_{i,d+1}}\ .
\end{equation}
The coordinates $\phi^i$ are defined in such a way that $e^{\phi^i M_{i,d+1}}$ maps the vector $e_{d+1}$ to $x_\perp/|x_\perp|$. Given two bulk insertion points $x_1$ and $x_2$, we can construct two elements $g_i = g_d(x_i)$ for $i=1,2$. In constructing the precise relation between $G_{2,0}(x_i)$ and spherical functions it will be important to perform the following decomposition of
the product
\begin{equation}\label{eq:defect20decomp}
    g_1^{-1} g_2 = g_d(x_1)^{-1} g_d(x_2) = b_l e^{\lambda D} b_r\ e^{\kappa M_{d,d+1}}\ .
\end{equation}
The last factor of our decomposition involves the generator $M_{d,d+1}$ of $\SO(2)$. The first three factors arise from the Cartan decomposition of $\SO(d-1,1)$ into a product of a left rotation $b_l\in\SO(d-1)$, a dilation and a right rotation $b_r\in\SO(d-1)$. Let us stress that all four factors in the decomposition \eqref{eq:defect20decomp} are functions of the insertion points $x_i$, with $b_{l,r}$ defined up to a defect rotation $r\in\SO(d-2)$ that commutes with dilations. For the coordinates $\lambda$ and $\kappa$ that parametrise the dilations and transverse rotations, the dependence on the insertion points takes the simple form
\begin{equation}\label{eq:lambdakappa}
    \cosh\lambda = \frac{x_{1\perp}^2 + x_{2\perp}^2 +\hat x_{12}^2}{2|x_{1\perp}| |x_{2\perp}|},
    \quad \cos\kappa = \frac{x_1^i x_2^i}{|x_{1\perp}| |x_{2\perp}|}\ .
\end{equation}
The precise coordinate dependence of the two factors $b_{l,r}$ can also be worked out, see \cite{Buric:2020buk} for similar calculations, but since we will not need explicit formulas below
we refrain from spelling them out.

So far we were concerned with the insertion points of the fields $\varphi$. Let us now turn our attention to the spaces $W_i$ the fields $\varphi_i$ take their values in. With the experience we have gathered, a natural impulse is to implement the choice of representations $\sigma$ through a covariance law for functions on the group. But this does not quite work since $K_d$ is not a subgroup of the defect conformal group $G_{d,p}$. Instead, it is easy to see that the relevant subgroup of $G_{d,p}$ that stabilises a point $x$ in the bulk is given by $S_{d,p} = \SO(p+1) \times \SO(q-1) \cong \SO(d-1)$. The idea now is to embed the stabiliser group $S_{d,p}$ into the group $K_d$ through some homomorphism
\begin{equation}
    \label{eq:iota}
    \iota: S_{d,p} \rightarrow K_d\ .
\end{equation}
This embedding then allows to pull any representation $\sigma$ of $K_d$ back to a representation $\mu$ of $S_{d,p}$, i.e. given $\iota$ we set
\begin{equation}\label{mu-iota}
\mu := \sigma \circ \iota\ .
\end{equation}
We can now use this representation $\mu$ to introduce a covariance law for
a function $f$ on the defect conformal group
\begin{equation} \label{eq:mufromrho}
    f(gs) = \mu(s)^{-1} f(g), \quad g\in G_{d,p},\ \ s\in S_{d,p}\ .
\end{equation}
Here, $f$ takes values in the space $W$, regarded however as the carrier space of $\mu$ rather than $\sigma$. Following \cite{Buric:2020zea}, we do not simply require that $f(g_d(x))$ coincides with $\varphi(x)$, but allow for the two to be related by a matrix factor $\Phi(x)$
\begin{equation}\label{eq:fPhiphi}
    f(g_d(x)) = \Phi(x)\varphi(x)\  \quad \textit{ where } \quad \Phi(x): W \rightarrow W\ .
\end{equation}
In \cite{Buric:2020zea}, we found a sufficient condition on $\mu$ and $\Phi$ which ensures that the map $\varphi\mapsto f$ is an intertwiner of $G_{d,p}$-representations. For representations $\mu$ constructed according to eq.\ \eqref{mu-iota}, this condition can be written as
\begin{equation}\label{eqn-iota-Phi}
    \Phi(hx)^{-1}\sigma(\iota(s_d(x,h)))\Phi(x) = \sigma(k(x,h)), \qquad \forall h\in G_{d,p}\,,
\end{equation}
where all the notation is defined in \cite{Buric:2020zea}. Equation \eqref{eqn-iota-Phi} is telling us how $\Phi$ and $\iota$ should be coordinated in order for the lift \eqref{eq:fPhiphi} to be consistent with defect conformal symmetry. A particular solution $(\Phi,\iota)$ to conditions \eqref{eqn-iota-Phi} may be chosen as follows. The matrix valued function $\Phi(x)$ takes the form
\begin{equation}\label{solution-to-lift-eqns}
    \Phi(x) = |x_\perp|^\Delta\sigma(e^{-\theta M_{d,d+1}})\,, \qquad \theta = \text{arg}(x_\perp)\ .
\end{equation}
In the last line, we made use of the fact that co-dimension of the defect is two and thus $x_\perp$ is a point on the plane. The angle $\theta$ is the polar coordinate of this point. To define $\iota$, recall that $S_{d,p}$ is generated by $L_{\alpha\beta}$ with $\alpha,\beta=1, \dots, d-1$, while the rotation factor in $K_d$ is generated by $L_{\mu\nu}$. We realise $\iota(S_{d,p})$ as a subgroup of $\SO(d)\subset K_d$ via the associated embedding of vector spaces
\begin{equation}
    \iota_\ast : \mathbb{R}^{d-1} \to \mathbb{R}^d, \qquad \iota_\ast(e_1) = e_d,\ \iota_\ast(e_2) = e_2,\ \dots,\ \iota_\ast(e_{d-1}) = e_{d-1}\ .
\end{equation}
To prove that the pair $(\Phi,\iota)$ satisfies eq.\ \eqref{eqn-iota-Phi}, it is sufficient to verify this condition for five different types of elements $h$
\begin{equation*}
    h = m(\hat x'), \quad h = e^{\lambda D}, \quad h = r_p\in SO(p), \quad h = r_q = e^{\phi M_{d,d+1}}, \quad h = w_p\,,
\end{equation*}
as such elements generate the whole defect conformal group. For the first four types, the equation \eqref{eqn-iota-Phi} follows immediately upon substituting all definitions. For the Weyl inversion $w_p$, the verification is somewhat more involved, but still straightforward.

Having obtained the pair $(\Phi,\iota)$, we follow the procedure from \cite{Buric:2020buk}, see also the section 4.1 of \cite{Buric:2020zea} for a detailed discussion of scalar two-point functions in the presence of a defect. Given the two-point function $G_{2,0}(x_i)$ we construct in turn functions $F_{2,0}$, $F$ and finally $\psi$ such that
\begin{equation} \label{eq:G20spherical}
    G_{2,0}(x_i) = \Phi_1^{-1}(x_1)\Phi^{-1}_2(x_2)\left(\mu_1(b_l(x_i))\otimes
    \mu_2(b_r(x_i))^{-1}\right) \psi(\lambda,\kappa)\ .
\end{equation}
Arguments $\lambda$ and $\kappa$ of $\psi$ are related to insertion points through eqs.\ \eqref{eq:lambdakappa}. The function
$$
\psi(\lambda,\kappa) = F(e^{\lambda D + \kappa M_{d,d+1}})\,,
$$
is directly related to a $W$-valued spherical function $F$ on the defect conformal group $G_{d,p}$ subject to the covariance laws
\begin{equation}\label{2pt-covariance}
    F\left(b_l e^{\lambda D} b_r\ e^{\kappa M_{d,d+1}} \right) =
    \left(\mu_1(b_l)\otimes\mu_2(b_r)^{-1}\right) F(e^{\lambda D + \kappa M_{d,d+1}})\,,
\end{equation}
where $b_{l,r} \in S_{d,p}$. Equation \eqref{eq:G20spherical} establishes a precise relation between spinning two-point functions in the presence of a co-dimension $q=2$ defect and $S_{d,p}$-spherical functions on $G_{d,p}$.\footnote{Note that covariance laws satisfied by $F$ make no reference to conformal weights $\Delta_i$ of the two bulk fields. Therefore, the conformal blocks are likewise independent of these quantum numbers, a fact that was observed in \cite{Lauria:2018klo}.}

\subsubsection{Conformal blocks}

Following our standard reasoning, the above motivates to study eigenfunctions of the group Laplacian within the space of $W$-valued functions $F$ with covariance laws given
by eq. \eqref{2pt-covariance}. For irreducible $\mu_1$ and $\mu_2$, eigenfunctions are labelled by a conformal weight $\hat\Delta$, a representation of $\SO(p)$ with Gelfand-Tsetlin labels $(\hat\ell_i)$, and a transverse spin $s$. These characterise the defect field that is exchanged after performing the bulk-defect operator product expansion. Notice however that in the context of CFT-s, the representations $\mu_i$ are typically not irreducible, since they arise as restrictions of irreducible representations of $\SO(d)$ to the subgroup $\SO(d-1)$. One can actually be more specific. If the bulk field is an $\SO(d)$ symmetric traceless tensor of spin $J$, then the associated representation $\mu$ is a direct sum of irreducibles $(l)$ for $\SO(d-1)$ where the $\SO(d-1)$-spin $l$ runs through $l=0, \dots, J$. In this decomposition, each irreducible component appears with multiplicity one. Our analysis of blocks proceeds by going through all admissible pairs $l,l'$ from the decomposition of $\mu_1$ and $\mu_2$, i.e. with $l \leq J_1$ and $l' \leq J_2$. For each such pair one can write down a Casimir equation and construct a set of blocks 
$$\psi^{l,l'}_{\hat\Delta,\hat\ell_i,s}(\lambda,\kappa) = \psi^{l,l'}_{\hat\Delta,\hat\ell_i}(\lambda) e^{is\kappa}\ . $$ 
Here we factorised the eigenfunctions in a way that reflects the direct product structure of the defect conformal group. Let us note that the superscript $l,l'$ labels bulk-defect two-point tensor structures. In particular, one can check that our enumeration of irreducible components agrees with the enumeration of two-point tensor structures in \cite{Lauria:2018klo}.

There are several ways to approach the computation of special functions $\psi$ one of which is by the radial component map. In the notation of section 2, the relevant groups are
\begin{equation}
    G = \SO(p+1,1), \quad K = \SO(p+1), \quad M = \SO(p)\ .
\end{equation}
For the remainder of this subsection, we will assume that the bulk fields are indeed symmetric traceless tensors and fix $J_1$, $J_2$, $l$ and $l'$. The left and right representations are $\rho_l = (l)$ and $\rho_r^\ast = (l')$. Following the general theory of section 2, the we start from the radial decomposition
\begin{equation}
    C_2^{\SO(p+1,1)} = L_{01}^2 + p\coth\lambda L_{01} + \frac{L'_{1a}L'_{1a} - 2\cosh\lambda L'_{1a}L_{1a} + L_{1a}L_{1a}}{\sinh^2\lambda} - \frac12 L^{ab}L_{ab}\ .
\end{equation}
In the second step, the reduced Laplacian $\Delta^{(p)}_{l,l'} = \Delta|_{\Gamma_{l,l'}}$ is found by replacing $L'_{\mu\nu}$ and $L_{\mu\nu}$ with partial derivatives and the appropriate representation operators. For this purpose, we use the function space realisation of symmetric traceless tensors,
\begin{align}  \label{eq:rhol1}
    & \rho_l(L_{12}) = -i(\xi_A\partial_A - l),\quad \rho_l(L_{1A}) = 
    -\frac12 \Big((1-\xi^2)\partial_A + 2\xi_A (\xi^B\partial_B - l)\Big),\\[2mm]
    & \rho_l(L_{2A}) = \frac{i}{2}\Big((1+\xi^2)\partial_A - 2x_A(\xi^B\partial_B-l)\Big) 
    ,\quad  \rho_l(L_{AB}) = \xi_A \partial_B - \xi_B\partial_A\ .\label{eq:rhol2}
\end{align}
Here, the indices $A$, $B$ range over $\{3,\dots,d-1\}$. The formulas resemble the ones for the action of 
the conformal group on scalar primary fields on $\mathbb{R}^p$, with the spin $l$ assuming the role of the 
conformal weight. The second representation $\rho_r$ of $\SO(d-1)$ is realised similarly through differential 
operators in a second set of $d-3$ variables $\xi'_A$. Initially, our $K$ spherical function take values in 
a space of polynomials. But, as we discussed before, consistency requires the $F$ to take values in the 
subspace of $M=\SO(d-2)$-invariant polynomials. From the components $\xi_A$ and $\xi'_A$ we can obviously form 
three $\SO(d-3)$ invariants, namely the scalar products $\xi^2, \xi^{'2}$ and $\xi\cdot \xi'$. Imposing full 
invariance under $M=\SO(d-2)$ we conclude that our $K$-spherical functions must take the form 
\begin{equation}
    F(\lambda,\xi_A,\xi'_A) = (1-\xi^2)^l (1-\xi'^2)^{l'} F\left(\lambda,\frac{(\xi^2+1)(\xi'^2+1)-4\xi\cdot\xi'}{(\xi^2-1)(\xi'^2-1)}\right) 
    \equiv (1-\xi^2)^l (1-\xi'^2)^{l'} F(\lambda,y)\ .
\end{equation}
Putting everything together, we arrive at the Laplacian
\begin{align}\label{defect-channel-Laplacian}
    \Delta^{(p)}_{l,l'} & = \partial_\lambda^2 + p\coth\lambda\ \partial_\lambda + \mathcal{D}^{(p)}_y +\frac{2\mathcal{D}^{(p)}_y - l(l+p-1) - l'(l'+p-1)}{\sinh^2\lambda} \\
    & - 2\cosh\lambda\ \frac{ y\mathcal{D}^{(p)}_y - (l+l'+p-2)(y^2-1)\partial_y + l l' y}{\sinh^2\lambda}\ .\nonumber
\end{align}
In the last expression, $\mathcal{D}^{(p)}_y$ is the Gegenbauer differential operator, given by
\begin{equation}\label{Gegenbauer-diff-op}
     \mathcal{D}^{(p)}_y = (y^2-1)\partial_y^2 + (p-1)y\partial_y\ .
\end{equation}
The number of independent components of the blocks, $N(l,l') = \text{min}(l,l')+1$, shows up in the fact that the operator \eqref{defect-channel-Laplacian} preserves the space of functions 
$F(t,y)$ which are polynomials in $y$ of degree not exceeding $N(l,l')-1$\footnote{For $p=2$, the space of functions is modified to polynomials in $y$ and $\sqrt{y^2-1}$.}. Having obtained 
the differential operator \eqref{defect-channel-Laplacian}, we turn to blocks, i.e. its eigenfunctions 
\begin{equation} \label{eq:defect_EVequation}
  \Delta^{(p)}_{l,l'}  \,  \psi_{p,\hat\Delta,\hat\ell}^{l,l'} (\lambda,y)
  = C_2(p,\hat\Delta,\hat \ell) \, \psi_{p,\hat\Delta,\hat\ell}^{l,l'}(\lambda,y), 
\end{equation}
with eigenvalues given by
\begin{equation}\label{eigenvalue-defect-Casimir}
    C_2(p,\hat\Delta,\hat\ell) = \hat\Delta(\hat\Delta-p) + \hat\ell(\hat\ell+p-2)\ .
\end{equation}
To find solutions, we supplement the Casimir equation by appropriate boundary conditions. 
These are found from from the bulk-to-defect limit $\lambda\to\infty$. In the said limit 
the differential operator \eqref{defect-channel-Laplacian} becomes separable
\begin{equation*}
    \Delta^{(p)}_{l,l'}\to\Delta^{(p)}_\infty = \partial_\lambda^2 + p \partial_\lambda 
    + \mathcal{D}^{(p)}_y\ .
\end{equation*}
Separated eigenfunctions are of the form $e^{k\lambda}C^{(\frac{p-2}{2})}_{\hat\ell}(y)$, where $C^{(\frac{p-2}{2})}_{\hat\ell}(y)$ are Gegenbauer polynomials. There are two solutions for $k$
\begin{equation}
    k(k+p) = \hat\Delta (\hat\Delta - p) \quad \implies \quad  k_1 = \hat\Delta - p, \quad k_2 = -\hat\Delta\ .
\end{equation}
Conformal blocks have the latter asymptotic behaviour. The asymptotic eigenfunctions of $\Delta^{(p)}_{l,l'}$ read
\begin{equation}\label{asymptotic-eigenfunctions}
    \psi_{p,\hat\Delta,\hat\ell}^\infty(\lambda,y) = e^{-\hat\Delta\lambda} C^{(\frac{p-2}{2})}_{\hat\ell}(y)\ .
\end{equation}
Conformal blocks with external and intermediate scalars, which play the role of zonal spherical functions\footnote{Note that honest zonal spherical functions for the Gelfand pair $(\SO(p+1,1),\SO(p+1))$ solves the same differential equation with different boundary conditions.}, are given by
\begin{equation}\label{zonal-spherical-defects}
    \psi_{p,\hat\Delta,0}^{0,0}(\lambda) = (\cosh\lambda)^{-\hat\Delta}\ _2 F_1\left(\frac{\hat\Delta+1}{2},\frac{\hat\Delta}{2};1+\hat\Delta-\frac{p}{2};1-\tanh^2\lambda\right)\ .
\end{equation}
We will compute general eigenfunctions by applying to these external and internal weight-shifting operators, whose general construction we described in section 2. In the case at hand, we have two external shifting operators
\begin{equation}\label{defect-channel-weight-shifting}
    q_{l,l'} = \partial_\lambda - l \coth\lambda + \frac{(y^2-1)\partial_y - l'y}{\sinh\lambda}, \quad \bar q_{l,l'} = \partial_\lambda - l' \coth\lambda + \frac{(y^2-1)\partial_y - ly}{\sinh\lambda}\ .
\end{equation}
These are readily observed to satisfy the relations
\begin{equation}\label{defect-channel-shift-equations}
    \Delta^{(p)}_{l+1,l'} q_{l,l'} = q_{l,l'} \Delta^{(p)}_{l,l'}, \quad \Delta^{(p)}_{l,l'+1} \bar q_{l,l'} = \bar q_{l,l'} \Delta^{(p)}_{l,l'}\ .
\end{equation}
Therefore, the operators $q_{l,l'}$ and $\bar q_{l,l'}$ map eigenfunctions of the Hamiltonian \eqref{defect-channel-Laplacian} with index $(l,l')$ to eigenfunctions of the Hamiltonians with index $(l+1,l')$ and $(l,l'+1)$, respectively. Furthermore, they preserve the asymptotic behaviour \eqref{asymptotic-eigenfunctions}. With the help of $q_{l,l'}$ and $\bar q_{l,l'}$, one obtains eigenfunctions with arbitrary $l$ and $l'$, however with internal representation being restricted to scalars. Let us now explain how to construct internal shifting operators, and thereby blocks with arbitrary exchanged representations.\footnote{As far as we are aware, in the context of dCFTs, such operators have not been constructed within other approaches.} We wish to compute the eigenfunctions $\psi$ in eq.\ \eqref{eq:defect_EVequation} for all $l,l' \geq \hat\ell$. External weight-shifting operators \eqref{defect-channel-shift-equations} can be used to 
produce these eigenfunctions from those with the minimal choices of $l = \hat\ell = l'$. 
Hence, we may focus on the construction of the latter. This function is obtained as
\begin{equation}\label{internal-projection}
    \psi^{\hat\ell,\hat\ell}_{p,\hat\Delta,\hat\ell}(\lambda,y) = \left(\Delta^{(p)}_{\hat\ell,\hat\ell} - C_2(\pi_1)\right)\dots\left(\Delta^{(p)}_{\hat\ell,\hat\ell} - C_2(\pi_{n-1})\right) \left(\psi^{\hat\ell,\hat\ell}_{p,\hat\Delta,0}(\lambda,y)\psi^{0,0}_{p,-\hat\ell,0}(\lambda)\right)\ .
\end{equation}
Here, $\{\pi_i\} = \{(\hat\Delta_i,\hat\ell_i)\}$ is a finite set of representation labels for the defect conformal group $\SO(d-1,1)$ to be determined as follows. Let us  
look at the following tensor product decomposition 
\begin{equation}\label{tensor-product}
    (\hat \jmath)\otimes(\hat\ell) = (\hat \jmath_1,\hat\ell_1) \oplus \dots \oplus (\hat \jmath_n,\hat\ell_n)\,,
\end{equation}
of representations of $\SO(p+2)$, for all positive integer values $\hat \jmath$. Of course this decomposition is well known. We order terms in the decomposition so that $(\hat \jmath_n,\hat\ell_n) = (\hat \jmath,\hat\ell)$, which is one of the irreducibles that appear for any values of $\hat\jmath$ and $\hat\ell$. The number of terms that appears on the right hand side stabilises for large values of $\hat \jmath$ and so we can think of the labels $\hat\jmath_i$ and $\hat\ell_i$ for mixed symmetry tensors as functions of $\hat\jmath$. After this, we can now continue $\hat\jmath$ through the complex plane to $\hat\jmath = - \hat \Delta$. After this we can define $\hat\Delta_i := -\hat \jmath_i(\hat\jmath=-\hat\Delta)$. To complete the description of solutions, one needs the wave function with upper labels $l,l' =\hat\ell$ and for scalar exchange which appears in the first factor in the argument of the differential operator on the right-hand side of eq.\ \eqref{internal-projection}. This function is obtained in the obvious way from eq.\ \eqref{zonal-spherical-defects} via external shifting operators \eqref{defect-channel-weight-shifting}. Our procedure \eqref{internal-projection} essentially takes spherical functions of the corresponding compact Gelfand pair and analytically continues them.
\smallskip

{\bf Example.} According to the above rules, we can compute the following wave function 
\begin{equation}\label{conjectured-block}
    \psi_{p,\hat\Delta,1}^{1,1}(\lambda,y) = (\Delta^{(p)}_{1,1}-C_2(\hat\Delta+1,0))(\Delta^{(p)}_{1,1}-C_2(\hat\Delta-1,0))\left(\psi_{p,-1,0}^{0,0}(\lambda) \psi_{p,\hat\Delta,0}^{1,1}(\lambda,y)\right)\ .
\end{equation}
In the last line, the function $\psi_{p,\hat\Delta,0}^{1,1}(\lambda,y)$ is defined via external weight-shifting operators as
\begin{equation}
    \psi_{p,\hat\Delta,0}^{1,1}(\lambda,y) = \bar q_{1,0}\left(q_{0,0}\left(\psi_{p,\hat\Delta,0}^{0,0}(\lambda)\right)\right)\ .
\end{equation}
It can be checked numerically that eq.\ \eqref{conjectured-block} solves the eigenvalue equation for the defect channel Laplacian \eqref{defect-channel-Laplacian} with the eigenvalue \eqref{eigenvalue-defect-Casimir}. Also, as $\lambda\to\infty$, it becomes proportional to the asymptotic eigenfunction \eqref{asymptotic-eigenfunctions}. 

This concludes our discussion of defect channel blocks. They are all obtained by successive applications of external \eqref{defect-channel-weight-shifting} and internal \eqref{internal-projection} weight-shifting operators to scalar blocks \eqref{zonal-spherical-defects}. The number of required applications of these operators scales with $l,l'$ and thus with spins of the two bulk fields involved. The only input from representation theory is the tensor product decomposition \eqref{tensor-product} - in particular, our construction does not rely on the knowledge of any Clebsch-Gordan coefficients.

\subsection{Bulk two-point functions - the bulk channel}

Having solved for the defect channel blocks, we turn to the bulk channel. Due to the fact that only symmetric traceless tensors $\mathcal{O}_{\Delta,J}$ can have a non-vanishing one-point functions in the presence of a co-dimension $q=2$ defect, \cite{Lauria:2018klo}, it is only these fields that contribute to the bulk channel expansion of the 
correlators \eqref{eq:2pointdefect}. Corresponding conformal blocks are eigenfunctions of quadratic and quartic Casimir operators constructed from sums of conformal generators at points $x_1$ and $x_2$. They are further labelled by tensor structures of the bulk three-point function $\langle\varphi_1\varphi_2\mathcal{O}_{\Delta,J}\rangle$. In this section, we will realise the correlation function \eqref{eq:2pointdefect} as a spherical function on $G_d$, covariant with respect to $K_d$. While such a relation is not surprising, writing out all the details requires some work which is carried out in the first subsection. The second subsection is devoted to the construction of conformal blocks.

\subsubsection{Lift of the correlator}

As we have explained above, the application of results on Laplace operators for the conformal group to Casimir operators requires to uplift correlation functions from functions of the fields' insertion points to functions on the group $G_d$. It is well known how to perform such an uplift for the individual fields. As before, the spin is described by a representation $\sigma$ of $K_d$ on some finite dimensional carrier space $W$. With these notations we can realise the representation of the conformal group that is associated with a local primary field $\varphi: \mathbb{R}^d
\to W$ on the subspace of functions $f:G_d \rightarrow W$ satisfying the covariance properties
\begin{equation}
    f(g k n) = \sigma(k^{-1}) f(g), \quad \mathit{ where } \quad k\in K_d,\ n = e^{y\cdot K}\,,
\end{equation}
are associated with rotations, dilations and special conformal transformations, respectively. These covariance properties imply that $f$ is completely characterised by the values it assumes on translations. The covariance properties are designed so that $f(e^{x\cdot P})$ transforms in the same way under conformal transformations as $\varphi(x)$.

After this preparation, it is now straightforward to promote the two-point function $G_{2,0}(x_i)$ to a function $F_{2,0}$ on the two-fold product $G_d \times G_d$ of the conformal group
\begin{equation} \label{eq:2ptF2G}
    F_{2,0} : G_d^2 \to W_1\otimes W_2, \quad F_{2,0}(e^{x_i\cdot P}) = G_{2,0}(x_i), \quad
    F_{2,0}(g_i k_i n_i) = (\sigma_1(k_1^{-1})\otimes \sigma_2(k_2^{-1})) F_{2,0}(g_i)\ .
\end{equation}
Ward identities satisfied by $G_{2,0}(x_i)$ translate into diagonal left invariance of $F_{2,0}$ with respect to $G_{d,p}$
\begin{equation}
    F_{2,0}(h g_i) = F_{2,0}(g_i), \quad h\in G_{d,p}\ .
\end{equation}
The relation between $G_{2,0}(x_i)$ and $F_{2,0}$ here, as well as the covariance properties, are the analogues of equations \eqref{eq:F4rightcov} in the case of bulk four-point functions. There is only one difference compared to that case that we need to discuss. Let us first stress that for defects of co-dimension $q=2$, the groups $K_d$ and $G_{d,p}$ have the same complexified Lie algebra. This implies that the two groups are conjugate under an element $g_0\in (G_d)_{\mathbb{C}}$ in the complexified conformal group. The appearance of this element $g_0$ in our analysis is essentially the only deviation from the analysis of bulk theory four-point functions.

Now we are ready to carry on with our uplift. To simplify notation, we introduce the representation $\sigma_{12}(k) =\sigma_1(k)\otimes\sigma_2(w k w^{-1})$, with $w\equiv w_{d+1}$, and denote its carrier space by $W_{12}$. Let us now define a new function
\begin{equation} \label{eq:firstg0}
    F: G_d \to W_1\otimes W_2, \quad  F(g) := F_{2,0}(g_0 g,g_0 g w^{-1})\ .
\end{equation}
It is not difficult to see that $F$ is a $K=K_d$-spherical function,
\begin{align}
    & F(g k) = F_{2,0}(g_0 g k,g_0 g k w^{-1}) = (\sigma_1(k^{-1})\otimes
    \sigma_2(w k^{-1} w^{-1})) F_{2,0}(g_0 g,g_0 g w^{-1}) = \sigma_{12}(k^{-1})F(g)\,,\\[2mm]
    & F(k g) = F_{2,0}(g_0 k g, g_0 k g w^{-1}) =
    F_{2,0}(g_0 k g_0^{-1} g_0 g, g_0 k g_0^{-1} g_0 g w^{-1}) = F_{2,0}(g_0 g, g_0 g w^{-1}) = F(g)\ .
\end{align}
Note that the covariance law under left rotations is trivial, while the one under right translations involves our representation $\rho_r = \sigma_{12}$, i.e. what we have shown is
that $F\in\Gamma_{1,\sigma_{12}}$. The restriction of $F$ to $A_p$ will be denoted by $F(t_1,t_2)$. Following the same steps as in the derivation of equation
\eqref{magic-formula}, one can show
\begin{equation*}
    G_{2,0}(x_i) = \left(1\otimes\sigma_2(k(t_{21})^{-1})\right) F_{2,0}(e^{x_1\cdot P}n(y_{21}),e^{x_1\cdot P}n(y_{21})w_d^{-1}) =
    \left(1\otimes\sigma_2(k(t_{21})^{-1})\right) F(g_0^{-1}e^{x_1\cdot P}n(y_{21}))\ .
\end{equation*}
Definitions of group elements $n(y_{ij})$ and $k(t_{ij})$ can be found in section 3, see also \cite{Buric:2020buk} for more detailed explanations. As in the case of four-point functions, the only remaining step is to $KA_pK$-decompose the argument $\tilde g$ of $F$ and use the covariance laws. Here we have
\begin{equation}\label{main-group-element}
    \tilde g \equiv g_0^{-1}e^{x_1\cdot P}n(y_{21}) = g_0^{-1}e^{x_1\cdot P} w^{-1} e^{y_{21}\cdot P} w = g_0^{-1} e^{x_1\cdot P} e^{Ix_{21}\cdot K} = \tilde k_l\,
    a(t_1,t_2)\, \tilde k_r\,,
\end{equation}
where $I$ denotes the conformal inversion. It is straightforward to work out the Cartan decomposition of $\tilde g$. In the process one finds in particular that the invariants
$t_i$ are given by
\begin{equation} \label{eq:ztcoordinates} 
    (\cosh t_1 + \cosh t_2)^2 = \frac{-4 z_1\bar z_2}{x_{12}^2}, \quad
    (\cosh t_1 - \cosh t_2)^2 = \frac{-4 \bar z_1 z_2}{x_{12}^2}\ .
\end{equation}
Here we have introduced the complex variable $z = x^d + ix^{d+1}$ to characterise the position of a point $x$ transverse to the defect. The relation between $t_i$ and cross ratios $\lambda$, $\kappa$ that we use in the defect channel analysis reads
\begin{equation}\label{coordinates-ti}
    (\cosh t_1 \pm \cosh t_2)^2 = \frac{-4e^{\mp i\kappa}}{\cosh\lambda - \cos\kappa}\ .
\end{equation}
After the insertion of the Cartan decomposition, one finally obtains the following concrete relation between the correlation function $G_{2,0}(x_i)$ and the restriction of $F$ to the
abelian subgroup $A_p$,
\begin{equation}\label{eq:G20sphericalbulk}
G_{2,0}(x_i) = \sigma_1(\tilde{k}_{r}^{-1})\sigma_2\left(k(t_{12})^{-1} 
(\tilde{k}^{-1}_{r})^w\right) F(a) \equiv \tilde \Xi(x_i;\sigma_i)f(t_1,t_2;\sigma_{12})\ .
\end{equation}
Once again we have now found the exact prefactor $\Xi$ that turns spherical functions $F$ 
into two-point functions in the presence of a defect of co-dimension $q=2$. But this time, 
the spherical function $F$ is a function on the bulk conformal group $G_d$. 

\subsubsection{Conformal blocks}

Under the mapping $G_{2,0}(x_i)\mapsto F$, conformal partial waves are carried to 
eigenfunctions of the Laplacian. After conjugation by the appropriate function 
$\delta$ given in eq.\ \eqref{Haar-measure-bulk-channel}, spherical harmonics become 
eigenfunctions of the spinning Calogero-Sutherland Hamiltonian \eqref{universal-Hamiltonian}, 
with the trivial left representation $\rho_l = 0$. From our discussion, it is 
clear that the Hamiltonian coincides with the one for a bulk four-point function 
of two scalars and two symmetric traceless tensors. The latter was computed in \cite{Buric:2021kgy}. 
Here we just give a lightening review. 

We focus on a single irreducible component $\rho_r^\ast$ of $\sigma_{12}$, which 
is in general a mixed symmetry tensor with Gelfand-Tsetlin labels $(j,\msj)$. The 
generators may be realised as differential operators
\begin{align}
    & \rho_r^\ast(L_{23}) = i\left(z^A \partial_A -j\right), \quad \rho_r^\ast(L_{AB}) =  
     z_A \partial_{z^B} - z_B \partial_{z^A} + w_A \partial_{w^B} - w_B \partial_{w^A}\,, \label{eq:rhobulk1}\\[2mm]
    & \rho_r^\ast(L_{2A}) = \frac{i}{2} \left((1+z^2)\partial_{z^A} - 2z_A (z^B \partial_{z^B} - j) + 2z^B(w_B \partial_{w^A} - w_A \partial_{w^B})\right),
    \label{eq:rhobulk2}\\[2mm] 
    & \rho_r^\ast(L_{3A}) = -\frac12 \left((1-z^2)\partial_{z^A} + 2z_A (z^B \partial_{z^B} - j) - 2z^B(w_B \partial_{w^A} - w_A \partial_{w^B})\right).
    \label{eq:rhobulk3}
\end{align}
%VS: any reason we use arrows here rather than \rho_r := ?
These operators act on an appropriate space of polynomials in $z^A$ and $w^A$, with 
the index running through $A = 4,\dots,d+1$. Such a realisation allows for a simple 
implementation of $M=\SO(d-2)$-invariance conditions, $\rho_r^\ast(L_{AB})f=0$. These are solved 
by functions of scalar products
\begin{equation}
    X = z^A z_A, \quad W = w^A w_A, \quad Y = z^A w_A\ .
\end{equation}
To get to the carrier space of the irreducible representation $\rho_r^\ast$ for spin $(j,\msj)$, one is further imposes the homogeneity $Y\partial_Y f=\msj f$ and restricts to the lightcone $\{W=0\}$. Individual pieces of the Hamiltonian \eqref{universal-Hamiltonian} restrict to well-defined operators on such functions. This gives the final operator
\begin{align}\label{eq:HCSMST3scalars} 
     H^{(b)}_{j,\msj} &= \partial_{t_1}^2 + \partial_{t_2}^2 +\frac{1 - (2b+\msj-j 
     + 2X\partial_X)^2}{2\sinh^2(t_1+t_2)} + 
     \frac{1 -(2b-\msj+j-2X\partial_X)^2}{2\sinh^2(t_1-t_2)}\nonumber\\[2mm]  
     & + \frac{L_{j,\msj}(X)-\frac14(d-2)(d-4)}{\sinh^2 t_1} +
     \frac{L_{j,\msj}(-X)- \frac14 (d-2)(d-4)}{\sinh^2 t_2} - \frac{d^2-2d+2}{2}\ .
\end{align}
Here, the operator $L_{j,\msj}$ is defined as
\begin{align}\label{eq:Lll}
    L_{j,\msj}(X) &= -X(1-X)^2 \partial_X^2 
    - \left(\msj(1-X)-2(1-j)X +
    \frac{d-2}{2}(1+X)\right)(1-X)\partial_X\nonumber\\[2mm] 
    & + \left(1-j-\frac{d-2}{2}\right)(\msj(1-X) + jX) 
    - \frac{j(d-2)}{2}\ .
\end{align}
Conformal blocks are eigenfunctions of the operator \eqref{eq:HCSMST3scalars} with the eigenvalue 
$C_2(d,\Delta,J)$, see eq.\ \eqref{eigenvalue-defect-Casimir},
\begin{equation} \label{eq:eigbulk2point}
   H^{(b)}_{j,\msj} \  \psi^{b,j,\msj}_{\Delta,J,m}(t_1,t_2,X)
   = C_2(d,\Delta,J) \, \psi^{b,j,\msj}_{\Delta,J,m}(t_1,t_2,X)\ .
\end{equation}
Here, $(\Delta,J)$ are the conformal weight and spin of the intermediate field. Furthermore, they carry a label $m$ for the three-point function of two external and the intermediate 
bulk field. This can also be understood as the multiplicity index, labelling different eigenfunctions of the spinning Calogero-Sutherland operator with the same eigenvalue.
\medskip 

Let us now solve for eigenfunctions of the operator \eqref{eq:HCSMST3scalars}. To begin with, notice that the Hamiltonian \eqref{eq:HCSMST3scalars} preserves the space of functions that are polynomials in $X$ of degree less than or equal to $j-\msj$. Indeed, the operator \eqref{eq:Lll} maps in general an $n$-th degree polynomial in $X$ to one of degree $n+1$ and is the only term in the Hamiltonian that potentially raises the degree. However, it is easy to see that $L_{j,\msj}(X^{j-\msj})$ is again a polynomial of degree $j-\msj$. In particular, the operator \eqref{eq:HCSMST3scalars} with $j=\msj$ is well-defined on the space of functions $\psi(t_1,t_2)$, a scalar $BC_2$ Calogero-Sutherland Hamiltonian\footnote{We are using conventions of \cite{Buric:2021yak}.}
\begin{equation} \label{eq:Hll}
    H^{(b)}_{\msj,\msj} \psi(t_1,t_2) = \left(-2 H_{cs}^{(0,b,\epsilon = 2\msj+d-2)} 
    - \frac{d^2-2d+2}{2}\right) \psi(t_1,t_2)\ .
\end{equation}
Eigenfunctions of this simpler operator with $j=\msj$ are well-known. Among various eigenfunctions with the same eigenvalue, conformal blocks are distinguished
by their asymptotic behaviour, as will be detailed below. For the moment, we regard these scalar functions as known. To obtain eigenfunctions when $j>\msj$, we use the weight-shifting 
operators constructed at the end of section 2. In the present case, we have external shifting operators $q$ and $p$ which satisfy the intertwining properties
\begin{equation}\label{bulk-channel-shifting}
    q^{(b)}_{j,\msj} H^{(b)}_{j,\msj} = H^{(b-1/2)}_{j+1,\msj} q^{(b)}_{j,\msj}, 
    \qquad p^{(b)}_{j,\msj} H^{(b)}_{j,\msj} = H^{(b+1/2)}_{j+1,\msj} p^{(b)}_{j,\msj}\ .
\end{equation}
Explicitly, the differential shifting operators read
\begin{align}
    q^{(b)}_{j,\msj} & = (1+X)\partial_{t_1} + \coth(t_1+t_2)(2X\partial_X + 2b-1-j+\msj)\nonumber \\[2mm]
    & + (1-X)\partial_{t_2} - \coth(t_1-t_2) X (2X\partial_X - 2b+1-j+\msj)\label{shift-q-bulk-channel}\\[2mm]
    & - \coth t_1 \left(2X(1-X)\partial_X +\frac12((d-2)(1+X)+4jX+2\msj(1-X))\right)\nonumber\\[2mm]
    & - \coth t_2 \left(2X(1+X)\partial_X + \frac12((d-2)(1-X)-4jX+2\msj(1+X))\right)\,,\nonumber
\end{align}
and 
\begin{align}
    p^{(b)}_{j,\msj} & = (1+X)\partial_{t_1} - \coth(t_1+t_2)X(2X\partial_X + 2b+1-j+\msj)\nonumber \\[2mm]
    & - (1-X)\partial_{t_2} + \coth(t_1-t_2) (2X\partial_X - 2b-1-j+\msj)\label{shift-p-bulk-channel}\\[2mm]
    & - \coth t_1 \left(2X(1-X)\partial_X +\frac12((d-2)(1+X)+4jX+2\msj(1-X))\right)\nonumber\\[2mm]
    & - \coth t_2 \left(2X(1+X)\partial_X + \frac12((d-2)(1-X)-4jX+2\msj(1+X))\right)\ .\nonumber
\end{align}
The shifting operators satisfy the relation
\begin{equation}\label{relation-shifting}
    p^{(b-1/2)}_{j+1,\msj} q^{(b)}_{j,\msj} = q^{(b+1/2)}_{j+1,\msj} p^{(b)}_{j,\msj}\ .
\end{equation}
Consequently, by applying either side of this equation to a wavefunction $\psi^{b,j,\msj}_{\Delta,J,m}$, one obtains the same eigenfunction of the
Hamiltonian \eqref{eq:HCSMST3scalars} with the first subscript $j$ shifted to $j+1$. Indeed, if no such commutativity relation between $p$ and $q$ existed, 
one could construct too many independent solutions to the Calogero-Sutherland 
problem.

As mentioned above, general conformal blocks in eq.\ \eqref{eq:eigbulk2point} are contain an additional "multiplicity index" $m$, i.e.\ we need to distinguish between 
eigenfunctions for different values of $m$ even though they possess the same eigenvalue. The multiplicity index ranges over at most $j-\msj+1$ different values, as this is the 
maximal number of solutions with appropriate boundary conditions. The eigenfunctions are obtained by applications of the shifting operators $q$ and $p$ to scalar wavefunctions, 
i.e.\ to wavefunctions with $j = \msj$. Relations \eqref{relation-shifting} imply that 
there are at most $j-\msj+1$ different functions obtained in this way. Distinct 
eigenfunctions of the Hamiltonian \eqref{eq:HCSMST3scalars} for generic $j \neq \msj$ 
are of the schematic form 
$$\psi^{b,j,\msj}_{\Delta,J,m} = p^m q^{j-\msj-m} \psi^{b+\frac{j-\msj}{2}-m,\msj,\msj}_{\Delta,J}\ . $$ 
"Commutativity" of $p$ and $q$ ensures there are no ordering issues, i.e. that 
$p^m q^{j-\msj-m}$ is uniquely defined. Thus, one obtains $j-\msj+1$ functions for 
$m=0,1,\dots j-\msj$. Let us illustrate this on a few simple examples. When 
$j=\msj+1$, we have two solutions
\begin{equation}
    \psi^{b,\msj+1,\msj}_{\Delta,J,0}(t_1,t_2,X) = q^{(b+1/2)}_{\msj,\msj} \psi^{b+1/2,\msj,\msj}_{\Delta,J}(t_1,t_2), \quad 
    \psi^{b,\msj+1,\msj}_{\Delta,J,1}(t_1,t_2,X) = p^{(b-1/2)}_{\msj,\msj} \psi^{b-1/2,\msj,\msj}_{\Delta,J}(t_1,t_2)\ .
\end{equation}
Going to $j = \msj+2$, we can construct four solutions, two of which coincide
\begin{align}
    & \psi^{b,\msj+2,\msj}_{\Delta,J,0}(t_1,t_2,X) = q^{(b+1/2)}_{\msj+1,\msj}\left(q^{(b+1)}_{\msj,\msj} \psi^{b+1,\msj,\msj}_{\Delta,J}(t_1,t_2)\right)\,,\label{solution1}\\[2mm]
    & \psi^{b,\msj+2,\msj}_{\Delta,J,2}(t_1,t_2,X) = p^{(b-1/2)}_{\msj+1,\msj}\left(p^{(b-1)}_{\msj,\msj} \psi^{b-1,\msj,\msj}_{\Delta,J}(t_1,t_2)\right)\,,\label{solution2}\\[2mm]
    & \psi^{b,\msj+2,\msj}_{\Delta,J,1}(t_1,t_2,X) = q^{(b+1/2)}_{\msj+1,\msj}\left(p^{(b)}_{\msj,\msj} \psi^{b,\msj,\msj}_{\Delta,J}(t_1,t_2)\right) = p^{(b-1/2)}_{\msj+1,\msj}\left(q^{(b)}_{\msj,\msj} \psi^{b,\msj,\msj}_{\Delta,J}(t_1,t_2)\right)\ . \label{solution3}
\end{align}
Continuing in the obvious way, one indeed obtains all conformal blocks. Hence, we have achieved our principal goal, namely to construct all bulk channel blocks for bulk fields in arbitrary STT representations and in any dimension $d$. 
\medskip

{\bf Example} The number of conformal blocks we constructed matches that of 
\cite{Lauria:2018klo}. We illustrate this on the example $J_1 = J_2 = 2$. 
In this case
\begin{equation*}
    \sigma_{12} = (2) \otimes (2) = (4) \oplus (2) \oplus (0) \oplus (3,1) \oplus (2,2) \oplus (1,1)\ .
\end{equation*}
Thus, we have six Calogero-Sutherland Hamiltonians with matrix sizes 5, 3, 1, 3, 1 and 1, respectively. This gives a total of 14 conformal blocks for each $(\Delta,J)$. Bearing in mind that only symmetric traceless tensors can couple to a co-dimension $q=2$ defect, this matches the number of blocks given in table (87) of \cite{Lauria:2018klo}. For all other cases from the same table, our counting again agrees with \cite{Lauria:2018klo}.
\medskip

The arguments given above are valid as long as weight-shifting operators preserve boundary conditions satisfied by the blocks. This is shown in the remainder of the subsection. The boundary conditions in the scalar case were studied in \cite{Liendo:2019jpu}. This work makes use of the coordinates $(x,\bar x)$ and $(\tau_1,\tau_2)$, which are related to our variables $(\lambda,\kappa)$ through
\begin{equation}
    \cosh\lambda = \frac{1+x\bar x}{2\sqrt{x\bar x}}, \quad \cos\kappa = \frac{x+\bar x}{2\sqrt{x\bar x}}\,, \qquad x = \frac{1}{\tanh^2 \frac{\tau_1 + \tau_2}{4}}, \quad \bar x = \frac{1}{\tanh^2 \frac{\tau_1 - \tau_2}{4}}\ .
\end{equation}
These imply the relation between $t_j$ and $\tau_j$, $2 t_j  = \tau_j + i \pi$. Such a change of variables preserves the Calogero-Sutherland form of the Hamiltonian. In fact, $(\tau_1,\tau_2)$ are closely related to the radial coordinates of \cite{Lauria:2017wav}, from which follows their physical range
\begin{equation}
    r = e^{-\frac{\tau_1}{2}}, \quad \theta = \frac{i\tau_2}{2}, \qquad \tau_1 \in (0,\infty), \quad \tau_2\in i[0,\pi]\ .
\end{equation}
Boundary conditions for blocks are determined from the OPE limit, $\tau_1\to\infty$. In this limit, the blocks decay asymptotically as $\exp((\frac{d}{4}-\frac{\Delta}{2})\tau_1)$ (the term $d/4$ comes from the Haar measure \eqref{Haar-measure-bulk-channel}). Therefore, the Hamiltonian \eqref{eq:HCSMST3scalars} in the OPE limit becomes
\begin{equation}\label{OPE-Hamiltonian}
    (1+X)^{\msj-j} H^{(b),\infty}_{j,\msj} (1+X)^{j-\msj} = -\partial_\theta^2 +\frac{D_y^{(2\msj+d-1)} + 
    \left(\msj+\frac{d-3}{2}\right)^2 -\frac14}{\cos^2\theta} + \Delta(\Delta-d) - \frac{(d-2)^2}{4}\,,
\end{equation}
where $y = (1-X)/(1+X)$ and the operator $D_y^{(a)}$ was defined in eq.\ \eqref{Gegenbauer-diff-op}. We have kept the parameter $b$ as a label of the Hamiltonian, even though the latter does not depend on $b$ after taking the OPE limit. It is a simple matter to verify that the operators $q$ and $p$ have well-defined OPE limits and that the resulting limiting operators $\{H^{(b),\infty}_{j,\msj},q^{(b),\infty}_{j,\msj},p^{(b),\infty}_{j,\msj}\}$ still satisfy the 
exchange relations \eqref{bulk-channel-shifting} and \eqref{relation-shifting}. From eq.\ \eqref{OPE-Hamiltonian}, one immediately finds asymptotic wavefunctions, solutions of \eqref{eq:eigbulk2point} in the limiting regime. They are given by
\begin{equation}
    \psi^\infty_{j,\msj,J,n}(\theta,y) = (1+X)^{j-\msj} (\cos\theta)^{n+\msj+\frac{d-2}{2}}\ C_{J-n-\msj}^{(n+\msj+\frac{d-2}{2})}(\sin\theta)\ C^{(\msj+\frac{d-3}{2})}_n(y),\quad n=0,
    \dots,j-\msj\ .
\end{equation}
It is not difficult to verify that weight-shifting operators map such asymptotic wavefunctions to other asymptotic wavefunctions, as required.

{\bf Example} Let us look at a few simple examples. For $j=\msj$, the asymptotic blocks are given by 
\begin{equation}
    \psi^\infty_{\msj,\msj,J,0}(\theta) = (\cos\theta)^{\msj+\frac{d-2}{2}}\ C_{J-\msj}^{(\msj+\frac{d-2}{2})}(\sin\theta)\ .
\end{equation}
For $j=\msj+1$, we construct two wavefunctions with given $(\Delta,J)$ by applications of $q$ and $p$. Their asymptotics are the following linear combinations of asymptotic wave 
functions with shifted labels, 
\begin{align*}
    & q^{(b+1/2)}_{\msj,\msj} \psi^\infty_{\msj,\msj,J,0}(\theta) = (1+2b-\msj-\Delta) \psi^\infty_{\msj+1,\msj,J,0} + i\frac{2\msj+d-2}{2\msj+d-3}\psi^\infty_{\msj+1,\msj,J,1}\,,\\
    & p^{(b-1/2)}_{\msj,\msj} \psi^\infty_{\msj,\msj,J,0}(\theta) = (1-2b-\msj-\Delta) \psi^\infty_{\msj+1,\msj,J,0} - i\frac{2\msj+d-2}{2\msj+d-3}\psi^\infty_{\msj+1,\msj,J,1}\ .
\end{align*}
Solutions for $j=\msj+2$ were written above in eqs.\ \eqref{solution1}-\eqref{solution3}. We denote the corresponding boundary conditions by
\begin{equation*}
    \varphi_0 = q^{(b+1/2)}_{\msj+1,\msj}\left(q^{(b+1)}_{\msj,\msj} \psi^\infty_{\msj,\msj,J,0}(\theta)\right), \quad \varphi_1 = q^{(b+1/2)}_{\msj+1,\msj}\left(p^{(b)}_{\msj,\msj} \psi^\infty_{\msj,\msj,J,0}(\theta)\right), \quad \varphi_2 = p^{(b-1/2)}_{\msj+1,\msj}\left(p^{(b-1)}_{\msj,\msj} \psi^\infty_{\msj,\msj,J,0}(\theta)\right)\ .
\end{equation*}
These boundary conditions are linear combinations of asymptotic wave functions 
\begin{equation}\label{matrix-A}
    \psi^\infty_{\msj+2,\msj,J,n} = A_{nm} \varphi_m\,,
\end{equation}
where the matrix $A_{nm}$ is given in appendix E. This concludes our discussion of conformal blocks in the bulk channel.

\subsection{Bulk-bulk-defect three-point functions}

In this final subsection we would like to extend the setup of the previous two and admit the insertion of an additional defect field. Our discussion here is geared to the case in which this additional field is scalar, but spinning defect field insertions can be treated analogously. Without the defect field insertion, the correlator $G_{2,0}(x_i)$ could be decomposed into blocks that solve the 1-sided spinning Calogero-Sutherland equations, see above. With the additional defect field inserted, it turns our that the relevant blocks solve 2-sided spinning Calogero-Sutherland equations. Our focus here will be to map the correlator to spherical functions on the $d$-dimensional conformal group and to identify the covariance conditions for the relevant spherical functions with those discussed in section 2. Detailed solution theory is left for future work. The defect channel decomposition of the three-point function $G_{2,1}(x_i)$ has been analysed in \cite{Buric:2020zea} and partial waves are known for the case in which the bulk fields are scalar. For completeness, we shall include a non-technical review of these defect channel results at the end of the subsection. 

The objects we wish to study in this subsection are the three-point functions $G_{2,1}(x_i)$ of two bulk and one defect field, 
\begin{equation} 
G_{2,1}(x_i) = \langle\varphi_1(x_1)\varphi_2(x_2)\hat\varphi_3(x_3)\rangle\ .
\end{equation} 
We shall assume for most of this section the bulk fields to be scalars and the defect field to be scalar and have vanishing transverse spin. For purposes of lifting the correlator to a spherical function, these assumptions may be dropped, as will be clear from the argument. The three-point function $G_{2,1}(x_i)$ is the next simplest correlator, after $G_{2,0}(x_i)$, for which one can formulate crossing equations. It admits three conformal invariants which we take to be 
\begin{equation}
v_0 = \cos\kappa = \frac{x_{1\perp}\cdot x_{2\perp}}{|x_{1\perp}| |x_{2\perp}|} \ , 
\quad  v_i  =  - x^2_{3-i,\perp}\frac{x_{i 3}^4}{\hat x_{12}^2 x_{13}^2 x_{23}^2 + 
(\hat x_{13}^2 x_{23}^2 - \hat x_{23}^2 x_{13}^2)(x_{23}^2 - x_{13}^2)},\quad i=1,2\ .
\end{equation}
In writing these expressions, we have set $x^2_{i3} = x^2_{i\perp} + \hat x^2_{i3}$ for $i=1,2$.  

In the bulk channel expansion, the fields $\varphi_1$ and $\varphi_2$ exchange symmetric traceless tensors $\mathcal{O}_{\Delta,J}$. Corresponding conformal blocks are eigenfunctions of quadratic and quartic Casimir operators of $\mathfrak{g}_d$ constructed out of sums of generators at points $x_1$ and $x_2$. They carry a further label which enumerates tensor structures of the two-point function $\langle\mathcal{O}_{\Delta,J}\hat\varphi_3\rangle$. It is possible to obtain the bulk channel Casimir equations straightforwardly by acting on the correlator of the form \eqref{bulk-bulk-defect-correlator}. This leads to a somewhat complicated operator
\begin{equation}\label{3pt-bulk-Casimir}
    \mathcal{D} = \sum_{a,b=0}^2 C_{ab} \partial_{v_a} \partial_{v_b} + \sum_{a=0}^2 C_a \partial_{v_a} + C\ .
\end{equation}
For completeness, we write the coefficients $C_{ab}$, $C_a$ and $C$, which are functions of conformal invariants, in the appendix F.

On the other hand, the above setup is expected to lead to spherical functions for the following reason. The stabiliser in $G_d$ of a $p$-dimensional defect together with a point on it (at which the defect field is inserted) is given by 
\begin{equation}
    \dot G_{d,p} \cong \left( \SO(1,1)\times \SO(p) \ltimes \mathbb{R}^p \right) \times \SO(q)\ .
\end{equation}
Representations $\hat\sigma$ of this group that are trivial on $\mathbb{R}^p$ characterise defect fields. We see that these are determined by a choice of a weight, a $p$-dimensional spin and a $q$-dimensional transverse spin. 

Before we continue our discussion, let us recall that given some representation $\sigma$ of a subgroup 
$K\subset G$ on a vector space $W$ we can induce a  representation $\pi$ of $G$. The carrier space of
$\pi$ can be build as  
\begin{equation} 
\Gamma^{G/K}_{\sigma} = \{ \, f: G \rightarrow W\, 
| \, f(gh)= \sigma(h^{-1})f(g)\ , \ h \in K \,\}\ . 
\end{equation} 
The action $\pi$ of $G$ on this space is through left multiplication on $g$. Let us note that 
induction can be performed in stages. Suppose that $P \subset K$ is a subgroup of $K$ and 
consider some representation $\sigma'$ of $P$. Then one has 
\begin{equation}\label{eq:indstages}
\Gamma^{G/P}_\sigma \cong \Gamma^{G/K}_\pi \quad \mathit{where} \quad 
W_{\pi} = \Gamma^{K/P}_\sigma \ \ . 
\end{equation}
With this preparation we are now ready to spell out our model for the space of correlation 
functions $G_{2,1}(x_i)$ which is given by 
\begin{equation}\label{space-3pt-functions}
    \left(\Gamma^{G_d/K_d}_{\sigma_{12}} \otimes \Gamma^{G_d /\dot G_{d,p}}_{\hat\sigma_3}\right)^{G_d} \cong \left(\Gamma^{G_d/G_{d,p}}_{\sigma_{12}} \otimes 
    \Gamma^{G_d/G_{d,p}}_{\hat\pi_3}\right)^{G_d}\cong \Gamma_{\sigma_{12},\hat\pi_3}\ .
\end{equation}
Here we used the property $K_d\sim G_{d,p}$ of the complexified groups that was described before eq.\eqref{eq:firstg0} as well as induction in stages, see eq.\ \eqref{eq:indstages}. 
The representation $\hat \pi_3$ of $G_{d,p}$ is obtained by induction from the representation $\hat\sigma_3$ that described the spin and weight of the defect local field. Note that $\hat\pi_3$ 
is infinite dimensional. It belongs to the principal series of $\SO(p+1,1)$ and coincides with $\hat\sigma_3$ on the group $\SO(q)$ of transverse rotations. In the second step, the 
identification of two spaces follows by the usual argument, see \cite{Schomerus:2016epl,Buric:2020buk}. In the remainder of this subsection, we will prove in detail that the above argument is indeed correct, deriving along the way the explicit map between position space three-point functions $G_{2,1}(x_i)$ and elements of $\Gamma_{\sigma_{12},\hat\pi_3}$. The corresponding CS Hamiltonian is readily written down using \eqref{universal-Hamiltonian}. We will verify that the more complicated operator $\mathcal{D}$ can be mapped to it.
\smallskip 

In constructing the map from correlation to spherical functions we shall follow the very 
same steps we have carried out in the previous cases. First, given the correlator 
$G_{2,1}(x_i)$, we can define a function $F_{2,1}: G_d^2\to V\otimes W_1\otimes W_2$ 
by assigning values on a particular section and extending covariantly to $G_d^2$ as 
follows
\begin{equation}\label{F21-right-covariance}
    F_{2,1}(e^{x_1\cdot P},e^{x_2\cdot P})(\hat x_3) = G_{2,1}(x_1,x_2,\hat x_3), 
    \quad F_{2,1}(g_i k_i n_i) = (\sigma_1(k_1^{-1})\otimes\sigma_2(k_2)^{-1}) F_{2,1}(g_i)\ .
\end{equation}
Here, $V$ is the carrier space of the principal series representation $\hat\pi_3$ of $\SO(p+1,1)$. This representation is realised on the space of functions $\mathbb{R}^p\to\hat 
W_3$ in the usual way. The coordinates on $\mathbb{R}^p$ are denoted by $\hat x_3$. In order 
to get a bit more acquainted with the construction of $F_{2,1}$ let us see how the Ward 
identities for the correlation function $G_{2,1}(x_i)$ are expressed in terms of $F_{2,1}$. 
For the correlator, the Ward identities read 
\begin{equation}\label{3pt-Ward-identities}
    G_{2,1}(hx_1,hx_2,h\hat x_3) = (\sigma_1(k(x_1,h))\otimes\sigma_2(k(x_2,h))
    \otimes\hat\sigma_3(k(\hat x_3,h))) G_{2,1}(x_1,x_2,\hat x_3), \quad h\in G_{d,p}\ .
\end{equation}
Elements $k(x,h)$ are defined through the Bruhat decomposition of $he^{x\cdot P}$, see \cite{Buric:2020buk} and eq.\ \eqref{matrix-identity} below. Using the covariance properties 
\eqref{F21-right-covariance} of $F_{2,1}$ we find
\begin{align*}
    & F_{2,1}(hg_j)(h\hat x_3) = F_{2,1}(he^{x_j\cdot P}k_j n_j)(h\hat x_3) 
    = F_{2,1}(e^{h x_j\cdot P} n(z(x_j,h)) k(x_j,h)k_j n_j)(h\hat x_3)\\[2mm]
    & = \bigotimes_{j=1}^2 \sigma_j(k(x_j,h)k_j)^{-1} F_{2,1}(e^{h x_j\cdot P})(h\hat x_3) 
    = \bigotimes_{j=1}^2 \sigma_j(k(x_j,h)k_j)^{-1} G_{2,1}(h x_i)\\[2mm]
    & = \left(\sigma_1(k_1)^{-1}\otimes\sigma_2(k_2)^{-1}\otimes\hat\sigma_3(k(\hat x_3,h))\right)G_{2,1}(x_i) = 
    \hat\sigma_3(k(\hat x_3,h))F_{2,1}(g_i)(\hat x_3)\ .
\end{align*}
To obtain the last line, we used the Ward identities \eqref{3pt-Ward-identities}. In the 
last equality, we substituted the definition of $F_{2,1}$ and again applied covariance 
properties \eqref{F21-right-covariance} to bring some of the prefactors into the argument 
of $F_{2,1}$. The final result can be written in terms of the field representation 
$\hat\pi_3$ as
\begin{equation}
    F_{2,1}(hg_j) = \hat\pi_3(h) F_{2,1}(g_j), \quad h\in G_{d,p}\ .
\end{equation}
At this point the setup looks very similar to the one we considered in the last subsection. Hence we can now replicate the analysis from there. In particular, we shall now define a new
function
\begin{equation}
    F : G_d \to V\otimes W_{12}, \quad F(g) = F_{2,1}(g_0 g,g_0 g w^{-1})\ .
\end{equation}
It is not difficult to determine how this function behaves under left and right action with 
elements $k \in K_d$. A short computation gives 
\begin{align*}
    & F(g k) = F_2(g_0 g k,g_0 g k w^{-1}) = (\sigma_1(k^{-1})\otimes\sigma_2(w k^{-1} w^{-1})) F_2(g_0 g,g_0 g w^{-1}) = \sigma_{12}(k^{-1})F(g)\,,\\[2mm]
    & F(k g) = F_2(g_0 k g, g_0 k g w^{-1}) = F_2(g_0 k g_0^{-1} g_0 g, g_0 k g_0^{-1} g_0 g w^{-1}) = \hat\pi'_3(k)F_2(g_0 g, g_0 g w^{-1}) =\hat\pi'_3(k) F(g)\ .
\end{align*}
Hence, we conclude that $F$ is a $K$-spherical function \eqref{K-spherical-functions} with the right representation $\rho_r = \sigma_{12}$ and left representation given by 
\begin{equation} 
\hat \pi_3' (k) \equiv \hat\pi_3(g_0 k g_0^{-1}) \ . 
\end{equation} 
Recall that $\hat\pi_3$ entered our discussion as a representation of $G_{d,d-2} =  \SO(d-1,1)\times\SO(2)$. Through the conjugation with $g_0$ it is indeed turned into a representation of $K_d = \SO(d) \times \SO(1,1)$. In conclusion we have shown that 
\begin{equation} \label{eq:G21F}
G_{2,1}(x_i) = \Xi(x_i) \, F(t_1,t_2)\,,
\end{equation} 
where $F(t_1,t_2)$ is the restriction of the spherical function $F \in \Gamma_{\hat \pi_3',\sigma_{12}}$ to the abelian group $A_p$. The map $\Xi$ is of the form given in eq.\ \eqref{eq:4ptGFnew} with $\sigma_1 \mapsto \hat \pi_3'$ and $\sigma_2$ the trivial representation while for the right translations we use $\sigma_3 \mapsto \sigma_1$ and $\sigma_4 \mapsto \sigma_2$. We can finally relate $F(t_i)$ to a Calogero-Sutherland wavefunction $\Psi$ by the standard gauge transformation
\begin{equation}\label{Haar-measure-bulk-channel}
    F = \delta(t_i)^{-1}\Psi, \quad \delta(t_i) = \sqrt{\sinh^{d-2} t_1 \sinh^{d-2} t_2 \sinh(t_1+t_2) \sinh(t_1-t_2)}\ .
\end{equation}
The map $G_{2,1}(x_i)\mapsto\Psi$ sends conformal partial waves to eigenfunctions of the Calogero-Sutherland Hamiltonian. The Hamiltonian is obtained readily by substituting generators 
in eq.\ \eqref{universal-Hamiltonian} by operators in representations $\hat\pi_3$ and $\sigma_{12}$ and imposing $M$-invariance. E.g. for the case of identical scalars $\varphi_1 = \varphi_2$ and $\hat\varphi_3$ of vanishing transverse spin in $d=4$, we get
\begin{align}
    H & = \partial_{t_1}^2 + \partial_{t_2}^2 + \frac{1-\left((1+x^2)\partial_x+\Delta_{\hat3} x\right)^2}{2\sinh^2(t_1+t_2)} + 
    \frac{1-\left((1+x^2)\partial_x+\Delta_{\hat3} x\right)^2}{2\sinh^2(t_1-t_2)} - \frac{(1+x^2)\partial_x^2 + 
    2x \partial_x}{\sinh^2 t_2}\nonumber\\[2mm]
    & - \frac{x^2(1+x^2)\partial_x^2 + 2x((\Delta_{\hat3}+1)x^2+\Delta_{\hat3})\partial_x + 
    \Delta_{\hat3}((\Delta_{\hat3}+1)x^2+\Delta_{\hat3}-1)}{\sinh^2 t_1} - 5\ .\label{final-spinning-CS}
\end{align}
To recover the correlation function from $F$, one can follow the same steps as for $G_{2,0}(x_i)$ in the previous section. We have already written coordinates $t_i$ in eq.\ \eqref{coordinates-ti}. 
The remaining coordinate $x$ is given in appendix F, equation \eqref{coordinate-x}. Finally, the prefactor that relates eigenfunctions of the Hamiltonian \eqref{final-spinning-CS} and 
the operator $\mathcal{D}$ is written in eq.\ \eqref{prefactor}. In the same appendix, we verify by a direct computation that the two operators are equivalent to one another. This represents 
a non-trivial consistency check of our formalism.

Before we conclude this subsection we want to briefly review the known results about the defect channel blocks expansions for $G_{2,1}(x_i)$, see \cite{Buric:2020zea}. This defect channel expansion is obtained by expanding both of the bulk fields into defect primaries $\hat\varphi$ and $\hat\varphi'$. The primaries that appear are necessarily scalar, of dimensions $\Delta$, $\Delta'$ and have the same transverse spin $s$. This leads to the conformal block decomposition
\begin{equation}\label{bulk-bulk-defect-correlator}
\langle \varphi_1(x_1) \varphi_2(x_2) \hat \varphi_3(\hat x_3) \rangle = 
\frac{1}{|x_{1\perp}|^{\Delta_1} |x_{2\perp}|^{\Delta_2}} \left(\frac{|x_{2\perp}|}{x_{23}^2}\right)^{\Delta_{\hat 3}}\,
 \sum_{\hat \Delta,\hat \Delta',s} b_{\varphi_1\hat\varphi}b_{\varphi_2\hat\varphi'} 
 \psi_{\hat \Delta,\hat\Delta',s}(v_1,v_2,\kappa)\ .
\end{equation}
Partial waves $\psi_{\hat\Delta,\hat\Delta',s}$ are products of Gegenbauer polynomials $C^{(d-p-2)/2}_s$ and Appell's functions
\begin{equation}\label{3pt-defect-channel-blocks}
    \psi_{\hat\Delta,\hat\Delta',s}(v_1,v_2,\kappa) = v_1^{\frac{\hat\Delta}{2}-\frac{\Delta_{\hat3}}{4}} v_2^{\frac{\hat\Delta'}{2}-\frac{\Delta_{\hat3}}{4}} \left(\frac{v_2}{v_1}\right)^{\frac{\Delta_{\hat 3}}{4}} \,
    F(v_1,v_2) \ C^{(d-p-2)/2}_s (\cos\kappa)\ .
\end{equation}
More precisely, $F$ is given in terms of Appell's hypergeometric function by
\begin{equation}\label{Appell-function}
    F(v_1,v_2) = F_4\left(\frac{\hat\Delta+\hat\Delta'-\Delta_{\hat3}}{2},
    \frac{\hat\Delta+\hat\Delta'-\Delta_{\hat3}+2-p}{2},\hat\Delta-
    \frac{p}{2}+1,\hat\Delta'-\frac{p}{2}+1;v_1,v_2\right)\ .
\end{equation}
The appearance of the Gegenbauer polynomials for the two-point function of spinning fields in transverse space is standard. What was new about the result \eqref{3pt-defect-channel-blocks} of \cite{Buric:2020zea} was the construction of the block in the variables $v_i$. Special cases of eq.\ \eqref{3pt-defect-channel-blocks} in restricted kinematics or for quantum numbers corresponding to free fields had been obtained previously in \cite{Lauria:2020emq,Behan:2020nsf}. 

\section{Conclusions and Outlook}

Let us summarise the results of this work and indicate some future directions. The main actor of our discussion was Harish-Chandra's radial component map. The map allows to systematically construct covariant differential operators acting on spherical functions. These spherical functions appear as partial waves in bulk and defect CFTs and as wavefunctions of spin Calogero-Sutherland integrable models. Universality of Harish-Chandra's map allows to derive compact expressions for Casimir equations (Calogero-Sutherland Hamiltonians) for arbitrary choices of spins (left and right representations). When applied to vector fields on the conformal group, rather than to the Laplacian, the map gives rise to equally compact expressions for differential shifting operators. With the help of function space realisation of finite-dimensional carrier spaces for spin degrees of freedom, the reduced Casimir and shifting operators become families of differential operators, parametrised by external quantum numbers and acting on the same space of functions of cross ratios and analogous spin invariants. The shifting property is then expressed through exchange relations such as eqs.\ \eqref{bulk-channel-shifting}, \eqref{relation-shifting}. We used these external shift operators together with a set of internal weight-shifting operators to compute bulk- and defect-channel conformal blocks of two spinning bulk fields in the presence of a co-dimension $q=2$ defect, simplifying and slightly extending results of \cite{Lauria:2018klo}. In the case of spinning four-point functions, the present work greatly extends \cite{Schomerus:2016epl,Schomerus:2017eny} and provides an alternative view on weight-shifting techniques. As far as we are aware, within the existing literature on spherical functions, as well as spinning Calogero-Sutherland models, our results are new.

We have described two applications of universal spinning Casimir equations to higher-point conformal blocks. The first concerned the reduction of six-point conformal blocks to products of spinning four-point blocks in a certain OPE limit, \cite{Buric:2021kgy}. Comparison of two sets of blocks goes via comparison of associated Casimir equations. Since intermediate operators in the six-point function can carry arbitrarily high spins, the comparison required knowledge of the four-point equations universally in external spins. As the second application, we derived new Casimir equations for three-point function conformal blocks of correlators involving one co-dimension two defect field and two bulk fields. The equation takes the form of the spinning $BC_2$ Calogero-Sutherland problem with an infinite dimensional spin representation, and was matched to the equation derived (also in this work) by more conventional means.

The most explicit results for Calogero-Sutherland wavefunctions above were given for systems of real rank (the number of particles) one or two. Given the similarity between treatements of these two cases, we expect the methods to apply with little difference to higher rank systems. Another measure of complexity of spin Calogero-Sutherland models is given by depths of left and right representations. In the present work, we studied two sided systems with depths one on each side (STTs), and one sided systems with an MST of depth two. In the upcoming work \cite{BuricSmat}, it will be shown, at least in several cases, that our techniques extend to two-sided MST systems.

The present work opens several directions. One is to analyse the bulk-bulk-defect three-point function crossing equations in the lightcone limit. The defect channel blocks are known for generic point configurations and the bulk channel Casimir equations simplify considerably and can be solved in this limit. The lightcone bootstrap with two bulk points was performed already in \cite{Lemos:2017vnx}. It is natural to ask whether the conclusions of that work can be improved upon by adding a third point on the defect. Away from the lightcone limit, the bulk channel blocks are yet to be computed. The solution may be attempted using results of the last subsection and trying to generalise Calogero-Sutherland solution techniques to infinite-dimensional external representations. The mentioned three-point function also admits an associated Gaudin model in the spirit of \cite{Buric:2020dyz}, as will be shown in \cite{Buricinprep}. In particular we shall explain how to construct a third {\it vertex} differential operator which commutes with the two Casimirs.

Let us point out that the Hamiltonians described above when combined with results of \cite{Buric:2019rms} lead to universal Casimir equations for superconformal blocks of type I supersymmetry (equivalently, spherical functions on supergroups of type I). While \cite{Buric:2019rms} also developed solution theory for these equations, assuming the knowledge of bosonic spherical functions, a more satisfactory method using weight-shifting operators is still missing. It was shown in \cite{Buric:2019rms} that spherical functions on supergroups can carry an action of non-trivial invariant operators (i.e. elements of the universal enveloping algebra of $\mathfrak{g}$) that are not present for the underlying Lie group (in \cite{Buric:2019rms}, these operators formed the Lie superalgebra $\mathfrak{gl}(1|1)$). Determining all invariant and covariant (shifting) operators, as well as their algebraic structure, remains an intriguing question for future research. It might also be interesting to study shifting operators for spinning extensions of the Casimir equations for superconformal primaries that we recently solved by \cite{Aprile:2021pwd}.

It turns out that spherical functions appear as partial waves in ordinary QFT without conformal symmetry, by choosing the appropriate Gelfand pair. Thus, our techniques can be used for computations of these waves as well, relevant for scattering of spinning particles, as will be shown in \cite{BuricSmat} (see \cite{Caron-Huot:2022jli} for an alternative approach).

As we have underlined, our methods for constructing spinning Calogero-Sutherland wavefunctions are self-sufficient. Still, it would be of value to explore relations between these methods and other treatments of spherical functions and Calogero-Sutherland integrable systems. For instance, our techniques allow to obtain explicit expressions for matrix spherical functions, see \cite{Koelink_2012,Tirao_2014} and references therein. In some cases, these works made a further connection to the matrix hypergeometric equation, \cite{10.2307/3139893}. Another question is to obtain the Dunkl operators and the Lax pair for spinning models (the two being closely related in the scalar case, \cite{Chalykh_2019}). To this end, the right place to start may be the classical models of \cite{Kharchev_2017,Kharchev_2018}, which are closely related to ours. There exists a general, if difficult to execute, method for constructing Calogero-Sutherland eigenfunctions, in terms of power series with coefficients (called Harish-Chandra coefficients) in the universal enveloping algebra of $\mathfrak{k}$, see \cite{Isachenkov:2018pef,Stokman:2020bjj} for references. A valid question is whether our solution generating techniques can be used for computations of Harish-Chandra coefficients, which would in turn be used to obtain solutions with different external representations. Other aspects of integrability of spinning Calogero-Sutherland models with less emphasis on explicit wavefunctions, such as the r-matrix, bi-Hamiltonian structure and superintegrability, have also been studied, see \cite{Feh_r_2008,Reshetikhin_2016,Feh_r_2021} and references therein.

Finally, let us mention a few systems where one might try to make progress along the lines we followed here. Scalar Calogero-Sutherland models arise not only from cosets $K\backslash G/K$, but also their asymmetric cousins $K_1\backslash G/K_2$, where both $K_1$ and $K_2$ are spherical subgroups of $G$, \cite{Flensted-Jensen,Molchanov1995}. We are not aware of an appropriate generalisation of Harish-Chandra's map that would allow for universal reductions of differential operators to such double quotients. A theory for these asymmetric reductions would e.g. allow to extend results of section 4 to defects of arbitrary co-dimension.

Perhaps one of the most interesting generalisations to be explored comes through the relation to Gaudin models. As we have touched upon several times, higher-point conformal blocks are can be characterised as eigenfunctions of Casimir and vertex operators, which together constitute Hamiltonians of a Gaudin integrable system, \cite{Buric:2021kgy,Buric:2021ttm,Buric:2021ywo}. Extension of these results to setups with defects is possible and will be explained in \cite{Buricinprep}. As we have seen on one example in this work, higher-point Casimir equations can sometimes also be put in spinning Calogero-Sutherland form. In fact, it is not difficult to see that constructions similar to that of section 4.3 are possible for bulk higher-point functions as well. These lead to an interplay between Gaudin and spinning Calogero-Sutherland models, which has been very little explored. An outstanding open question would be to uplift our weight-shifting algebra to the full Gaudin model, reversing the process described in section 3. If achieved, this would provide a systematic construction of six-point conformal blocks.
\bigskip 

\vskip0.1cm \textbf{Acknowledgements:} We wish to thank Evgeny Sobko and Edwin Langmann for discussions on matrix Calogero-Sutherland models and Nikolai Reshetikhin for pointing out to us the universal character of Harish-Chandra's map. Several aspects of weight-shifting were understood in discussions with Francesco Russo. We have also benefited from discussions with Sylvain Lacroix, Jeremy Mann, Lorenzo Quintavalle, Jasper Stokman and Alessandro Vichi. This project received funding from the German Research Foundation DFG under Germany’s Excellence Strategy -- EXC  2121 Quantum Universe -- 390833306. I.B. is funded by a research grant under the project H2020 ERC STG 2017 G.A. 758903 "CFT-MAP".

\appendix

\section{Cartan involutions}

In this appendix we give some details about Cartan decompositions of real simple Lie algebras. Also, we introduce some standard terminology which is used in the main text.
\smallskip

Let $\mathfrak{g}_c$ be a complex simple Lie algebra and $\mathfrak{h}_c$ a Cartan subalgebra of $\mathfrak{g}_c$. By the standard theory, $\mathfrak{g}_c$ decomposes as
\begin{equation}\label{root-decomposition}
    \mathfrak{g}_c = \mathfrak{h}_c \oplus \sum_{\alpha\in\Phi} \mathfrak{g}_c^\alpha\ .
\end{equation}
Here, $\alpha$ are roots of $\mathfrak{g}_c$ and the set of all roots is denoted by $\Phi$. We regard roots as linear maps $\mathfrak{h}_c\xrightarrow{}\mathbb{C}$ such that $\text{ad}_H = \alpha(H)$ on the space $\mathfrak{g}_c^\alpha$. Root spaces are one-dimensional, $\mathfrak{g}_c^\alpha = \text{span}\{e_\alpha\}$. We normalise root vectors $e_\alpha$ so that the following relations are satisfied
\begin{equation}
    [e_\alpha,e_{-\alpha}] = h_\alpha, \quad \kappa(h_\alpha,H) = \alpha(H), \quad \forall H\in\mathfrak{h}_c\ .
\end{equation}
In the last equality, $\kappa$ is the Killing form of $\mathfrak{g}_c$. The real vector space spanned by $\{h_\alpha\}$ is denoted by $\mathfrak{h}$.
\smallskip

The complex Lie algebra $\mathfrak{g}_c$ can be also regarded as a real Lie algebra, written $\mathfrak{g}_c^{\mathbb{R}}$. A Lie subalgebra $\mathfrak{g}$ of $\mathfrak{g}_c^{\mathbb{R}}$ is said to be a real form of $\mathfrak{g}_c$ if the following direct sum decomposition is valid
\begin{equation}
    \mathfrak{g}_c^{\mathbb{R}} = \mathfrak{g} \oplus i \mathfrak{g}\ .
\end{equation}
That is, any element $X$ of $\mathfrak{g}_c^{\mathbb{R}}$ can be uniquely decomposed as $X = Y + i Z$, with $Y,Z\in\mathfrak{g}$. The map $\sigma:\mathfrak{g}_c\xrightarrow{}\mathfrak{g}_c$ defined by $\sigma(Y+iZ) = Y-iZ$ is an automorphism of $\mathfrak{g}_c^\mathbb{R}$ called conjugation with respect to $\mathfrak{g}$.
\smallskip

It is well-known that any complex simple Lie algebra $\mathfrak{g}_c$ has a compact real form $\mathfrak{u}$ on which the Killing form is negative-definite. In terms of the basis introduced above, we have $\mathfrak{u} = \text{span}_{\mathbb{R}}\{i h_\alpha,e_{\alpha}-e_{-\alpha},i(e_\alpha + e_{-\alpha})\}$. Conjugation with respect to $\mathfrak{u}$ will be denoted by $\tau$
\begin{equation}
    \tau(h_\alpha) = -h_\alpha, \quad \tau(e_\alpha) = -e_{-\alpha}, \quad \alpha\in\Phi\ .
\end{equation}
Let now $\mathfrak{g}$ be a real simple Lie algebra and $\mathfrak{g}_c$ its complexification. A decomposition $\mathfrak{g} = \mathfrak{k} \oplus \mathfrak{p}$, where $\mathfrak{k}$ is a subalgebra and $\mathfrak{p}$ is a linear subspace of $\mathfrak{g}$ is called a Cartan decomposition if, in the above notation
\begin{equation}\label{Cartan-decomposition-conditions-1}
    \sigma(\mathfrak{u}) \subset \mathfrak{u}, \quad \mathfrak{u} \cap \mathfrak{g} = \mathfrak{k}, \quad i\mathfrak{u} \cap \mathfrak{g} = \mathfrak{p}\ .
\end{equation}
These conditions imply that, for any $k\in\mathfrak{k}$ and $p\in\mathfrak{p}$
\begin{equation}
    \sigma(\tau(k)) = \sigma(k) = k, \quad \tau(\sigma(k)) = \tau(k) = k, \quad \sigma(\tau(p)) = \sigma(-p) = -p, \quad \tau(\sigma(p)) = \tau(p) = -p\,,
\end{equation}
that is, $\tau\sigma = \sigma\tau = 1_\mathfrak{k} - 1_\mathfrak{p}$. An equivalent definition of a Cartan decomposition of $\mathfrak{g}$ is that into a direct sum of a subalgebra $\mathfrak{k}$ and a subspace $\mathfrak{p}$ subject to conditions
\begin{equation}
    \kappa|_\mathfrak{k} < 0 ,\quad \kappa|_\mathfrak{p} > 0, \quad \theta = 1_\mathfrak{k} - 1_\mathfrak{p} \ \ \text{is an automorphism of } \mathfrak{g}\ .
\end{equation}
A real form $\mathfrak{g}$ of $\mathfrak{g}_c$ is said to be split if $\mathfrak{p}$ contains a maximal (ad-diagonalisable) abelian subalgebra of $\mathfrak{g}$. It is known that any complex simple Lie algebra has a unique (up to isomorphism) split real form. In fact, this real form is the span of $\{h_\alpha,e_\alpha\}$. For a real form that is not necessarily split, we denote by $\mathfrak{a}_p$ a maximal abelian subspace of $\mathfrak{p}$. Let $\mathfrak{a}$ be a maximal abelian subalgebra of $\mathfrak{g}$ that contains $\mathfrak{a}_p$. Then we have the direct sum decomposition
\begin{equation}
    \mathfrak{a} = (\mathfrak{a} \cap \mathfrak{k}) \oplus (\mathfrak{a} \cap \mathfrak{p}) \equiv \mathfrak{a}_k \oplus \mathfrak{a}_p\ .
\end{equation}
The complexification $\mathfrak{a}_c$ of $\mathfrak{a}$ is a Cartan subalgebra of $\mathfrak{g}_c$, which is stable under $\theta$. On the other hand
\begin{equation}
    \mathfrak{h} = \mathfrak{a}_p \oplus i \mathfrak{a}_k\ .
\end{equation}
Let us split the positive roots $\Phi_+$ into a disjoint union of two subsets $P_\pm$. Elements of $P_+$ are positive roots that do not vanish identically on $\mathfrak{a}_p$ and elements of $P_-$ are the ones that do. Any linear functional on $\mathfrak{h}$ is a linear functional on $\mathfrak{a}_p$ by restriction. By restricting elements of $P_+\cup(-P_+)$ to $\mathfrak{a}_p$ we arrive at the set of restricted roots $\Sigma$ from the main text. We will adopt the notation to write $\tilde\alpha$ for the restriction of $\alpha\in \Phi$ to $\mathfrak{a}_p$. Elements of $\Sigma_+$ are restrictions of $\alpha\in P_+$.
\smallskip

Vectors $e_\alpha$ in the Cartan-Weyl basis of $\mathfrak{g}_c$ may be chosen such that $\tau(e_\alpha)=-e_{-\alpha}$. Since $\mathfrak{h}_c$ is $\theta$-stable, we may regard $\theta$ as a linear involution of $\mathfrak{h}_c^\ast$. This map restricts to an involution of the root system $\Phi$ such that
\begin{equation*}
    \theta\alpha|_{\mathfrak{a}_p} = -\alpha|_{\mathfrak{a}_p}\ .
\end{equation*}
We have
\begin{equation}
    \sigma(e_{\alpha_+}) = k_{\alpha_+} e_{\alpha_+}, \quad \sigma(e_{\alpha_-}) = - e_{\alpha_-}, \quad \alpha_\pm\in P_\pm \cup (-P_\pm)\ .
\end{equation}
The first equation defines numbers $k_{\alpha_+}$. From $\sigma\tau = \tau\sigma$, we get $\bar k_{\alpha_+} = k_{-\alpha_+}$. The second equation follows from the fact that $e_{-\alpha}\in\mathfrak{k}_c$.

\vskip0.1cm Another space of some importance in our analysis is the centraliser of $\mathfrak{a}_p$ in $\mathfrak{k}$, denoted $\mathfrak{m}$. From above, we see that
\begin{equation}
    \mathfrak{m}_c = \mathfrak{a}_{k_c} \oplus \sum_{\alpha\in P_-} \mathfrak{g}_c^\alpha \oplus \sum_{\alpha\in P_-} \mathfrak{g}_c^{-\alpha}\ .
\end{equation}
Notice that spaces $\mathfrak{g}^\lambda$ are $\text{ad}_{\mathfrak{m}}$-stable. Indeed
\begin{equation*}
    [h,[m,x]] = -[m,[x,h]]-[x,[h,m]] = \lambda(x)[m,x], \quad x\in\mathfrak{g}^\lambda \ .
\end{equation*}
The sum of positive restricted root spaces is denoted by $\mathfrak{n}_c$
\begin{equation}
    \mathfrak{n}_c = \sum_{\alpha\in P_+} \mathfrak{g}_c^\alpha, \quad \mathfrak{n} = \mathfrak{g} \cap \mathfrak{n}_c\ .
\end{equation}
Then $\mathfrak{n}$ is a maximal nilpotent subalgebra of $\mathfrak{g}$. By the above remarks, both $\mathfrak{n}$ and $\mathfrak{n}_c$ are $\text{ad}_{\mathfrak{m}}$-stable. The direct sum decomposition
\begin{equation}
    \mathfrak{g} = \mathfrak{k} \oplus \mathfrak{a}_p \oplus \mathfrak{n}\,,
\end{equation}
is called the Iwasawa decomposition of $\mathfrak{g}$. The projection of $\mathfrak{n}_c$ to $\mathfrak{k}_c$ is denoted by $\mathfrak{q}_c$
\begin{equation}
    \mathfrak{q}_c = (1+\theta)(\mathfrak{n}_c) = \text{span}\{y_\alpha\ |\ \alpha\in P_+\}\ .
\end{equation}

\section{Lorentzian conformal group}

Conventions for the Lorentzian conformal group have been given in the main text. Here we add several useful formulas. Matrices in the $(d+2)$-dimensional representation are written in the block form in the obvious way. Then
\begin{equation}
    x^i L_{0i} = \begin{pmatrix}
    0 & -x^T & 0\\
    -x & 0 & 0 \\
    0 & 0 & 0\end{pmatrix}, \quad  x^i L_{i,d+1} = \begin{pmatrix}
    0 & 0 & 0\\
    0 & 0 & x \\
    0 & x^T & 0\end{pmatrix}, \quad w = \begin{pmatrix}
    -1 & 0 & 0\\
    0 & 1 & 0 \\
    0 & 0 & -1\end{pmatrix}\ .
\end{equation}
In accordance with the general theory, we have
\begin{equation}\label{compact-conformal-group}
    \mathfrak{u} = \mathfrak{k} \oplus i \mathfrak{p} = \text{span}_{\mathbb{R}}\{L_{0,d+1},L_{ij},i L_{0i},i L_{i,d+1}\} \cong \mathfrak{so}(d+2)\ .
\end{equation}
From here, one sees how the conjugation $\tau$ acts on vectors $e^{(a)}_\lambda$
\begin{equation}
    \tau\left(e^{(a)}_\lambda\right) = -e^{(a)}_{-\lambda}\ .
\end{equation}
The last relation is also clear from the corresponding equations for $\theta$, keeping in mind that $\theta = \sigma\tau$ and that $\sigma$ preserves $e^{(a)}_\lambda$-s. To elaborate on this point, notice that the Cartan subalgebra $\mathfrak{h}$ can be chosen in the case at hand as
\begin{equation}
    \mathfrak{h} = \text{span}_{\mathbb{R}}\{L_{01},i L_{23},...,L_{d,d+1}\}\ .
\end{equation}
Here, one distinguishes two cases, of even and odd $d$. For both $d=2n$ or $d=2n+1$, the second to last term in the set above is $i L_{2n-2,2n-1}$. There are $n+1$ elements in total, which is indeed the rank of $\mathfrak{g}$. In this basis, clearly $e^a_\lambda$ are not root vectors, e.g.
\begin{equation*}
    [L_{23},e^2_{(1,0)}] = [L_{23},L_{02}+L_{12}] = -L_{03} -L_{13}\ .
\end{equation*}
However, we will continue to use vectors $e^a_\lambda$ and associated objects such as $y^a_\lambda$, $z^a_\lambda$ etc. They are linear combinations of objects labelled by roots $\alpha\in P_+\cup(-P_+)$ and satisfy most of their properties.
\smallskip

In accord with the above remarks, we have
\begin{align}
    & L_{1a} = y_{(1,0)}^a, \quad L_{ad} = y^a_{(0,1)}, \quad L_{1d} = \frac{1}{\sqrt{2}}(y_{(1,-1)}+y_{(1,1)}), \quad L_{0,d+1} = \frac{1}{\sqrt{2}}(y_{(1,-1)}-y_{(1,1)}),\\[2mm]
    & L_{0a} = z_{(1,0)}^a, \quad L_{a,d+1} = -z^a_{(0,1)}, \quad L_{0d} = \frac{1}{\sqrt{2}} (z_{(1,-1)}+z_{(1,1)}),\quad L_{1,d+1} = \frac{1}{\sqrt{2}} (z_{(1,-1)}-z_{(1,1)}),
\end{align}
with
\begin{equation}\label{eqn-z-appendix}
    z^{(a)}_\alpha = \coth(\alpha\cdot t) y^{(a)}_\alpha - \frac{1}{\sinh(\alpha\cdot t)}y'^{(a)}_\alpha\ .
\end{equation}
The quadratic Casimir $C_2 = -\frac12 L^{\alpha\beta} L_{\alpha\beta}$ can be written as
\begin{align}
    C_2 & = -\frac12 L^{ab}L_{ab} - \Big( L^{0a} L_{0a} + L^{1a} L_{1a} + L^{ad} L_{ad} + L^{a,d+1} L_{a,d+1} \\
    & + L^{0d} L_{0d} + L^{1d} L_{1d} + L^{1,d+1} L_{1,d+1} + L^{0,d+1} L_{0,d+1} + L^{01}L_{01} + L^{d,d+1}L_{d,d+1}\Big)\ .
\end{align}
By expressing $L_{ij}$-s in terms of generators $H$, $y$, $z$ we get
\begin{equation}
    C_2 = -\frac12 L^{ab}L_{ab} + H_1^2 + H_2^2 - \sum_{\lambda\in\Sigma_+} \left( y^{(a)}_\lambda y^{(a)}_\lambda - z^{(a)}_\lambda z^{(a)}_\lambda \right),
\end{equation}
where the summation over $a$ is understood. Finally, by eq. \eqref{eqn-z-appendix}
\begin{equation}\label{Casimir-so(d,2)-next-to-last}
    C_2 = -\frac12 L^{ab}L_{ab} + H_1^2 + H_2^2 + \sum_{\lambda\in\Sigma_+} \frac{y^{(a)}_\lambda y^{(a)}_\lambda + y'^{(a)}_\lambda y'^{(a)}_\lambda}{\sinh^2(\lambda\cdot t)} - \frac{\cosh(\lambda\cdot t)}{\sinh^2(\lambda\cdot t)}\{y^{(a)}_\lambda,y'^{(a)}_\lambda\}\ .
\end{equation}
One can check explicitly that with our normalisations, the following brackets hold (no sum)
\begin{equation}
    [y^{(a)}_\lambda,y'^{(a)}_\lambda] = -\sinh(\lambda\cdot t) \lambda \cdot H\ .
\end{equation}
Upon substitution into eq. \eqref{Casimir-so(d,2)-next-to-last} we get the expression for the quadratic Casimir element written in eq. \eqref{Casimir-so(d,2)}.

\section{Euclidean conformal group}

Here we spell out our conventions for the Euclidean conformal group $G=\SO(d+1,1)$ and its Lie algebra. The non-vanishing Lie brackets in $\mathfrak{g}=\mathfrak{so}(d+1,1)$ read
\begin{align}
    & [M_{\mu\nu},P_\rho] = \delta_{\nu\rho} P_\mu - \delta_{\mu\rho} P_\nu,\quad [M_{\mu\nu},K_\rho] = \delta_{\nu\rho} K_\mu - \delta_{\mu\rho} K_\nu,\\
    & [M_{\mu\nu},M_{\rho\sigma}] = \delta_{\nu\rho} M_{\mu\sigma} - \delta_{\mu\rho} M_{\nu\sigma} + \delta_{\nu\sigma} M_{\rho\mu} - \delta_{\mu\sigma} M_{\rho\nu},\\
    & [D,P_\mu] = P_\mu,\quad [D,K_\mu]=-K_\mu,\quad [K_\mu,P_\nu] = 2(M_{\mu\nu} - \delta_{\mu\nu}D)\ .
\end{align}
In the Lorentz-like notation, we write the generators as $\{L_{\alpha\beta}\}$, $\alpha,\beta = 0,1,...,d+1$. These obey the relations
\begin{equation}
    [L_{\alpha\beta},L_{\gamma\delta}] = \eta_{\beta\gamma} L_{\alpha\delta} - \eta_{\alpha\gamma} L_{\beta\delta} + \eta_{\beta\delta} L_{\gamma\alpha} - \eta_{\alpha\delta} L_{\gamma\beta}\,,
\end{equation}
where $\eta$ is the mostly-positive Minkowski metric. The relation between conformal and Lorentz generators reads
\begin{equation}
    L_{01} = D, \quad L_{0\mu} = \frac12(P_\mu + K_\mu), \quad L_{1\mu} = \frac12(P_\mu - K_\mu), \quad L_{\mu\nu} = M_{\mu\nu}\ .
\end{equation}
The quadratic Casimir is given by
\begin{align}
    C_2 &= -\frac12 L^{\alpha\beta} L_{\alpha\beta} = -L^{01}L_{01} - L^{0\mu}L_{0\mu} - L^{1\mu}L_{1\mu} - \frac12 L^{\mu\nu}L_{\mu\nu} = D^2 +\frac12 \{P_\mu,K^\mu\} -\frac12 M^{\mu\nu} M_{\mu\nu}\\
    & = L_{01}^2 + L_{02}^2 + L_{03}^2 - L^{0a}L_{0a} - L_{12}^2 - L_{13}^2 - L^{1a}L_{1a} - L_{23}^2 - L^{2a}L_{2a} - L^{3a}L_{3a} - \frac12 L^{ab}L_{ab}\ .
\end{align}
The maximal compact subgroup of $G$ is $K=\SO(d+1)$. The Cartan decomposition of $\mathfrak{g}$ reads
\begin{equation}
    \mathfrak{g} = \mathfrak{k} \oplus \mathfrak{p}, \quad \mathfrak{k} = \text{span}\{L_{ij}\}, \quad  \mathfrak{p} = \{L_{0i}\}, \quad i,j=1,...,d+1\ .
\end{equation}
The Cartan involution $\theta$ is conjugation by $\text{diag}(-1,1,1,...,1)$. Obviously, the real rank of $G$ is equal to one. We will choose $\mathfrak{a}_p$ to be spanned by $D = L_{01}$. Its stabiliser in $\mathfrak{k}$ is
\begin{equation}
    \mathfrak{m} = \text{span}\{L_{\mu\nu}\}\cong \mathfrak{so}(d), \quad \mu,\nu = 2,...,d+1\ .
\end{equation}
There are two root spaces spanned by
\begin{equation}
    \mathfrak{g}^{(1)} = \text{span}\{P_\mu\}, \quad \mathfrak{g}^{(-1)} = \text{span}\{K_\mu\}\ .
\end{equation}
In terms of these, root vectors are $y_\mu = \frac12(P_\mu-K_\mu) = L_{1\mu}$ and $z_\mu=\frac12(P_\mu+K_\mu) = L_{0\mu}$. According to the general theory, the quadratic Casimir of $\mathfrak{g}$ can be written as
\begin{equation}
    C_2 = L_{01}^2 + d \coth t\ L_{01} + \sum_{\mu=2}^{d+1} \frac{L'^2_{1\mu}- 2\cosh t\ L'_{1\mu} L_{1\mu} + L^2_{1\mu}}{\sinh^2 t} - \frac12 L^{\mu\nu} L_{\mu\nu}\ .
\end{equation}
where, as usual, $h = e^{t D}$ and $L'_{1\mu} = h^{-1} L_{1\mu} h$.

\section{Proof of equation \eqref{magic-formula}} 

In this appendix we would like to briefly sketch the proof of our formula \eqref{magic-formula}. The derivation we provide here follows the one given in \cite{Buric:2020buk} in the context of superconformal symmetry. It is conceptually clearer than the original in \cite{Buric:2019dfk} and extends also to the defect setups addressed in section 4. Our proof here is
based on an extension of the factorisation formula \eqref{eq:factorization}
\begin{equation}
    h m(x) = m(y(x,h))\, n(z(x,h))\, k(t(x,h)) \,, \label{matrix-identity}
\end{equation}
from the Weyl inversion $h = w_{d+1}$ to arbitrary elements $h$ of the conformal group. The extended factorisation formula involves three sets of functions $y(x,h)
= (y(x,h)_a)$, $z(x,h) = (z(x,h)_a)$ and $t(x,h) = (t(x,h)_\varrho)$. Let us note that e.g. $y(x,h)$ describes the usual global action of the conformal group element $h$ on the insertion
point $x$. In case of $z(x,h)$, this action is conjugated with the Weyl inversion.

A four-point correlation function $G_4(x_i)$ satisfies a set of Ward identities. For global conformal transformations $h$ these may be written in the form
\begin{equation} \label{eq:G4Wardid}
    G_4 (x_i^h) = \Big(\bigotimes_{i=1}^4 \rho_i (k(t(x_i,h)))\Big) G_4(x_i) \ .
\end{equation}
Note that correlation functions are essentially invariant under these transformations except some factors depending in the weights and spins of the fields. This dependence is encoded in the
choice of representations $\sigma_i$, as we explained above. In the first step we wish to lift the correlator $G_4(x_i)$ to a function $F_4$ on four copies of $G_d$ valued in the vector 
space $W_1\otimes\dots\otimes W_4$. This lifting can be achieved in a unique way if we require
\begin{equation}\label{eq:F4rightcov}
    F_4(m(x_i)) = G_4(x_i)\  ,\quad \quad F_4 (g_i n_i k_i) = \bigotimes_{i=1}^4 \sigma_i (k_i^{-1}) F_4(g_i) \ .
\end{equation}
The Ward identities \eqref{eq:G4Wardid} satisfied by $G_4(x_i)$ imply the following invariance conditions satisfied by $F_4$ under simultaneous left multiplication of its four arguments
by an element $h$ of the conformal group,
\begin{eqnarray}\label{eq:F4leftinv}
    F_4(h m(x_i)) & = & F_4\Big( m(x_i^h) n(z(x_i,h)) k(t(x_i,h)) \Big) \\[2mm]
    & = & \Big( \bigotimes_{i=1}^4 \sigma_i (k(t(x_i,h))^{-1}) \Big) G_4(x_i^h) =
    G_4(x_i) = F(m(x_i)) \ .
\end{eqnarray}
Other than the Ward identity, we have used the definitions \eqref{matrix-identity} and \eqref{eq:F4rightcov}. Given $F_4$ and the Weyl inversion $w = w_{d+1}$, we can construct a function $F:G_d \to W_1\otimes\dots\otimes W_4$ through the prescription
\begin{equation}\label{eq:FfromF4}
    F(g) := F_4 (e,w^{-1},g,gw^{-1}) \ .
\end{equation}
While this might look a bit odd at first, it is easy to verify that $F$ is a $K$-spherical function, i.e. that it satisfies the covariance laws \eqref{K-spherical-functions} for appropriate $\rho_l$ and $\rho_r$. Indeed, from the definition \eqref{eq:FfromF4} of $F$, the left invariance condition \eqref{eq:F4leftinv} and the right covariance law in eq.\ \eqref{eq:F4rightcov} of $F_4$ we obtain
\begin{align*}
    F(k_l g k_r) &= F_4(e,w^{-1},k_l g k_r, k_l g k_r w^{-1}) =
    F_4(k_l^{-1},w^{-1} w k_l^{-1}w^{-1},g k_r, g w^{-1} w k_r w^{-1} )\\[2mm]
                 &= \Big(\sigma_1(k_l)\otimes\sigma_2(w k_l w^{-1})\otimes\sigma_3(k_r^{-1})\otimes\sigma_4(wk_r^{-1}w^{-1})\Big) F(g) \ .
\end{align*}
In conclusion we have shown that a correlation function $G_4(x_i)$ provides us with a $K$-spherical function $F$. It is not difficult to invert the map and recover $G_4(x_i)$ from $F$. Suppressing the last two arguments and their corresponding prefactors for simplicity, we have
\begin{eqnarray*}
F_4(m(x_1),m(x_2)) & = & \left(1 \otimes \sigma_2(k(t_{21})^{-1})\right)F_4\left(m(x_1) n(y_{21}), m(x_2) k(t_{21})^{-1} n(z_{21})^{-1} \right) \\[2mm]
& = & \left(1 \otimes \sigma_2(k(t_{21})^{-1})\right)F_4\left(m(x_1) n(y_{21}), m(x_1) m(x_{21}) k(t_{21})^{-1} n(z_{21})^{-1} \right) \\[2mm]
& = & \left(1 \otimes \sigma_2(k(t_{21})^{-1})\right)F_4\left(m(x_1) n(y_{21}), m(x_1) w^{-1} m(y_{21}) \right) \\[2mm]
& = & \left(1 \otimes \sigma_2(k(t_{21})^{-1})\right)F_4\left(m(x_1) n(y_{21}), m(x_1) n(y_{21}) w^{-1} \right)\ .
\end{eqnarray*}
In the first step we used the covariance property \eqref{eq:F4rightcov} of $F_4$ in the
first two arguments to multiply the first argument with $n(y_{21})$ and the second with
$k(t_{21})^{-1} n(z_{21})^{-1}$. Since the latter contains a factor $k$ it needed to
be compensated by a rotation in the second factor. Then we inserted the definition of
$m(x_{21})$ and used that
$$ m(x_{21}) = w^{-1} m(y_{21}) n(z_{21}) k(t_{21})\ .  $$
This factorisation formula is essentially the definition of $y_{21}, z_{21}$ and $t_{21}$.
Finally we moved the Weyl element $w^{-1}$ through $m$ using that $n = w^{-1} m w$. We
can now apply the same steps to the third and fourth argument to obtain
\begin{equation}\label{eq:F4ggwggw}
F_4(m(x_i)) = \left(1 \otimes \sigma_2(k(t_{21})^{-1}) \otimes 1 \otimes \sigma_4(k(t_{43})^{-1})\right)
F_4\left(g_{12}(x_i),g_{12}(x_i) w^{-1}, g_{34}(x_i), g_{34}(x_i) w^{-1} \right),
\end{equation}
where we introduced the elements
$$ g_{ij}= m(x_i) n(y_{ji})\ .  $$
Finally, we can use the invariance property \eqref{eq:F4leftinv} of $F$ for $h = g_{12}^{-1}$ to obtain
$$ F_4(m(x_i)) = \left(1 \otimes \sigma_2(k(t_{21})^{-1})\otimes 1 \otimes \sigma_4(k(t_{43})^{-1})\right)F_4\left(e, w^{-1}, g(x_i), g(x_i) w^{-1}\right) \ .$$
Here $g(x_i)$ is the element we introduced in eq.\ \eqref{eq:gxi}. Using our definition of the functional $F$ in eq.\ \eqref{eq:FfromF4} and the relation between $F_4$ and $G_4(x_i)$
we have thereby established the lifting formula \eqref{magic-formula}.

\section{Asymptotic solutions}

In this appendix, we give the matrix $A_{nm}$ from \eqref{matrix-A}. The matrix reads
\begin{align*}
    A = \frac{1}{\msj+\Delta-1} &\begin{pmatrix}
    \frac{1}{4(\msj+\Delta)} & 0 & 0\\
    0 & \frac{i(2\msj+d-3)}{4(2\msj+d-2)} & 0\\
    0 & 0 & \frac{2\msj+d-3}{(2\msj+d)(2\msj+d-2)}
    \end{pmatrix}\begin{pmatrix}
    1 & 2 & 0\\
    \frac{2b}{\msj+\Delta} & \frac{4b}{\msj+\Delta} & 1\\
    C & 2C + \frac{(\msj+\Delta-1)(2\msj+d-1)}{2} & -\frac{b(2\msj+d-1)}{2}
    \end{pmatrix}\begin{pmatrix}
    1 & 0 & 1\\
    0 & 1 & 0\\
    1 & 0 & -1
    \end{pmatrix}\,,
\end{align*}
where 
\begin{equation*}
    C = \frac{(J-\msj)(J+\msj+d-2) - (2\msj+d-1)\left(4b^2+(\msj+\Delta)^2\right)}{8(\msj+\Delta)}\ .
\end{equation*}

\section{Three-point bulk Casimir equations}

In this appendix, give the Casimir \eqref{3pt-bulk-Casimir} and its map to the Calogero-Sutherland Hamiltonian \eqref{final-spinning-CS}. For concreteness, we will put $d=4$. Coefficients of the bulk Casimir operator in variables $v_0=\cos\kappa$, $v_1$, $v_2$ read
\begin{align*}
   & C = \Delta_1(\Delta_1 - 4) + \Delta_2 (\Delta_2 - 4) - \Delta_1 \frac{\Delta_2(v_1+v_2-1) + \Delta_{\hat3}(v_1-v_2+1)}{\sqrt{v_1 v_2}}\cos\kappa,\\
   & C_0 = \frac{1}{2\sqrt{v_1 v_2}} \Big((\Delta_1+\Delta_2-1)(v_1+v_2-1) + \Delta_{\hat3}(v_1-v_2+1)\\
   & -\big(\Delta_{\hat3}(v_1-v_2+1)+(\Delta_1+\Delta_2+1)(v_1+v_2-1)\big)\cos 2\kappa \Big) - 2\cos\kappa,\\
   & C_1 = 4 v_1 (v_1+v_2-\Delta_{\hat3})- 2\sqrt{\frac{v_1}{v_2}}\Big(4v_1 v_2 + \Delta_1 v_1+(\Delta_1+2\Delta_2-2\Delta_{\hat3}-4)v_2 - \Delta_1\Big)\cos\kappa,\\
   & C_2 = 4 v_2 (v_1+v_2)-2\sqrt{\frac{v_2}{v_1}}\Big(4v_1 v_2 + (2 \Delta_1+\Delta_2+\Delta_{\hat3}-4)v_1+(\Delta_2-\Delta_{\hat3})v_2 - \Delta_2 + \Delta_{\hat3}\Big)\cos\kappa,\\
   & C_{00} = \left(2+\frac{(v_1+v_2-1)\cos\kappa}{\sqrt{v_1 v_2}}\right)\sin^2\kappa, \quad C_{ii} = 4v_i^2(v_1+v_2-1) - 8(v_1-1)v_i^{3/2}v_{3-i}^{1/2}\cos\kappa,\\
   & C_{0i} = 2\sqrt{\frac{v_i}{v_{3-i}}}(v_i + 3 v_{3-i} - 1)\sin^2\kappa, \quad C_{12} = 8v_1 v_2 (v_1+v_2)+4\sqrt{v_1 v_2}(4 v_1 v_2 - v_1 - v_2 +1)\cos\kappa\ .
\end{align*}
After introducing coordinates $t_1$ and $t_2$ as in the main text, there is an essentially unique third coordinate
\begin{equation}\label{coordinate-x}
    x = i \frac{\sqrt{-v_1} - \sqrt{-v_2}}{\sqrt{-v_1} + \sqrt{-v_2}} \sqrt{\frac{1+(\sqrt{-v_1}+\sqrt{-v_2})^2}{1+(\sqrt{-v_1}-\sqrt{-v_2})^2}}\,,
\end{equation}
such that the transformed operator in $(t_1,t_2,x)$ has vanishing coefficients with mixed partial derivatives $\partial_{t_i}\partial_x$. The freedom in $x$ is that of a redefinition $x'=x'(x)$ which does not involve variables $t_i$. Next, there is an essentially unique factor $\omega$
\begin{equation}\label{prefactor}
    \omega(t_1,t_2,x) = \frac{(\cosh 2t_1 - \cosh 2t_2)^{\frac12 (1-\Delta_1 - \Delta_2 + \Delta_{\hat3})}}{\sinh^{-1} t_1 \sinh^{-1} t_2} \left(\frac{2}{x^2+1}\sinh t_1 + \frac{2ix}{x^2+1}\sinh t_2\right)^{-\Delta_{\hat3}},
\end{equation}
such that $H = \omega\mathcal{D}\omega^{-1}$ has vanishing coefficients multiplying $\partial_{t_i}$. Similarly as above, the choice of $\omega$ is unique up to a multiplication by a function that depends only on $x$. With choices made here, a direct, if somewhat tedious, computation shows that the operator $\mathcal{D}$ is transformed into the spinning Hamiltonian \eqref{final-spinning-CS}.

\bibliographystyle{JHEP}
\bibliography{bibliography}

\end{document}